\documentclass[journal]{IEEEtran}
\usepackage[cmex10]{amsmath}
\usepackage{amsthm,amsfonts,amssymb}
\usepackage{hyperref}
\usepackage{array}    % recommended by IEEE to improve display of tables
\usepackage{fixltx2e} % fix LaTeX bug regarding ordering of floated figures
\usepackage{graphicx}
\usepackage{cite}
\usepackage{enumerate}
\usepackage{multicol}
\usepackage{nicefrac}

\newtheorem{theorem}{Theorem}
\newtheorem{corollary}{Corollary}
\newtheorem{lemma}{Lemma}
\newtheorem{proposition}{Proposition}
\newtheorem{conjecture}{Conjecture}
\theoremstyle{definition}
\newtheorem{definition}{Definition}
\newtheorem{example}{Example}
\theoremstyle{remark}
\newtheorem{remark}{Remark}

\DeclareMathOperator*{\essentialsup}{ess\,sup}  % Essential supremum
\DeclareMathOperator*{\essentialinf}{ess\,inf}  % Essential infimum

\DeclareMathOperator*{\argmin}{arg\,min}
\DeclareMathOperator{\E}{\mathbf{E}}            % Expectation
\DeclareMathOperator{\ind}{\mathbf{1}}          % Indicator function
\DeclareMathOperator{\hel}{Hel}                 % Hellinger distance

\DeclareMathOperator{\bin}{Bin}                 % Binomial distribution

\DeclareMathOperator{\union}{\cup}

\DeclareMathOperator{\intersection}{\cap}

\newcommand{\e}{\textnormal{e}}                 % Constant e
\newcommand{\der}{\mathrm{d}}
\newcommand{\intder}{\,\der}
\newcommand{\samplespace}{\mathcal{X}}
\newcommand{\salgebra}{\mathcal{F}}       % Sigma-algebra
\newcommand{\salgebraAlt}{\mathcal{G}}    % Alternative sigma-algebra
\newcommand{\partition}{\mathcal{P}}      % Partition of \samplespace
\newcommand{\partitionAlt}{\mathcal{Q}}   % Alternative partition
\newcommand{\reals}{\mathbb{R}}
\newcommand{\distset}{\mathcal{P}}        % Set of distributions
\newcommand{\alphaset}{\mathcal{A}}       % Set of alphas
\newcommand{\half}{\frac{1}{2}}
\newcommand{\thalf}{\tfrac{1}{2}}         % tiny half
\newcommand{\nicehalf}{\nicefrac{1}{2}}
\newcommand{\Qopt}{Q_\textnormal{opt}}    % capacity achieving output distribution
\newcommand{\Qoptalpha}{Q_{\textnormal{opt}(\alpha)}}    % capacity achieving output distribution,
\newcommand{\Qoptzero}{Q_{\textnormal{opt}(0)}}
\newcommand{\Qoptone}{Q_{\textnormal{opt}(1)}}
\newcommand{\Qoptinfty}{Q_{\textnormal{opt}(\infty)}}
\newcommand{\piopt}{\pi_\textnormal{opt}} % capacity achieving input distribution
\newcommand{\mltheta}{\hat{\theta}}
\newcommand{\simplex}{\Delta}
\newcommand{\normal}{\mathcal{N}}         % Normal distribution
\newcommand{\contiguous}{\vartriangleleft}
\newcommand{\mutcontiguous}{\vartriangleleft\vartriangleright}   % Mutually contiguous
\newcommand{\entiresep}{\vartriangle}     % Entirely separated
\newcommand{\conv}{\circ}                 % Convolution
\newcommand{\symdiff}{\triangle}          % Symmetric set difference
\newcommand{\sublevelset}{\mathcal{S}}
\newcommand{\comp}[1]{#1^\textnormal{c}}           % Set complement

\usepackage{color}
\title{R\'enyi Divergence and Kullback-Leibler Divergence}

\author{Tim~van~Erven\qquad\qquad 
        Peter Harremo\"es,~\IEEEmembership{Member,~IEEE}%
\thanks{Tim van Erven (tim@timvanerven.nl) is with the D\'epartement de
Math\'ematiques, Universit\'e Paris-Sud, France. Peter Harremo\"es
(harremoes@ieee.org) is with the Copenhagen Business College, Denmark.
Some of the results in this paper have previously been presented at the
ISIT 2010 conference.}
}

\begin{document}
\maketitle

\begin{abstract}
R\'{e}nyi divergence is related to R\'{e}nyi entropy much like
Kullback-Leibler divergence is related to Shannon's entropy, and comes
up in many settings. It was introduced by R\'{e}nyi as a measure of
information that satisfies almost the same axioms as Kullback-Leibler
divergence, and depends on a parameter that is called its order. In
particular, the R\'enyi divergence of order $1$ equals the
Kullback-Leibler divergence.

We review and extend the most important properties of R\'{e}nyi
divergence and Kullback-Leibler divergence, including convexity,
continuity, limits of $\sigma$-algebras and the relation of the special
order $0$ to the Gaussian dichotomy and contiguity. We also show how to
generalize the Pythagorean inequality to orders different from $1$, and
we extend the known equivalence between channel capacity and minimax
redundancy to continuous channel inputs (for all orders) and present
several other minimax results.
\end{abstract}


\IEEEpeerreviewmaketitle

\section{Introduction}

\IEEEPARstart{S}{hannon} entropy and Kullback-Leibler divergence (also
known as information divergence or relative entropy) are perhaps the two
most fundamental quantities in information theory and its applications.
Because of their success, there have been many attempts to generalize
these concepts, and in the literature one will find numerous entropy and
divergence measures. Most of these quantities have never found any
applications, and almost none of them have found an interpretation in
terms of coding. The most important exceptions are the R\'{e}nyi entropy
and R\'{e}nyi divergence \cite{Renyi1961}. Harremo\"es
\cite{Harremoes2006} and Gr\"unwald \cite[p.\,649]{Grunwald2007} provide
an operational characterization of R\'{e}nyi divergence as the number of
bits by which a mixture of two codes can be compressed; and Csisz\'ar
\cite{Csiszar1995} gives an operational characterization of R\'{e}nyi
divergence as the cut-off rate in block coding and hypothesis testing.

R\'enyi divergence appears as a crucial tool in proofs of convergence of
minimum description length and Bayesian estimators, both in parametric
and nonparametric models \cite{Zhang2006,HausslerOpper1997},
\cite[Chapter~5]{VanErven2010}, and one may recognize it implicitly in
many computations throughout information theory. It is also closely
related to Hellinger distance, which is commonly used in the analysis of
nonparametric density estimation
\cite{LeCam1973,Birge1986,VanDeGeer1993}. R\'enyi himself used his
divergence to prove the convergence of state probabilities in a
stationary Markov chain to the stationary distribution \cite{Renyi1961},
and still other applications of R\'{e}nyi divergence can be found, for
instance, in hypothesis testing \cite{MoralesPardoVajda2000}, in
multiple source adaptation \cite{MansourMohriRostamizadeh2009} and in
ranking of images \cite{HeroMaMichelGorman2003}.

Although the closely related R\'{e}nyi entropy is well studied
\cite{AczelDaroczy1975,Ben-BassatRaviv1978}, the properties of R\'{e}nyi
divergence are scattered throughout the literature and have often only
been established for finite alphabets. This paper is intended as a
reference document, which treats the most important properties of
R\'{e}nyi divergence in detail, including Kullback-Leibler divergence as
a special case. Preliminary versions of the results presented here can
be found in \cite{VanErvenHarremoes2010:isit} and \cite{VanErven2010}.
During the preparation of this paper, Shayevitz has independently
published closely related work \cite{Shayevitz2010,Shayevitz2011}.

\subsection{R\'enyi's Information Measures}\label{sec:RenyisInformationMeasures}

For finite alphabets, the \emph{R\'{e}nyi divergence} of positive order
$\alpha\neq1$ of a probability distribution $P=(p_{1},\ldots,p_{n})$
from another distribution $Q=(q_{1},\ldots,q_{n})$ is
\begin{equation}\label{eqn:sumformula}
  D_{\alpha}(P\Vert Q)
    =\frac{1}{\alpha-1}
        \ln\sum_{i=1}^{n}p_{i}^{\alpha} q_{i}^{1-\alpha},
\end{equation}
where, for $\alpha>1$, we read $p_{i}^{\alpha}q_{i}^{1-\alpha}$ as
$p_{i}^{\alpha}/q_{i}^{(\alpha-1)}$ and adopt the conventions that
$\nicefrac{0}{0}=0$ and $\nicefrac{x}{0}=\infty$ for $x>0$. As described in
Section~\ref{sec:definition}, this definition generalizes to continuous
spaces by replacing the probabilities by densities and the sum by an
integral. If $P$ and $Q$ are members of the same exponential family,
then their R\'enyi divergence can be computed using a formula by
Huzurbazar \cite{Huzurbazar1955} and Liese and Vajda
\cite[p.\,43]{LieseVajda1987}, \cite{MoralesPardoVajda2000}. Gil
provides a long list of examples \cite{Gil2011,GilAlajajiLinder2013}.

\begin{example}
Let $Q$ be a probability distribution and $A$ a set with positive
probability. Let $P$ be the conditional distribution of $Q$ given $A$.
Then
\begin{equation*}
  D_{\alpha}(P\Vert Q)=-\ln Q(A).
\end{equation*}
We observe that in this important special case the factor $\frac{1}{\alpha-1}$
in the definition of R\'{e}nyi divergence has the effect that the value of
$D_{\alpha}(P\Vert Q)$ does not depend on $\alpha$.
\end{example}

The \emph{R\'{e}nyi entropy}
\begin{equation*}
  H_{\alpha}(P)=\frac{1}{1-\alpha}\ln\sum_{i=1}^{n}p_{i}^{\alpha}
\end{equation*}
can be expressed in terms of the R\'{e}nyi divergence of $P$ from the uniform
distribution $U=(\nicefrac{1}{n},\ldots,\nicefrac{1}{n})$:
\begin{equation}\label{eqn:entropyUniform}
  H_{\alpha}(P)=H_{\alpha}(U)-D_{\alpha}(P\Vert U)=\ln n-D_{\alpha}(P\Vert U).
\end{equation}
As $\alpha$ tends to $1$, the R\'{e}nyi entropy tends to the Shannon
entropy and the R\'{e}nyi divergence tends to the Kullback-Leibler
divergence, so we recover a well-known relation. The \emph{differential
R\'{e}nyi entropy} of a distribution $P$ with density $p$ is given by
\begin{equation*}
  h_{\alpha}(P)
    =\frac{1}{1-\alpha}
      \ln\int\big(p(x)\big)^\alpha \intder x
\end{equation*}
whenever this integral is defined. If $P$ has support in an interval $I$ of length
$n$ then
\begin{equation}\label{eqn:differentialentropy}
  h_{\alpha}(P)=\ln n-D_{\alpha}(P\Vert U_{I}),
\end{equation}
where $U_{I}$ denotes the uniform distribution on $I$, and $D_\alpha$ is
the generalization of R\'enyi divergence to densities, which will be
defined formally in Section~\ref{sec:definition}. Thus the
properties of both the R\'{e}nyi entropy and the differential R\'{e}nyi
entropy can be deduced from the properties of R\'enyi divergence as long
as $P$ has compact support.

There is another way of relating R\'{e}nyi entropy and R\'{e}nyi divergence,
in which entropy is considered as self-information. Let $X$ denote a discrete
random variable with distribution $P$, and let $P_{\text{diag}}$ be the
distribution of $(X,X)$. Then
\begin{equation}\label{eqn:entropyDiag}
  H_{\alpha}(P)
    = D_{2-\alpha}(P_{\text{diag}}\Vert P\times P).
\end{equation}
For $\alpha$ tending to $1$, the right-hand side tends to the mutual
information between $X$ and itself, and again a well-known formula is recovered.

\subsection{Special Orders}

Although one can define the R\'{e}nyi divergence of any order, certain values
have wider application than others. Of particular interest are the values $0$,
$\nicehalf$, $1$, $2$, and $\infty$.

\begin{figure}[ptb]
\centering
\includegraphics{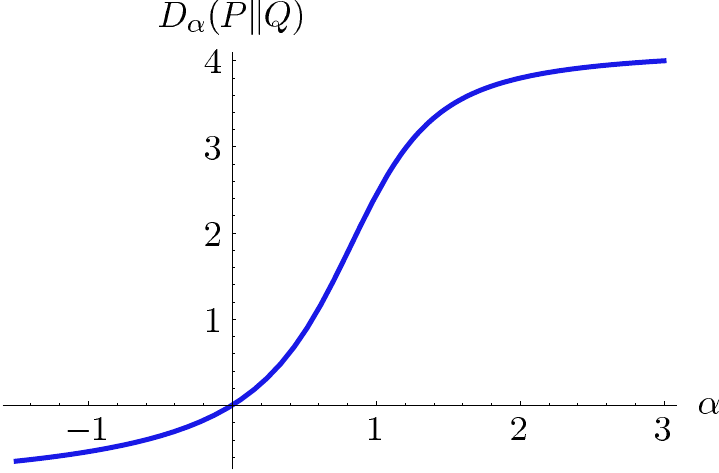}
\caption{R\'enyi divergence as a function of its order for fixed
distributions}%
\label{fig:VaryAlpha}
%$P=(2/3,1/3)$ and $Q=(1/100,99/100)$
\end{figure}

The values $0,$ $1,$ and $\infty$ are \emph{extended orders} in the
sense that R\'{e}nyi divergence of these orders cannot be calculated by
plugging into \eqref{eqn:sumformula}. Instead, their definitions are
determined by continuity in $\alpha$ (see Figure~\ref{fig:VaryAlpha}).
This leads to defining R\'{e}nyi divergence of order $1$ as the
Kullback-Leibler divergence. For order $0$ it becomes $-\ln Q(\{i\mid
p_{i}>0\}),$ which is closely related to absolute continuity and
contiguity of the distributions $P$ and $Q$ (see
Section~\ref{sec:absolutecontinuity}). For order $\infty$, R\'{e}nyi
divergence is defined as $\ln\max_i\frac{p_i}{q_i}$. In the literature
on the \emph{minimum description length principle} in statistics, this
is called the \emph{worst-case regret} of coding with $Q$ rather than
with $P$ \cite{Grunwald2007}. The R\'{e}nyi divergence of order $\infty$
is also related to the \emph{separation distance}, used by Aldous and
Diaconis \cite{AldousDiaconis1987} to bound the rate of convergence to
the stationary distribution for certain Markov chains.

Only for $\alpha=\nicefrac{1}{2}$ is R\'{e}nyi divergence symmetric in its
arguments. Although not itself a metric, it is a function of the squared
\emph{Hellinger distance}
$\hel^{2}(P,Q)=\sum_{i=1}^{n}\big(p_i^{\nicefrac{1}{2}}-q_i^{\nicefrac{1}{2}}\big)^2$
\cite{GibbsSu2002}:
\begin{equation}\label{eqn:relationHellinger}
  D_{\nicefrac{1}{2}}(P\Vert Q)
    =-2\ln\left(1-\frac{\hel^{2}(P,Q)}{2}\right). 
\end{equation}
Similarly, for $\alpha=2$ it satisfies
\begin{equation}\label{eqn:relationChiSquared}
  D_2(P\Vert Q)=\ln\left(1+\chi^{2}(P,Q)\right),
\end{equation}
where $\chi^{2}(P,Q)=\sum_{i=1}^{n}\frac{(p_{i}-q_{i})^{2}}{q_{i}}$
denotes the \emph{$\chi^2$-divergence}
%(with the convention that $0/0=0$)
\cite{GibbsSu2002}. It will be shown that R\'{e}nyi divergence is
nondecreasing in its order. Therefore, by $\ln t\leq t-1,$
\eqref{eqn:relationHellinger} and \eqref{eqn:relationChiSquared} imply
that
\begin{multline}\label{eqn:relations}
  \hel^{2}(P,Q)
    \leq D_{\nicefrac{1}{2}}(P\Vert Q)
    \leq D_{1}(P\Vert Q)\\
    \leq D_{2}(P\Vert Q)
    \leq\chi^{2}(P,Q).
\end{multline}
Finally, Gilardoni \cite{Gilardoni2010} shows that R\'enyi divergence is
related to the \emph{total variation distance}\footnote{N.B. It is also common to define the total
variation distance as $\half V(P,Q)$. See the discussion by Pollard
\cite[p.\,60]{Pollard2002}. Our definition is consistent with the
literature on Pinsker's inequality.} $V(P,Q) = \sum_{i=1}^n \lvert
p_i - q_i \rvert$ by a generalization of \emph{Pinsker's
inequality}:
\begin{equation}\label{eqn:relationVariation}
  \frac{\alpha}{2} V^2(P,Q) \leq D_\alpha(P\|Q)
  \qquad \text{for $\alpha \in (0,1]$.}
\end{equation}
(See Theorem~\ref{thm:pinsker} below.) For $\alpha=1$ this is the normal
version of Pinsker's inequality, which bounds total variation distance
in terms of the square root of the Kullback-Leibler divergence.

\subsection{Outline}

The rest of the paper is organized as follows. First, in
Section~\ref{sec:definition}, we extend the definition of R\'{e}nyi
divergence from formula \eqref{eqn:sumformula} to continuous spaces. One can
either define R\'{e}nyi divergence via an integral or via
discretizations. We demonstrate that these definitions are equivalent.
Then we show that R\'{e}nyi divergence extends to the extended orders
$0$, $1$ and $\infty$ in the same way as for finite spaces. Along the
way, we also study its behaviour as a function of $\alpha$.
By contrast, in Section~\ref{sec:ExtendedOrdersFixedOrder} we study
various convexity and continuity properties of R\'{e}nyi divergence as a
function of $P$ and $Q$, while $\alpha$ is kept fixed. We also
generalize the Pythagorean inequality to any order $\alpha \in
(0,\infty)$.
Section~\ref{sec:minimax} contains several minimax results, and treats
the connection to Chernoff information in hypothesis testing, to which
many applications of R\'{e}nyi divergence are related. We also discuss
the equivalence of channel capacity and the minimax redundancy for all
orders $\alpha$.
Then, in Section~\ref{sec:negativeorders}, we show how R\'{e}nyi
divergence extends to negative orders. These are related to the orders
$\alpha>1$ by a negative scaling factor and a reversal of the arguments
$P$ and $Q$. Finally, Section~\ref{sec:counterexamples} contains a
number of counterexamples, showing that properties that hold for certain
other divergences are violated by R\'{e}nyi divergence.

For fixed $\alpha$, R\'{e}nyi divergence is related to various forms of
\emph{power divergences}, which are in the well-studied class of
\emph{$f$-divergences} \cite{LieseVajda2006}. Consequently, several of
the results we are presenting for fixed $\alpha$ in
Section~\ref{sec:ExtendedOrdersFixedOrder} are equivalent to known
results about power divergences. To make this presentation
self-contained we avoid the use of such connections and only use general
results from measure theory.

% Per request of a referee, add a box that summarizes the results of the
% paper
\begin{table*}[tbp]
  % Prevent stretching above displays inside summary box
  \setlength{\abovedisplayskip}{1.5ex plus0pt minus1pt}
  \setlength{\belowdisplayskip}{\abovedisplayskip}
  % Set-up box around summary, height is calibrated manually
  \fbox{%
  \begin{minipage}[t][68\baselineskip][t]{\textwidth}
    \centerline{{\large\textbf{Summary}}}
    \vspace{1.5\baselineskip}

    \begin{multicols}{2}
      \textit{Definition for the \emph{simple orders}
      $\alpha \in (0,1) \union\, (1,\infty)$:}
      \begin{equation*}
        D_\alpha(P\|Q)
          = \tfrac{1}{\alpha-1}\ln\int
          p^{\alpha}q^{1-\alpha}\intder\mu.
      \end{equation*}
      \textit{For the extended orders (Thms~\ref{thm:AlphaZero}--\ref{thm:AlphaInfty}):}
      \begin{align*}
        D_0(P\|Q)&= -\ln Q(p > 0)\\
        D_1(P\|Q) &= D(P\|Q) = \text{ Kullback-Leibler divergence}\\
        D_\infty(P\|Q) &= \ln \Big(\essentialsup_P \frac{p}{q}\Big)
        = \text{worst-case regret.}
      \end{align*}
      \textit{Equivalent definition via discretization
      (Thm~\ref{thm:supfinite}):}
      \begin{equation*}
        D_{\alpha}(P\Vert Q)
          =\sup_{\partition \in \text{finite partitions}}
          \!\!\!\!\!\!\!\!D_\alpha(P_{\lvert\partition}\Vert Q_{\lvert\partition}).
      \end{equation*}
      
      \textit{Relations to (differential) R\'enyi entropy
      % NB Expressing the following equations as a range is a recipe for
      % mistakes, because there are unnumbered equations in between,
      % which we may decide to give a number later on.
      \big(\eqref{eqn:entropyUniform}, \eqref{eqn:differentialentropy},
      \eqref{eqn:entropyDiag}\big):} For $\alpha \in
      [0,\infty]$,
      \begin{align*}
        H_{\alpha}(P)&=\ln |\samplespace|-D_{\alpha}(P\Vert U)
          = D_{2-\alpha}(P_{\text{diag}}\Vert P\times P)
        \qquad \text{for finite $\samplespace$,}\\
        h_{\alpha}(P)&=\ln n -D_{\alpha}(P\Vert U_{I})
        \qquad \text{if $\samplespace$ is an interval $I$ of length
        $n$.}
      \end{align*}

      \textit{Relations to other divergences
      \big(\eqref{eqn:relationHellinger}--\eqref{eqn:relations},
      Remark~\ref{rem:relations}\big) and
      Pinsker's inequality (Thm~\ref{thm:pinsker}):}
        \begin{align*}
        \hel^2 \leq D_{\nicefrac{1}{2}} &\leq D \leq D_2 \leq \chi^2\\
          \frac{\alpha}{2} V^2 &\leq D_\alpha
          \qquad \text{for $\alpha \in (0,1]$.}
        \end{align*}

      \textit{Relation to Fisher information
      (Section~\ref{sec:taylorparametric}):} For a parametric
      statistical model $\{P_\theta \mid \theta \in \Theta \subseteq
      \reals\}$ with ``sufficiently regular'' parametrisation,
      \begin{equation*}
        \lim_{\theta^{\prime}\rightarrow\theta}
          \frac{1}{(\theta-\theta^{\prime})^{2}}
          D_{\alpha}(P_{\theta}\Vert P_{\theta^{\prime}})
        =\frac{\alpha}{2}J(\theta)
        \qquad \text{for $\alpha \in (0,\infty)$.}
      \end{equation*}

      \textit{Varying the order (Thms~\ref{thm:increasinginorder},
      \ref{thm:extendedcontinuity}, Corollary~\ref{cor:concavealpha}):}
      \begin{itemize}
      \item $D_\alpha$ is nondecreasing in $\alpha$, often strictly so.
      \item $D_\alpha$ is continuous in $\alpha$ on $[0,1] \union
      \{\alpha \in (1,\infty] \mid D_\alpha < \infty\}$.
      \item $(1-\alpha)D_\alpha$ is concave in $\alpha$ on $[0,\infty]$.
      \end{itemize}

      \textit{Positivity (Thm~\ref{thm:positivity}) and skew symmetry
      (Proposition~\ref{prop:simplesymmetry}):}
      \begin{itemize}
      \item $D_\alpha \geq 0$ for $\alpha \in [0,\infty]$, often
      strictly so.
      \item $D_{\alpha}(P\Vert Q)=\tfrac{\alpha}{1-\alpha}
      D_{1-\alpha}(Q\Vert P)$ for $0 < \alpha < 1$.
      \end{itemize}

      \textit{Convexity
      (Thms~\ref{thm:PQconvex}--\ref{thm:PQquasiconvex}):} $D_\alpha(P\|Q)$ is
      \begin{itemize}
      \item jointly convex in $(P,Q)$ for $\alpha \in [0,1]$,
      \item convex in $Q$ for $\alpha \in [0,\infty]$,
      \item jointly quasi-convex in $(P,Q)$ for $\alpha
      \in [0,\infty]$.
      \end{itemize}

      \textit{Pythagorean inequality (Thm~\ref{thm:pythagoras}):}
      For $\alpha \in (0,\infty)$, let $\distset$ be an $\alpha$-convex
      set of distributions and let $Q$ be an arbitrary distribution.
      If the \emph{$\alpha$-information projection} $P^\ast = \argmin_{P
      \in \distset} D_\alpha(P\|Q)$ exists, then
      \begin{equation*}
        D_\alpha(P\|Q) \geq D_\alpha(P\|P^\ast) + D_\alpha(P^\ast\|Q)
        \qquad \text{for all $P \in \distset$.}
      \end{equation*}

      \textit{Data processing (Thm~\ref{thm:extendeddataprocessing},
      Example~\ref{ex:dataprocessing}):} If we fix the transition
      probabilities $A(Y|X)$ in a Markov chain $X \to Y$, then
      \begin{equation*}
        D_\alpha(P_Y \| Q_Y) \leq D_\alpha\big(P_X \| Q_X\big)
        \qquad \text{for $\alpha \in [0,\infty]$.}
      \end{equation*}

      \begin{samepage}
      \textit{The topology of setwise convergence
      (Thms~\ref{thm:lowersemicontinuity}, \ref{thm:Qcontinuity}):}
      \begin{itemize}
      \item $D_\alpha(P\|Q)$ is lower semi-continuous in the pair $(P,Q)$ for
      $\alpha \in (0,\infty]$.
      \item If $\samplespace$ is finite, then $D_\alpha(P\|Q)$ is
      continuous in $Q$ for $\alpha \in [0,\infty]$.
      \end{itemize}
      \end{samepage}

      \begin{samepage}
      \textit{The total variation topology (Thm~\ref{thm:continuity},
      Corollary~\ref{cor:uppersemicontinuity0}):}
      \begin{itemize}
      \item $D_\alpha(P\|Q)$ is uniformly continuous in $(P,Q)$ for $\alpha \in (0,1)$.
      \item $D_0(P\|Q)$ is upper semi-continuous in $(P,Q)$.
      \end{itemize}
      \end{samepage}

      \begin{samepage}
      \textit{The weak topology
      (Thms~\ref{thm:weaklowersemicontinuity},
      \ref{thm:compactsublevelsets}):} Suppose $\samplespace$ is a
      Polish space. Then
      \begin{itemize}
      \item $D_\alpha(P\|Q)$ is lower semi-continuous in the pair
      $(P,Q)$ for $\alpha \in (0,\infty]$;
      \item The sublevel set $\{P \mid D_\alpha(P\|Q) \leq c\}$ is
      convex and compact for $c \in [0,\infty)$ and $\alpha \in
      [1,\infty]$.
      \end{itemize}
      \end{samepage}

      \textit{Orders $\alpha \in (0,1)$ are all equivalent
      (Thm~\ref{thm:equivalentopologies}):}
      \begin{equation*}
        \tfrac{\alpha}{\beta} \tfrac{1-\beta}{1-\alpha} D_\beta
          \leq D_\alpha \leq D_\beta
          \qquad \text{for $0 < \alpha \leq \beta < 1$.}
      \end{equation*}

      \begin{samepage}
      \textit{Additivity and other consistent sequences of distributions
      (Thms~\ref{thm:consistentdistributions}, \ref{thm:additivity}):}
      \begin{itemize}
      \item For arbitrary distributions $P_1,P_2,\ldots$ and
      $Q_1,Q_2,\ldots$, let $P^N = P_{1}\times\cdots\times P_{N}$ and
      $Q^N = Q_{1}\times\cdots\times Q_{N}$. Then
      \begin{equation*}
      \sum_{n=1}^{N} D_{\alpha}(P_{n}\Vert Q_{n})
        =D_{\alpha}(P^N\Vert Q^N)
        \quad \begin{cases}
          \text{for $\alpha \in [0,\infty]$} & \text{if $N < \infty$,}\\
          \text{for $\alpha \in (0,\infty]$} & \text{if $N = \infty$.}
        \end{cases}
      \end{equation*}
      \item Let $P^1,P^2,\ldots$ and $Q^1,Q^2,\ldots$ be consistent
      sequences of distributions on $n = 1,2,\ldots$ outcomes. Then
      \begin{equation*}
        D_{\alpha}(P^{n}\Vert Q^{n})
          \to D_{\alpha}(P^{\infty}\Vert Q^{\infty})
          \qquad \text{for $\alpha \in (0,\infty]$.}
      \end{equation*}
      \end{itemize}
      \end{samepage}

      \textit{Limits of $\sigma$-algebras
      (Thms~\ref{thm:LimitAlgebraIncreasing},
      \ref{thm:LimitAlgebraDecreasing}):} 
      \begin{itemize}
      \item For $\sigma$-algebras $\salgebra_1
      \subseteq\salgebra_2\subseteq\cdots \subseteq \salgebra$ and
      $\salgebra_\infty=\sigma\left(\bigcup_{n=1}^{\infty}\salgebra_{n}\right)$,
      \begin{equation*}
        \lim_{n\rightarrow\infty} D_{\alpha}(P_{\lvert\salgebra_{n}}\Vert Q_{\lvert\salgebra_{n}})
          =D_{\alpha}(P_{\lvert\salgebra_{\infty}}\Vert Q_{\lvert\salgebra_{\infty}})
          \qquad \text{for $\alpha \in (0,\infty]$.}
      \end{equation*}
      \item For $\sigma$-algebras
      $\salgebra \supseteq \salgebra_{1}\supseteq\salgebra_{2}\supseteq\cdots$ and 
      $\salgebra_{\infty}=\bigcap _{n=1}^{\infty}\salgebra_{n}$,
      \begin{equation*}
        \lim_{n\rightarrow\infty} D_{\alpha}(P_{\lvert\salgebra_{n}}\Vert
        Q_{\lvert\salgebra_{n}}) = D_{\alpha}(P_{\lvert\salgebra_{\infty}} \Vert
        Q_{\lvert\salgebra_{\infty}})
        \qquad \text{for $\alpha \in [0,1)$}
      \end{equation*}
      and also for $\alpha \in [1,\infty)$ if
      $D_{\alpha}(P_{\lvert\salgebra_{m}}\Vert Q_{\lvert \salgebra_{m}})
      < \infty$ for some $m$.
      \end{itemize}

      \textit{Absolute continuity and mutual singularity
      (Thms~\ref{thm:abscont}, \ref{thm:mutsing}, \ref{thm:contiguity},
      \ref{thm:entireseparation}):}
      \begin{itemize}
      \item $P \ll Q$ if and only if $D_0(P\|Q) = 0$.
      \item $P \perp Q$ if and only if $D_\alpha(P\|Q) = \infty$ for
      some/all $\alpha \in [0,1)$.
      \item These properties generalize to contiguity and entire
      separation.
      \end{itemize}

      \textit{Hypothesis testing and Chernoff information
      (Thms~\ref{thm:MomentGeneratingFunction},
      \ref{thm:ChernoffInformationDivergence}):} If
      $\alpha$ is a simple order, then
      \begin{equation*}
        (1-\alpha)D_{\alpha}(P\Vert Q)
          =\inf_{R}\left\{  \alpha D(R\Vert
          P)+(1-\alpha)D(R\Vert Q)\right\}.
      \end{equation*}
      Suppose $D(P\|Q) < \infty$. Then the Chernoff information
      satisfies
      \begin{multline*}
        \sup_{\alpha \in (0,\infty)}\inf_{R}\left\{\alpha D(R\Vert
        P)+(1-\alpha)D(R\Vert Q)\right\}\\
        =\inf_{R}\sup_{\alpha \in (0,\infty)}\left\{  \alpha D(R\Vert
        P)+(1-\alpha)D(R\Vert Q)\right\},
      \end{multline*}
      and, under regularity conditions, both sides equal
      $D(P_{\alpha^*}\|P) = D(P_{\alpha^*}\|Q)$.

      \vspace{\belowdisplayskip}
      \textit{Channel capacity and minimax redundancy
      (Thms~\ref{thm:radius}, \ref{thm:center},
      \ref{thm:capacityequalityInfty},
      \ref{thm:piopt},
      Lemma~\ref{lem:supredundancyfunction},
      Conjecture~\ref{conj:capacityinequality}):} Suppose $\samplespace$ is
      finite. Then, for $\alpha \in [0,\infty]$,
      \begin{itemize}
      \item The channel capacity $C_\alpha$ equals the minimax
      redundancy $R_\alpha$;
      \item There exists $\Qopt$ such that $\sup_\theta D(P_\theta \|
      \Qopt) = R_\alpha$;
      \item If there exists a capacity achieving input distribution
      $\piopt$, then $D(P_\theta \| \Qopt) = R_\alpha$ almost surely for
      $\theta$ drawn from $\piopt$;
      \item If $\alpha = \infty$ and the maximum likelihood is 
      achieved by $\mltheta(x)$, then $\piopt(\theta) = \Qopt(\{x \mid
      \mltheta(x) = \theta\})$ is a capacity achieving input
      distribution;
      \end{itemize}
      Suppose $\samplespace$ is countable and $R_\infty < \infty$. Then,
      for $\alpha = \infty$, $\Qopt$ is the Shtarkov distribution
      defined in \eqref{eqn:shtarkov} and
      \begin{equation*}
        \sup_\theta D_\infty(P_\theta\| Q)
          = R_\infty
           +D_\infty(\Qopt\| Q)
        \qquad \text{for all $Q$.}
      \end{equation*}
      We conjecture that this generalizes to a one-sided inequality for
      any $\alpha > 0$.

      \vspace{\belowdisplayskip}
      \textit{Negative orders (Lemma~\ref{lem:symmetry},
      Thms~\ref{thm:increasinginordernegative},
      \ref{thm:negativecontinuity}):}
      \begin{itemize}
      \item Results for positive $\alpha$ carry over, but
      often with reversed properties.
      \item $D_\alpha$ is nondecreasing in $\alpha$ on
      $[-\infty,\infty]$.
      \item $D_\alpha$ is continuous in $\alpha$ on $[0,1] \union
      \{\alpha \mid -\infty < D_\alpha < \infty\}$.
      \end{itemize}

      \textit{Counterexamples (Section~\ref{sec:counterexamples}):}
      \begin{itemize}
      \item $D_\alpha(P\|Q)$ is not convex in $P$ for $\alpha > 1$.
      \item For $\alpha \in (0,1)$, $D_\alpha(P\|Q)$ is not continuous
      in $(P,Q)$ in the topology of setwise convergence.
      \item $D_\alpha$ is not (the square of) a metric.
      \end{itemize}
    \end{multicols}
  \end{minipage}
  }
\end{table*}

\section{Definition of R\'{e}nyi divergence}

\label{sec:definition}

Let us fix the notation to be used throughout the paper. We consider
(probability) measures on a measurable space $(\samplespace,\salgebra)$.
If $P$ is a measure on $(\samplespace,\salgebra)$, then we write
$P_{\lvert\salgebraAlt}$ for its restriction to the sub-$\sigma$-algebra
$\salgebraAlt\subseteq\salgebra$, which may be interpreted as the
marginal of $P$ on the subset of events $\salgebraAlt$. A measure $P$ is
called \emph{absolutely continuous} with respect to another measure $Q$
if $P(A)=0$ whenever $Q(A)=0$ for all events $A\in\salgebra$. We will
write $P\ll Q$ if $P$ is absolutely continuous with respect to $Q$ and
$P\not \ll Q$ otherwise. Alternatively, $P$ and $Q$ may be
\emph{mutually singular}, denoted $P\perp Q,$ which means that there
exists an event $A\in\salgebra$ such that $P(A)=0$ and $Q(\samplespace
\setminus A)=0$. We will assume that all (probability) measures are
absolutely continuous with respect to a common $\sigma$-finite measure
$\mu,$ which is arbitrary in the sense that none of our definitions or
results depend on the choice of $\mu$. As we only consider (mixtures of)
a countable number of distributions, such a measure $\mu$ exists in all
cases, so this is no restriction. For measures denoted by capital
letters (e.g.\ $P$ or $Q$), we will use the corresponding lower-case
letters (e.g.\ $p,q$) to refer to their densities with respect to $\mu$.
This includes the setting with a finite alphabet from the introduction
by taking $\mu$ to be the counting measure, so that $p$ and $q$ are
probability mass functions. Using that densities are random variables,
we write, for example, $\int p^\alpha q^{1-\alpha} \der \mu$ instead of
its lengthy equivalent $\int p(x)^\alpha q(x)^{1-\alpha} \der\mu(x)$.
For any event $A\in\salgebra$, $\text{\textbf{1}}_{A}$ denotes its
indicator function, which is $1$ on $A$ and $0$ otherwise. Finally, we
use the natural logarithm in our definitions, such that information is
measured in nats ($1$ bit equals $\ln 2$ nats). 

We will often need to distinguish between the orders for which R\'{e}nyi
divergence can be defined by a generalization of formula
\eqref{eqn:sumformula} to an integral over densities, and the other orders.
This motivates the following definitions.

\begin{definition}
We call a (finite) real number $\alpha$ a \emph{simple order} if $\alpha
> 0$ and $\alpha\neq1$. The values $0$, $1$, and $\infty$ are called
\emph{extended orders}.
\end{definition}

\subsection{Definition by Formula for Simple Orders}

Let $P$ and $Q$ be two arbitrary distributions on
$(\samplespace,\salgebra)$. The formula in \eqref{eqn:sumformula}, which
defines R\'enyi divergence for simple orders on finite sample spaces,
generalizes to arbitrary spaces as follows:
\begin{definition}
[Simple Orders]\label{def:renyidivergence} For any simple order
$\alpha,$ the \emph{R\'{e}nyi divergence} of \emph{order} $\alpha$ of
$P$ from $Q$ is defined as
\begin{equation}\label{eqn:commondefinition}%
  D_{\alpha}(P\Vert Q)
    =\frac{1}{\alpha-1}\ln\int p^{\alpha}q^{1-\alpha}\intder\mu,
\end{equation}
where, for $\alpha>1,$ we read $p^{\alpha}q^{1-\alpha}$ as
$\frac{p^{\alpha}}{q^{\alpha-1}}$ and adopt the conventions that
$\nicefrac{0}{0}=0$ and $\nicefrac{x}{0}=\infty$ for $x>0$.
\end{definition}

For example, for any simple order $\alpha$, the R\'enyi divergence
of a normal distribution (with mean $\mu_0$ and positive variance
$\sigma_0^2$) from another normal distribution (with mean $\mu_1$ and
positive variance $\sigma_1^2$) is
\begin{multline}\label{eqn:renyinormal}
  D_\alpha\Big(\normal(\mu_0,\sigma_0^2)\|\normal(\mu_1,\sigma_1^2)\Big)\\
    = \frac{\alpha(\mu_1 - \mu_0)^2}{2\sigma_\alpha^2}
      + \frac{1}{1-\alpha} \ln
      \frac{\sigma_\alpha}{\sigma_0^{1-\alpha}\sigma_1^\alpha},
\end{multline}
provided that $\sigma_\alpha^2 = (1-\alpha)\sigma_0^2 + \alpha
\sigma_1^2 > 0$ \cite[p.\,45]{LieseVajda1987}.

\begin{remark}\label{rem:relations}
The interpretation of $p^{\alpha}q^{1-\alpha}$ in
Definition~\ref{def:renyidivergence} is such that the \emph{Hellinger
integral} $\int p^{\alpha}q^{1-\alpha}\intder\mu$ is an $f$-divergence
\cite{LieseVajda2006}, which ensures that the relations from the
introduction to squared Hellinger distance
\eqref{eqn:relationHellinger} and $\chi^{2}$-distance
\eqref{eqn:relationChiSquared} hold in general, not just for finite
sample spaces.
\end{remark}

For simple orders, we may always change to integration with respect to
$P$:
\begin{equation*}
  \int p^{\alpha}q^{1-\alpha}\intder\mu
    =\int\left(\frac{q}{p}\right)^{1-\alpha}\der P,
\end{equation*}
which shows that our definition does not depend on the choice of
dominating measure $\mu$. In most cases it is also equivalent to
integrate with respect to $Q$:
\begin{equation*}
  \int p^{\alpha}q^{1-\alpha}\intder\mu
    =\int\left(\frac{p}{q}\right)^{\alpha}\der Q
     \qquad (0<\alpha<1 \text{ or } P\ll Q).
\end{equation*}
However, if $\alpha>1$ and $P\not \ll Q,$ then $D_{\alpha}(P\Vert
Q)=\infty,$ whereas the integral with respect to $Q$ may be finite. This
is a subtle consequence of our conventions. For example, if
$P=(\nicefrac{1}{2},\nicefrac{1}{2})$, $Q = (1,0)$ and $\mu$ is the counting measure, then for
$\alpha > 1$
%
% NB I really prefer to write (1/2)^alpha instead of 2^{-alpha}, because
% we are explaining a very subtle notational convention related to
% exponents
\begin{equation}
  \int p^{\alpha}q^{1-\alpha}\intder\mu
    = \frac{(\nicefrac{1}{2})^\alpha}{1^{\alpha-1}} +
      \frac{(\nicefrac{1}{2})^\alpha}{0^{\alpha-1}}
    = \infty,
\end{equation}
but
\begin{equation}
  \int\left(\frac{p}{q}\right)^{\alpha}\der Q
    = \int_{q > 0} \left(\frac{p}{q}\right)^{\alpha}\der Q
    = \frac{(\nicefrac{1}{2})^\alpha}{1^{\alpha-1}} = 2^{-\alpha}.
\end{equation}

\subsection{Definition via Discretization for Simple Orders}

We shall repeatedly use the following result, which is a direct
consequence of the Radon-Nikod\'{y}m theorem \cite{Shiryaev1996}:
\begin{proposition}
\label{prop:SubAlgebraExpectation} Suppose $\lambda\ll\mu$ is a probability
distribution, or any countably additive measure such that $\lambda
(\samplespace)\leq1$. Then for any sub-$\sigma$-algebra $\salgebraAlt%
\subseteq\salgebra$
\begin{equation*}
\frac{\der\lambda_{|\salgebraAlt}}{\der\mu_{|\salgebraAlt}}
  =\E\left[  \left.  \frac{\der\lambda}{\der\mu}\right\vert
   \salgebraAlt\right]  \qquad\text{($\mu$-a.s.)}
\end{equation*}
\end{proposition}

It has been argued that grouping observations together (by considering a
coarser $\sigma$-algebra), should not increase our ability to
distinguish between $P$ and $Q$ under any measure of divergence
\cite{AliSilvey1966}. This is expressed by the \emph{data processing
inequality}, which R\'{e}nyi divergence satisfies:
\begin{theorem}[Data Processing Inequality]\label{thm:simpledataprocessing}
For any simple order $\alpha$ and any sub-$\sigma$-algebra $\salgebraAlt
\subseteq\salgebra$
\begin{equation*}
  D_{\alpha}(P_{\lvert\salgebraAlt}\Vert Q_{\lvert\salgebraAlt})
    \leq D_\alpha(P\Vert Q).
\end{equation*}
\end{theorem}

Theorem~\ref{thm:extendeddataprocessing} below shows that the data
processing inequality also holds for the extended orders.

\begin{example}\label{ex:dataprocessing}
  The name ``data processing inequality'' stems from the following
  application of Theorem~\ref{thm:simpledataprocessing}. Let $X$ and $Y$
  be two random variables that form a Markov chain
  \begin{equation*}
    X \to Y,
  \end{equation*}
  where the conditional distribution of $Y$ given $X$ is $A(Y|X)$. Then
  if $Y = f(X)$ is a deterministic function of $X$, we may view $Y$ as
  the result of ``processing'' $X$ according to the function $f$. In
  general, we may also process $X$ using a nondeterministic function,
  such that $A(Y|X)$ is not a point-mass.

  Suppose $P_X$ and $Q_X$ are distributions for $X$. Let $P_X \conv A$
  and $Q_X \conv A$ denote the corresponding joint distributions, and
  let $P_Y$ and $Q_Y$ be the induced marginal distributions for $Y$.
  Then the reader may verify that $D_\alpha(P_X \conv A \| Q_X \conv A)
  = D_\alpha(P_X\| Q_X)$, and consequently the data processing
  inequality implies that processing $X$ to obtain $Y$ reduces R\'enyi
  divergence:
  \begin{equation}\label{eqn:exampledataprocessing}
    D_\alpha(P_Y \| Q_Y)
      \leq D_\alpha(P_X \conv A \| Q_X \conv A)
      = D_\alpha(P_X \| Q_X).
  \end{equation}
\end{example}

\begin{IEEEproof}[Proof of Theorem~\ref{thm:simpledataprocessing}]
Let $\tilde{P}$ denote the absolutely continuous component of $P$ with respect
to $Q$. Then by Proposition~\ref{prop:SubAlgebraExpectation} and Jensen's
inequality for conditional expectations
\begin{equation}\label{eqn:dataprocessingderivation}
  \begin{split}
  \frac{1}{\alpha-1}\ln\int
    &\left(\frac{\der\tilde{P}_{|\salgebraAlt}}{\der Q_{|\salgebraAlt}}\right)^{\alpha}\der Q\\
    &=\frac{1}{\alpha-1}\ln\int
     \left(\E\left[ \left. \frac{\der\tilde{P}}{\der Q}
      \right\vert \salgebraAlt\right] \right)^{\alpha}\der Q\\
  &\leq\frac{1}{\alpha-1}\ln\int\E\left[  \left.  \left(  \frac
{\der\tilde{P}}{\der Q}\right)  ^{\alpha}\right\vert \salgebraAlt\right]  \der Q\\
  &=\frac{1}{\alpha-1}\ln\int\left(\frac{\der \tilde{P}}
  {\der Q}\right)^{\alpha}\der Q.
  \end{split}
\end{equation}
If $0<\alpha<1$, then $p^{\alpha}q^{1-\alpha}=0$ if $q=0$, so the
restriction of $P$ to $\tilde{P}$ does not change the R\'{e}nyi
divergence, and hence the theorem is proved. Alternatively, suppose
$\alpha>1$. If $P\ll Q,$ then $\tilde{P}=P$ and the theorem again
follows from \eqref{eqn:dataprocessingderivation}. If $P\not \ll Q$,
then $D_{\alpha }(P\Vert Q)=\infty$ and the theorem holds as well.
\end{IEEEproof}

The next theorem shows that if $\samplespace$ is a continuous space,
then the R\'enyi divergence on $\samplespace$ can be arbitrarily well
approximated by the R\'enyi divergence on finite partitions of
$\samplespace$. For any finite or countable partition
$\partition=\left\{  A_{1},A_{2},\ldots\right\}$ of $\samplespace$, let
$P_{\lvert\partition}\equiv P_{\lvert\sigma(\partition)}$ and
$Q_{\lvert\partition}\equiv Q_{\lvert \sigma(\partition)}$ denote the
restrictions of $P$ and $Q$ to the $\sigma $-algebra generated by
$\partition$.
\begin{theorem}\label{thm:simplesupfinite}
For any simple order $\alpha$
\begin{equation}\label{eqn:supfinite}
  D_{\alpha}(P\Vert Q)
    =\sup_{\partition} D_\alpha(P_{\lvert\partition}\Vert Q_{\lvert\partition}),
\end{equation}
where the supremum is over all finite partitions $\partition \subseteq
\salgebra$.
\end{theorem}

It follows that it would be equivalent to first define R\'{e}nyi
divergence for finite sample spaces and then extend the definition to
arbitrary sample spaces using \eqref{eqn:supfinite}.

The identity \eqref{eqn:supfinite} also holds for the extended orders
$1$ and $\infty$. (See Theorem~\ref{thm:supfinite} below.)

\begin{IEEEproof}
[Proof of Theorem~\ref{thm:simplesupfinite}] By the data processing inequality
\begin{equation*}
  \sup_{\partition}D_{\alpha}(P_{\lvert\partition}\Vert Q_{\lvert\partition})
    \leq D_{\alpha}(P\Vert Q).
\end{equation*}
To show the converse inequality, consider for any $\varepsilon>0$ a
discretization of the densities $p$ and $q$ into a countable number of
bins
\begin{align*}
  B_{m,n}^{\varepsilon}
    =\{x\in\samplespace\mid \e^{m\varepsilon}&\leq p(x)<\e^{(m+1)\varepsilon},\\
    \e^{n\varepsilon}&\leq q(x)<\e^{(n+1)\varepsilon}\},
\end{align*}
where $n,m\in\{-\infty,\ldots,-1,0,1,\ldots\}$. Let
$\partitionAlt^{\varepsilon}=\{B_{m,n}^{\varepsilon}\}$ and
$\salgebra^{\varepsilon}=\sigma(\partitionAlt^{\varepsilon})\subseteq\salgebra$
be the corresponding partition and $\sigma$-algebra, and let
$p_{\varepsilon}=\der P_{\lvert\partitionAlt^{\varepsilon}}/\der\mu$ and
$q_{\varepsilon}=\der Q_{\lvert\partitionAlt^{\varepsilon}}/\der\mu$ be
the densities of $P$ and $Q$ restricted to $\salgebra^{\varepsilon}$.
Then by Proposition~\ref{prop:SubAlgebraExpectation}
\begin{equation*}
  \frac{q_{\varepsilon}}{p_{\varepsilon}}
    =\frac{\E[q\mid\salgebra^{\varepsilon}]}{\E[p\mid\salgebra^{\varepsilon}]}
    \leq\frac{q}{p}\e^{2\varepsilon} \qquad\text{($P$-a.s.)}
\end{equation*}
It follows that
\begin{equation*}
  \frac{1}{\alpha-1}\ln\int
    \left(\frac{q_{\varepsilon}}{p_{\varepsilon}}\right)^{1-\alpha}\der P
    \geq\frac{1}{\alpha-1}\ln\int\left(
      \frac{q}{p}\right)^{1-\alpha}\der P - 2\varepsilon,
\end{equation*}
and hence the supremum over all countable partitions is large enough:
\begin{equation*}
  \sup_{\overset{\text{countable }\partitionAlt}{\sigma(Q)\subseteq\salgebra}}
    D_{\alpha}(P_{\lvert\partitionAlt}\Vert Q_{\lvert\partitionAlt})
  \geq \sup_{\varepsilon>0}
    D_{\alpha}(P_{\lvert\partitionAlt^{\varepsilon}}\Vert Q_{\lvert\partitionAlt^{\varepsilon}})
  \geq D_{\alpha}(P\Vert Q).
\end{equation*}
It remains to show that the supremum over finite partitions is at least as
large. To this end, suppose $\partitionAlt=\{B_{1},B_{2},\ldots\}$ is
any countable partition and let
$\partition_{n}=\{B_{1},\ldots,B_{n-1},\bigcup_{i\geq n}B_{i}\}$. Then
by
\begin{align*}
  P\Big(\bigcup_{i\geq n}B_{i}\Big)^\alpha
    Q\Big(\bigcup_{i\geq n}B_{i}\Big)^{1-\alpha}
    & \geq 0 \qquad (\alpha>1),\\
  \lim_{n\rightarrow\infty}P\Big(\bigcup_{i\geq n}B_{i}\Big)^\alpha
    Q\Big(\bigcup_{i\geq n}B_{i}\Big)^{1-\alpha}
    &=0 \qquad (0<\alpha<1),
\end{align*}
we find that
\begin{align*}
  \lim_{n\rightarrow\infty}
    &D_{\alpha}(P_{\lvert\partition_{n}}\Vert Q_{\lvert\partition_{n}})
  =\lim_{n\rightarrow\infty}\frac{1}{\alpha-1}\ln
    \sum_{B\in\partition_{n}}P(B)^{\alpha}Q(B)^{1-\alpha}\\
  &\geq\lim_{n\rightarrow\infty}\frac{1}{\alpha-1}\ln
    \sum_{i=1}^{n-1} P(B_{i})^{\alpha}Q(B_{i})^{1-\alpha}\\
  &=D_{\alpha}(P_{\lvert\partitionAlt}\Vert Q_{\lvert\partitionAlt}),
\end{align*}
where the inequality holds with equality if $0 < \alpha < 1$.
\end{IEEEproof}

\subsection{Extended Orders: Varying the Order}
\label{sec:ExtendedOrdersVaryingOrder}

As for finite alphabets, continuity considerations lead to the following
extensions of R\'enyi divergence to orders for which it cannot be
defined using the formula in \eqref{eqn:commondefinition}.
\begin{definition}
[Extended Orders]\label{def:extendedorders} The \emph{R\'{e}nyi divergences}
of orders $0$ and $1$ are defined as
\begin{align*}
D_{0}(P\Vert Q)  &= \lim_{\alpha\downarrow0}D_{\alpha}(P\Vert Q),\\
D_{1}(P\Vert Q)  &= \lim_{\alpha\uparrow1}D_{\alpha}(P\Vert Q),
\end{align*}
and the R\'enyi divergence of order $\infty$ is defined as
\begin{equation*}
  D_{\infty}(P\Vert Q)
    = \lim_{\alpha\uparrow\infty}D_{\alpha}(P\Vert Q).
\end{equation*}
\end{definition}

Our definition of $D_{0}$ follows Csisz\'ar \cite{Csiszar1995}. It
differs from R\'{e}nyi's original definition \cite{Renyi1961}, which
uses \eqref{eqn:commondefinition} with $\alpha=0$ plugged in and is
therefore always zero. As illustrated by
Section~\ref{sec:absolutecontinuity}, the present definition is more
interesting.

The limits in Definition~\ref{def:extendedorders} always exist, because
R\'enyi divergence is nondecreasing in its order:
\begin{theorem}[Increasing in the Order]\label{thm:increasinginorder}
For $\alpha \in [0,\infty]$ the R\'{e}nyi divergence $D_{\alpha}(P\Vert
Q)$ is nondecreasing in $\alpha$. On $\alphaset=\{\alpha \in
[0,\infty]\mid 0 \leq\alpha\leq1 \text{ or } D_{\alpha}(P\Vert
Q)<\infty\}$ it is constant if and only if $P$ is the conditional
distribution $Q(\cdot \mid A)$ for some event $A \in \salgebra$.
\end{theorem}

\begin{IEEEproof}
Let $\alpha<\beta$ be simple orders. Then for $x\geq0$ the function $x\mapsto
x^{\frac{(\alpha-1)}{(\beta-1)}}$ is strictly convex if $\alpha<1$ and
strictly concave if $\alpha>1$. Therefore by Jensen's inequality
\begin{align*}
  \frac{1}{\alpha-1}\ln\int p^{\alpha}q^{1-\alpha}\intder\mu
    &=\frac{1}{\alpha-1}\ln\int
      \left(\frac{q}{p}\right)^{(1-\beta)
      \frac{\alpha-1}{\beta-1}}\der P\\
    &\leq\frac{1}{\beta-1}\ln\int
      \left(\frac{q}{p}\right)^{1-\beta}\der P.
\end{align*}
On $\alphaset$, $\int(\nicefrac{q}{p})^{1-\beta}\der P$ is finite.
As a consequence, Jensen's inequality holds with equality if and only if
$(\nicefrac{q}{p})^{1-\beta}$ is constant $P$-a.s., which is
equivalent to $\nicefrac{q}{p}$ being constant $P$-a.s., which in turn means
that $P = Q(\cdot \mid A)$ for some event $A$.

From the simple orders, the result extends to the extended orders by the
following observations:
\begin{align*}
D_{0}(P\|Q)  &  = \inf_{0 < \alpha< 1} D_{\alpha}(P\|Q),\\
D_{1}(P\|Q)  &  = \sup_{0 < \alpha< 1} D_{\alpha}(P\|Q) \leq\inf_{\alpha> 1}
D_{\alpha}(P\|Q),\\
D_{\infty}(P\|Q)  &  = \sup_{\alpha> 1} D_{\alpha}(P\|Q).\qedhere
\end{align*}
\end{IEEEproof}

Let us verify that the limits in Definition~\ref{def:extendedorders} can
be expressed in closed form, just like for finite alphabets. We require
the following lemma:
\begin{lemma}\label{lem:dominatedconvergence}
Let $\alphaset = \{\alpha \text{ a simple order}
\mid 0 < \alpha <1$  or $D_\alpha(P\|Q) < \infty\}$. Then,
for any sequence $\alpha_1,\alpha_2,\ldots\in \alphaset$ such that
$\alpha_n\rightarrow\beta\in \alphaset \union\{0,1\}$,
\begin{equation}\label{eqn:dominatedconvergence}
  \lim_{n \to\infty} \int p^{\alpha_n}q^{1-\alpha_n}\intder\mu
  = \int\lim_{n \to\infty} p^{\alpha_n}q^{1-\alpha_n}\intder\mu.
\end{equation}
\end{lemma}

Our proof extends a proof by Shiryaev \cite[pp.\ 366--367]{Shiryaev1996}.

\begin{IEEEproof}
We will verify the conditions for the dominated convergence theorem
\cite{Shiryaev1996}, from which \eqref{eqn:dominatedconvergence}
follows. First suppose $0\leq\beta<1$. Then $0<\alpha_{n}<1$ for all
sufficiently large $n$. In this case $p^{\alpha_{n}}q^{1-\alpha_{n}}$,
which is never negative, does not exceed
$\alpha_{n}p+(1-\alpha_{n})q\leq p+q$, and the dominated convergence
theorem applies because $\int(p+q)\intder\mu=2<\infty$. Secondly,
suppose $\beta\geq1$. Then there exists a $\gamma\geq\beta$ such that $\gamma\in
\alphaset\cup\{1\}$ and $\alpha_n\leq\gamma$ for all sufficiently large
$n$. If $\gamma=1$, then $\alpha_{n}<1$ and we are done by the same
argument as above. So suppose $\gamma>1$. Then convexity of
$p^{\alpha_{n}}q^{1-\alpha_{n}}$ in $\alpha_{n}$ implies that for
$\alpha_{n}\leq\gamma$
\begin{equation*}
  p^{\alpha_{n}}q^{1-\alpha_{n}}
    \leq (1-\frac{\alpha_{n}}{\gamma}) p^0 q^1
        +\frac{\alpha_n}{\gamma}p^{\gamma}q^{1-\gamma}
    \leq q+p^\gamma q^{1-\gamma}.
\end{equation*}
Since $\int q\intder\mu=1$, it remains to show that $\int p^\gamma
q^{1-\gamma}\intder\mu<\infty$, which is implied by $\gamma>1$ and
$D_{\gamma}(P\Vert Q)<\infty$.
\end{IEEEproof}

The closed-form expression for $\alpha = 0$ follows immediately:
\begin{theorem}[$\alpha=0$]\label{thm:AlphaZero}
\begin{equation*}
  D_0(P\Vert Q)=-\ln Q(p>0).
\end{equation*}
\end{theorem}

\begin{IEEEproof}[Proof of Theorem~\ref{thm:AlphaZero}]
By Lemma~\ref{lem:dominatedconvergence} and the fact that $\lim_{\alpha
\downarrow0}p^{\alpha}q^{1-\alpha}=\text{\textbf{1}}_{\{p>0\}}q$.
\end{IEEEproof}

For $\alpha = 1$, the limit in Definition~\ref{def:extendedorders} equals
the \emph{Kullback-Leibler divergence} of $P$ from $Q$, which is defined
as
\begin{equation*}
  D(P\Vert Q)=\int p\ln\frac{p}{q}\intder\mu,
\end{equation*}
with the conventions that $0\ln(\nicefrac{0}{q})=0$ and $p\ln
(\nicefrac{p}{0})=\infty$ if $p>0$. Consequently, $D(P\Vert Q)=\infty$ if
$P\not \ll Q$.

\begin{theorem}[$\alpha=1$]\label{thm:AlphaOne}
\begin{equation}\label{eqn:AlphaUpToOne}
D_1(P \Vert Q) =D(P\Vert Q).
\end{equation}
Moreover, if $D(P\Vert Q)=\infty$ or there exists a $\beta>1$ such that
$D_{\beta}(P\Vert Q)<\infty$, then also
\begin{equation}\label{eqn:AlphaDownToOne}
  \lim_{\alpha\downarrow1} D_{\alpha}(P\Vert Q) = D(P\Vert Q).
\end{equation}
\end{theorem}

For example, by letting $\alpha \uparrow 1$ in \eqref{eqn:renyinormal}
or by direct computation, it can be derived \cite{LieseVajda1987} that
the Kullback-Leibler divergence between two normal distributions with
positive variance is
\begin{multline*}
  D_1\Big(\normal(\mu_0,\sigma_0^2)\|\normal(\mu_1,\sigma_1^2)\Big)\\
    = \half\Big(\frac{(\mu_1 - \mu_0)^2}{\sigma_1^2}
      + \ln \frac{\sigma_1^2}{\sigma_0^2}
      + \frac{\sigma_0^2}{\sigma_1^2} - 1\Big).
\end{multline*}

It is possible that $D_{\alpha}(P\Vert Q)=\infty$ for all $\alpha>1$,
but $D(P\Vert Q)<\infty$, such that \eqref{eqn:AlphaDownToOne} does not
hold. This situation occurs, for example, if $P$ is doubly exponential
on $\samplespace=\reals$ with density $p(x)=\e^{-2\lvert x\rvert}$ and
$Q$ is standard normal with density $q(x)=\e^{-x^2/2}/\sqrt{2 \pi}$.
(Liese and Vajda \cite{LieseVajda2006} have previously used these
distributions in a similar example.) In this case there is no way to
make R\'{e}nyi divergence continuous in $\alpha$ at $\alpha=1$, and we
opt to define $D_{1}$ as the limit from below, such that it always
equals the Kullback-Leibler divergence.

The proof of Theorem~\ref{thm:AlphaOne} requires an intermediate lemma:

\begin{lemma}\label{lem:logtaylor}
For any $x>\nicefrac{1}{2}$
\begin{equation*}
  (x-1)\left(  1+\frac{1-x}{2}\right)  \leq\ln x\leq x-1.
\end{equation*}
\end{lemma}

\begin{IEEEproof}
By Taylor's theorem with Cauchy's remainder term we have for any positive $x$
that
\begin{align*}
\ln x  &  =x-1-\frac{(x-\xi)(x-1)}{2\xi^{2}}\\
&  =(x-1)\left(  1+\frac{\xi-x}{2\xi^{2}}\right)
\end{align*}
for some $\xi$ between $x$ and $1$. As $\frac{\xi-x}{2\xi^{2}}$ is increasing
in $\xi$ for $x > \nicehalf$, the lemma follows.
\end{IEEEproof}

\begin{IEEEproof}
[Proof of Theorem~\ref{thm:AlphaOne}]Suppose $P\not \ll Q$. Then $D(P\Vert
Q)=\infty=D_{\beta}(P\Vert Q)$ for all $\beta>1$, so
\eqref{eqn:AlphaDownToOne} holds. Let $x_{\alpha}=\int
p^{\alpha}q^{1-\alpha}\intder\mu$. Then
$\lim_{\alpha\uparrow1}x_{\alpha}=P(q > 0)$ by
Lemma~\ref{lem:dominatedconvergence}, and hence
\eqref{eqn:AlphaUpToOne} follows by
\begin{multline*}
  \lim_{\alpha\uparrow1}\frac{1}{\alpha-1}\ln\int
  p^{\alpha}q^{1-\alpha}\intder\mu\\
  =\lim_{\alpha\uparrow1}\frac{1}{\alpha-1}\ln P(q>0)=\infty=D(P\Vert
Q).
\end{multline*}
Alternatively, suppose $P\ll Q$. Then
$\lim_{\alpha\uparrow1}x_{\alpha}=1$ and therefore
Lemma~\ref{lem:logtaylor} implies that
\begin{multline} \label{eqn:hellingerdivergenceDown}%
\lim_{\alpha\uparrow1} D_{\alpha}(P\Vert Q)
  =\lim_{\alpha\uparrow1}\frac {1}{\alpha-1}\ln x_{\alpha}\\
  =\lim_{\alpha\uparrow1}\frac{x_{\alpha}-1}{\alpha-1}
  =\lim_{\alpha\uparrow1}\int_{p,q>0}
   \frac{p-p^{\alpha}q^{1-\alpha}}{1-\alpha}\intder\mu,
\end{multline}
where the restriction of the domain of integration is allowed because $q=0$
implies $p=0$ ($\mu$-a.s.) by $P\ll Q$. Convexity of $p^{\alpha}q^{1-\alpha}$
in $\alpha$ implies that its derivative, $p^{\alpha}q^{1-\alpha}\ln\frac{p}{q}$, is nondecreasing and therefore for $p,q>0$
\[
\frac{p-p^{\alpha}q^{1-\alpha}}{1-\alpha}=\frac{1}{1-\alpha}\int_{\alpha}%
^{1}p^{z}q^{1-z}\ln\frac{p}{q}\intder z
\]
is nondecreasing in $\alpha$, and $\frac{p-p^{\alpha}q^{1-\alpha}}{1-\alpha
}\geq\frac{p-p^{0}q^{1-0}}{1-0}=p-q$. As $\int_{p,q>0}(p-q)$~\textrm{d}%
$\mu>-\infty$, it follows by the monotone convergence theorem that
\begin{align*}
\lim_{\alpha\uparrow1}\int_{p,q>0}\frac{p-p^{\alpha}q^{1-\alpha}}{1-\alpha
}\intder\mu &  =\int_{p,q>0}\lim_{\alpha\uparrow1}\frac{p-p^{\alpha
}q^{1-\alpha}}{1-\alpha}\intder\mu\\
&  =\int_{p,q>0}p\ln\frac{p}{q}\intder\mu=D(P\Vert Q),
\end{align*}
which together with \eqref{eqn:hellingerdivergenceDown} proves
\eqref{eqn:AlphaUpToOne}. If $D(P\Vert Q)=\infty$, then $D_{\beta}(P\Vert
Q)\geq D(P\Vert Q)=\infty$ for all $\beta>1$ and \eqref{eqn:AlphaDownToOne}
holds. It remains to prove \eqref{eqn:AlphaDownToOne} if there exists a
$\beta>1$ such that $D_{\beta}(P\Vert Q)<\infty$. In this case, arguments
similar to the ones above imply that
\begin{equation}
\lim_{\alpha\downarrow1}D_{\alpha}(P\Vert Q)=\lim_{\alpha\downarrow1}%
\int_{p,q>0}\frac{p^{\alpha}q^{1-\alpha}-p}{\alpha-1}\intder%
\mu\label{eqn:hellingerdivergenceUp}%
\end{equation}
and $\frac{p^{\alpha}q^{1-\alpha}-p}{\alpha-1}$ is nondecreasing in $\alpha$.
Therefore $\frac{p^{\alpha}q^{1-\alpha}-p}{\alpha-1}\leq\frac{p^{\beta
}q^{1-\beta}-p}{\beta-1}\leq\frac{p^{\beta}q^{1-\beta}}{\beta-1}$ and, as
$\int_{p,q>0}\frac{p^{\beta}q^{1-\beta}}{\beta-1}$~\textrm{d}$\mu<\infty$ is
implied by $D_{\beta}(P\Vert Q)<\infty$, it follows by the monotone
convergence theorem that
\begin{align*}
\lim_{\alpha\downarrow1}\int_{p,q>0}\frac{p^{\alpha}q^{1-\alpha}-p}{\alpha
-1}\intder\mu &  =\int_{p,q>0}\lim_{\alpha\downarrow1}\frac{p^{\alpha
}q^{1-\alpha}-p}{\alpha-1}\intder\mu\\
&  =\int_{p,q>0}p\ln\frac{p}{q}\intder\mu=D(P\Vert Q),
\end{align*}
which together with \eqref{eqn:hellingerdivergenceUp} completes the proof.
\end{IEEEproof}

For any random variable $X$, the \emph{essential supremum} of $X$ with
respect to $P$ is $\essentialsup_P X = \sup \{c \mid P(X > c) > 0\}$.
\begin{theorem}
[$\alpha=\infty$]\label{thm:AlphaInfty}
\[
D_{\infty}(P\Vert Q)
  = \ln\sup_{A \in \salgebra} \frac{P(A)}{Q(A)}
  = \ln\left( \essentialsup_P \frac{p}{q}\right),
\]
with the conventions that $\nicefrac{0}{0} = 0$ and $\nicefrac{x}{0} = \infty$ if $x > 0$.
\end{theorem}

If the sample space $\samplespace$ is countable, then with the
notational conventions of this theorem the essential supremum reduces to
an ordinary supremum, and we have $D_\infty(P\|Q) = \ln \sup_x
\frac{P(x)}{Q(x)}$.

\begin{IEEEproof}
If $\samplespace$ contains a finite number of elements $n$, then
\begin{align*}
D_{\infty}(P\Vert Q)
  &=\lim_{\alpha\uparrow\infty}\frac{1}{\alpha-1}\ln
\sum_{i=1}^{n} p_{i}^{\alpha}q_{i}^{1-\alpha}\\
  &=\ln\max_{i}\frac{p_{i}}{q_{i}}
  =\ln\max_{A\subseteq\samplespace}\frac{P(A)}{Q(A)}.
\end{align*}
This extends to arbitrary measurable spaces $(\samplespace,\salgebra)$ by
Theorem~\ref{thm:simplesupfinite}:
\begin{align*}
D_{\infty}(P\Vert Q)
  &=\sup_{\alpha<\infty}\sup_{\partition}
    D_{\alpha}(P_{\lvert\partition}\Vert Q_{\lvert\partition})\\
  &=\sup_{\partition} \sup_{\alpha<\infty}
    D_{\alpha}(P_{\lvert\partition}\Vert Q_{\lvert \partition})\\
  &=\sup_{\partition}\ln\max_{A\in\partition}\frac{P(A)}{Q(A)}
  =\ln\sup_{A\in\salgebra}\frac{P(A)}{Q(A)},
\end{align*}
where $\partition$ ranges over all finite partitions in $\salgebra$.

Now if $P \not \ll Q$, then there exists an event $B \in \salgebra$ such
that $P(B) > 0$ but $Q(B) = 0$, and
\begin{equation*}
  P\Big( \frac{p}{q} = \infty\Big) = P(q = 0) \geq P(B) > 0
\end{equation*}
implies that $\essentialsup \nicefrac{p}{q} = \infty = \sup_A
\frac{P(A)}{Q(A)}$. Alternatively, suppose that $P \ll Q$. Then
\begin{equation*}
P\left(  A\right)
  =\int_{A \intersection \{q > 0\}} \kern-2em p\intder\mu
  \leq\int_{A \intersection \{q > 0\}}\kern -2em \essentialsup\frac{p}{q}\cdot q\intder\mu
  =\essentialsup\frac{p}{q}\cdot Q\left(  A\right)
\end{equation*}
for all $A\in\salgebra$ and it follows that
\begin{equation}\label{eqn:esssupbigger}
  \sup_{A\in\salgebra}\frac{P(A)}{Q(A)}\leq \essentialsup\frac{p}{q}.
\end{equation}
Let $a< \essentialsup \nicefrac{p}{q}$ be arbitrary. Then there exists a set
$A\in\salgebra$ with $P\left( A\right)  >0$ such that $\nicefrac{p}{q}\geq
a$ on $A$ and therefore%
\begin{equation*}
  P\left(  A\right)
    = \int_{A}p\intder\mu
    \geq\int_{A}a\cdot q\intder\mu
    =a\cdot Q\left(A\right).
\end{equation*}
Thus $\sup_{A\in\salgebra}\frac{P(A)}{Q(A)}\geq a$ for
any $a<\essentialsup\nicefrac{p}{q}$, which implies that
\[
\sup_{A\in\salgebra}\frac{P(A)}{Q(A)}\geq \essentialsup\frac{p}{q}.
\]
In combination with \eqref{eqn:esssupbigger} this completes the proof.
\end{IEEEproof}

Taken together, the previous results imply that R\'enyi divergence is a
continuous function of its order $\alpha$ (under suitable conditions):
\begin{theorem}
[Continuity in the Order]\label{thm:extendedcontinuity} The R\'{e}nyi
divergence $D_{\alpha}(P\Vert Q)$ is continuous in $\alpha$ on
$\alphaset=\{\alpha \in [0,\infty]\mid0\leq\alpha\leq1\text{ or }D_{\alpha}(P\Vert Q)<\infty\}$.
\end{theorem}

\begin{IEEEproof}
  Continuity at any simple order $\beta$ follows by
  Lemma~\ref{lem:dominatedconvergence}. It extends to the extended
  orders $0$ and $\infty$ by the definition of R\'enyi divergence at
  these orders. And it extends to $\alpha=1$ by
  Theorem~\ref{thm:AlphaOne}.
\end{IEEEproof}

\section{Fixed Nonnegative Orders}
\label{sec:ExtendedOrdersFixedOrder}

In this section we fix the order $\alpha$ and study properties of
R\'{e}nyi divergence as $P$ and $Q$ are varied. First we prove
nonnegativity and extend the data processing inequality and the relation
to a supremum over finite partitions to the extended orders. Then we study
convexity, we prove a generalization of the Pythagorean inequality to
general orders, and finally we consider various types of continuity.

\subsection{Positivity, Data Processing and Finite Partitions}

\begin{theorem}
[Positivity]\label{thm:positivity} For any order $\alpha \in [0,\infty]$
\[
D_{\alpha}(P\|Q)\geq0.
\]
For $\alpha> 0$, $D_{\alpha}(P\|Q) = 0$ if and only if $P = Q$. For $\alpha=
0$, $D_{\alpha}(P\|Q) = 0$ if and only if $Q \ll P$.
\end{theorem}

\begin{IEEEproof}
  Suppose first that $\alpha$ is a simple order. Then by Jensen's
  inequality
  \begin{align*}
  \frac{1}{\alpha-1}\ln\int p^{\alpha}q^{1-\alpha}\intder\mu
    &=\frac {1}{\alpha-1}\ln\int\left(
      \frac{q}{p}\right)^{1-\alpha}\intder P\\
    &\geq\frac{1-\alpha}{\alpha-1}\ln\int\frac{q}{p}\intder P
    \geq0.
  \end{align*}
  Equality holds if and only if $q/p$ is constant $P$-a.s.\ (first inequality)
  and $Q\ll P$ (second inequality), which together is equivalent to $P=Q$.

  The result extends to $\alpha\in\{1,\infty\}$ by $D_{\alpha}(P\Vert
  Q)=\sup_{\beta<\alpha}D_{\beta}(P\Vert Q)$. For $\alpha=0$ it can be
  verified directly that $-\ln Q(p>0)\geq0$, with equality if and only
  if $Q\ll P$.
\end{IEEEproof}

\begin{theorem}
[Data Processing Inequality]\label{thm:extendeddataprocessing} For any
order $\alpha \in [0,\infty]$ and any sub-$\sigma$-algebra $\salgebraAlt
\subseteq\salgebra$
\begin{equation}
\label{eqn:extendeddataprocessing}D_{\alpha}(P_{\lvert\salgebraAlt}\Vert
Q_{\lvert\salgebraAlt}) \leq D_{\alpha}(P\Vert Q).
\end{equation}
\end{theorem}

Example~\ref{ex:dataprocessing} also applies to the extended orders
without modification.

\begin{IEEEproof}
By Theorem~\ref{thm:simpledataprocessing}, \eqref{eqn:extendeddataprocessing}
holds for the simple orders. Let $\beta$ be any extended order and let
$\alpha_{n} \to\beta$ be an arbitrary sequence of simple orders that converges
to $\beta$, from above if $\beta= 0$ and from below if $\beta\in\{1,\infty\}$.
Then
\begin{align*}
D_{\beta}(P_{\lvert\salgebraAlt}\Vert Q_{\lvert\salgebraAlt})
  &= \lim_{n \to\infty} D_{\alpha_{n}}(P_{\lvert\salgebraAlt}\Vert
  Q_{\lvert\salgebraAlt})\\
  &\leq\lim_{n \to\infty} D_{\alpha_{n}}(P\Vert Q)
  = D_{\beta}(P\Vert Q).\qedhere
\end{align*}
\end{IEEEproof}

\begin{theorem}\label{thm:supfinite}
For any $\alpha \in [0,\infty]$
\begin{equation*}
  D_{\alpha}(P\Vert Q)
    =\sup_{\partition} D_\alpha(P_{\lvert\partition}\Vert Q_{\lvert\partition}),
\end{equation*}
where the supremum is over all finite partitions $\partition \subseteq
\salgebra$.
\end{theorem}

\begin{IEEEproof}
  For simple orders $\alpha$, the result holds by
  Theorem~\ref{thm:simplesupfinite}. This extends to $\alpha \in
  \{1,\infty\}$ by monotonicity and left-continuity in $\alpha$:
  \begin{align*}
    D_\alpha(P\Vert Q)
      &= \sup_{\beta < \alpha} D_\beta(P\Vert Q)
      = \sup_{\beta < \alpha} \sup_{\partition}
      D_\beta(P_{\lvert\partition}\Vert Q_{\lvert\partition})\\
      &= \sup_{\partition}\sup_{\beta < \alpha} D_\beta(P_{\lvert\partition}\Vert Q_{\lvert\partition})
      = \sup_{\partition} D_\alpha(P_{\lvert\partition}\Vert
      Q_{\lvert\partition}).\qedhere
  \end{align*}
  For $\alpha = 0$, the data processing inequality implies that
  \begin{equation*}
  D_{\alpha}(P\Vert Q)
    \geq \sup_{\partition} D_\alpha(P_{\lvert\partition}\Vert Q_{\lvert\partition}),
  \end{equation*}
  and equality is achieved for the partition $\partition = \{p > 0, p =
  0\}$.
\end{IEEEproof}

\subsection{Convexity}
\label{sec:convexity}

\begin{figure}[ptb]
\centering
\includegraphics{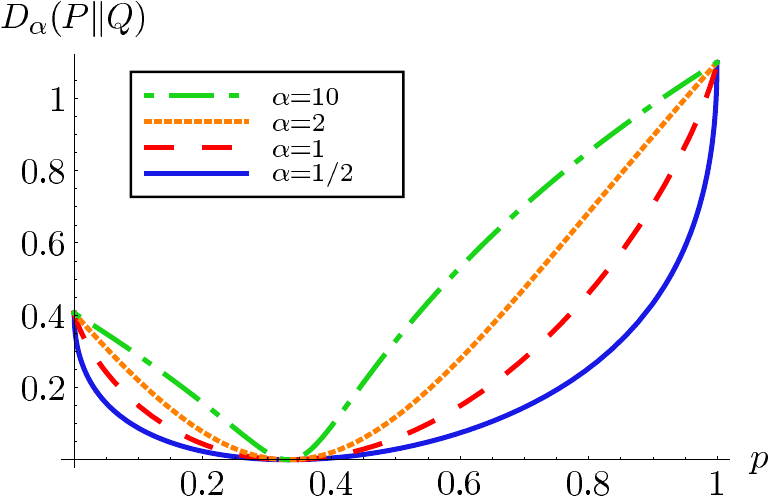}
\caption{R\'enyi divergence as a function of $P=(p,1-p)$ for
$Q=(\nicefrac{1}{3},\nicefrac{2}{3})$}%
\label{fig:RenyiBinarySpace}%
\end{figure}

\begin{figure}[ptb]
\centering
\includegraphics{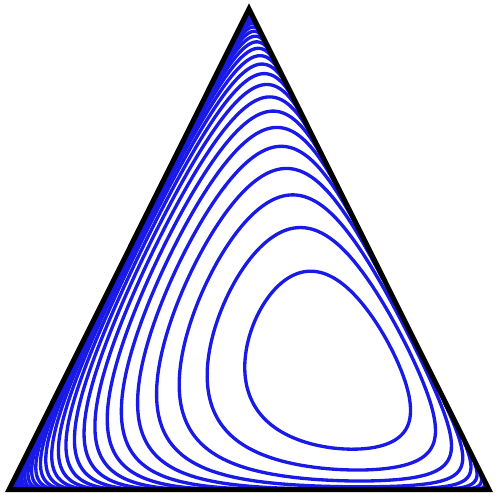}
\caption{Level curves of $D_{\nicefrac{1}{2}}(P\Vert Q)$ for fixed $Q$ as $P$ ranges over
the simplex of distributions on a three-element set}%
\label{fig:RenyiTernarySpace}
%$Q=(1/5,5/9,11/45)$
\end{figure}

Consider Figures~\ref{fig:RenyiBinarySpace} and
\ref{fig:RenyiTernarySpace}. They show $D_{\alpha}(P\Vert Q)$ as a
function of $P$ for sample spaces containing two or three elements.
These figures suggest that R\'{e}nyi divergence is convex in its first
argument for small $\alpha$, but not for large $\alpha$. This is in
agreement with the well-known fact that it is jointly convex in the pair
$(P,Q)$ for $\alpha=1$. It turns out that joint convexity extends to
$\alpha<1$, but not to $\alpha>1$, as noted by Csisz\'ar
\cite{Csiszar1995}. Our proof generalizes the proof for $\alpha=1$ by
Cover and Thomas \cite{CoverThomas1991}.

\begin{theorem}\label{thm:PQconvex}
For any order $\alpha \in [0,1]$ R\'{e}nyi divergence is jointly convex in its
arguments. That is, for any two pairs of probability distributions
$(P_{0},Q_{0})$ and $(P_{1},Q_{1})$, and any $0<\lambda<1$
\begin{equation}\label{eqn:convexitysmallalpha}
  \begin{split}
D_{\alpha}\big(  (1-\lambda)&P_{0}+\lambda P_{1}\Vert(1-\lambda)Q_{0}+\lambda
Q_{1}\big)\\
&\leq(1-\lambda)D_{\alpha}(P_{0}\Vert Q_{0})+\lambda D_{\alpha}(P_{1}\Vert Q_{1}).
  \end{split}
\end{equation}
Equality holds if and only if
\begin{flalign*}
  \text{$\alpha = 0$: } & 
  \text{$D_0(P_{0}\Vert Q_{0}) = D_0(P_{1}\Vert Q_{1})$,} & \\
  & \quad\text{$p_0 = 0 \Rightarrow p_1 = 0$ ($Q_0$-a.s.) and} \\
  & \quad\text{$p_1 = 0 \Rightarrow p_0 = 0$ ($Q_1$-a.s.);} & \\
  \text{$0 < \alpha < 1$: } &
  \text{$D_{\alpha}(P_{0}\Vert Q_{0})
    = D_{\alpha}(P_{1}\Vert Q_{1})$ and} \\
  & \quad\text{$p_0 q_1 = p_1 q_0$ ($\mu$-a.s.);}\\
  \text{$\alpha = 1$: } & \text{$p_0 q_1 = p_1 q_0$ ($\mu$-a.s.)}
\end{flalign*}
\end{theorem}

\begin{IEEEproof}
Suppose first that $\alpha=0$, and let
$P_{\lambda}=(1-\lambda)P_{0}+\lambda P_{1}$ and $Q_{\lambda}
=(1-\lambda)Q_{0}+\lambda Q_{1}$. Then
\begin{align*}
  (1-\lambda)&\ln Q_{0}(p_{0}>0)+\lambda\ln Q_{1}(p_{1}>0)\\
  &\leq\ln\left( \left(  1-\lambda\right)  Q_{0}(p_{0}>0)+\lambda
  Q_{1}\left(  p_{1}>0\right) \right) \\
  &\leq\ln Q_{\lambda}\big(p_{0}>0\text{ or }p_{1}>0\big)
    =\ln Q_{\lambda}(p_{\lambda}>0).
\end{align*}
Equality holds if and only if, for the first inequality, $Q_{0}(p_{0}%
>0)=Q_{1}(p_{1}>0)$ and, for the second inequality, $p_{1}>0\Rightarrow
p_{0}>0$ ($Q_{0}$-a.s.) and $p_{0}>0\Rightarrow p_{1}>0$ ($Q_{1}$-a.s.) These
conditions are equivalent to the equality conditions of the theorem.

Alternatively, suppose $\alpha>0$. We will show that point-wise
\begin{equation}\label{eqn:pointwise}
\begin{aligned}
  (1-\lambda) p_0^\alpha q_0^{1-\alpha} + \lambda p_1^\alpha q_1^{1-\alpha}
    &\leq p_\lambda^\alpha q_\lambda^{1-\alpha} &(0<\alpha<1);\\
  (1-\lambda) p_0 \ln \frac{p_0}{q_0} + \lambda p_1 \ln \frac{p_1}{q_1}
    &\geq p_\lambda \ln \frac{p_\lambda}{q_\lambda} &(\alpha = 1),
    \end{aligned}
\end{equation}
where $p_{\lambda}=(1-\lambda)p_{0}+\lambda p_{1}$ and $q_{\lambda}%
=(1-\lambda)q_{0}+\lambda q_{1}$. For $\alpha=1$,
\eqref{eqn:convexitysmallalpha} then follows directly; for $0<\alpha<1$,
\eqref{eqn:convexitysmallalpha} follows from \eqref{eqn:pointwise} by Jensen's
inequality:
\begin{multline}
(1-\lambda)\ln\int p_{0}^{\alpha}q_{0}^{1-\alpha}\intder\mu+\lambda
\ln\int p_{1}^{\alpha}q_{1}^{1-\alpha}\intder\mu\label{eqn:jumpthelog}\\
\leq\ln\left(  (1-\lambda)\int p_{0}^{\alpha}q_{0}^{1-\alpha}\intder%
\mu+\lambda\int p_{1}^{\alpha}q_{1}^{1-\alpha}\intder\mu\right).
\end{multline}
If one of $p_{0},p_{1},q_{0}$ and $q_{1}$ is zero, then \eqref{eqn:pointwise}
can be verified directly. So assume that they are all positive. Then for
$0<\alpha<1$ let $f(x)=-x^{\alpha}$ and for $\alpha=1$ let $f(x)=x\ln x$, such
that \eqref{eqn:pointwise} can be written as
\[
\frac{(1-\lambda)q_{0}}{q_{\lambda}}f\left(  \frac{p_{0}}{q_{0}}\right)
+\frac{\lambda q_{1}}{q_{\lambda}}f\left(  \frac{p_{1}}{q_{1}}\right)  \geq
f\left(  \frac{p_{\lambda}}{q_{\lambda}}\right).
\]
\eqref{eqn:pointwise} is established by recognising this as an application of
Jensen's inequality to the strictly convex function $f$. Regardless of whether
any of $p_{0},p_{1},q_{0}$ and $q_{1}$ is zero, equality holds in
\eqref{eqn:pointwise} if and only if $p_{0}q_{1}=p_{1}q_{0}$. Equality holds
in \eqref{eqn:jumpthelog} if and only if $\int p_{0}^{\alpha}q_{0}^{1-\alpha
}\intder\mu=\int p_{1}^{\alpha}q_{1}^{1-\alpha}\intder\mu$, which is
equivalent to $D_{\alpha}(P_{0}\Vert Q_{0})=D_{\alpha}(P_{1}\Vert Q_{1})$.
\end{IEEEproof}

Joint convexity in $P$ and $Q$ breaks down for $\alpha>1$ (see
Section~\ref{sec:notconvex}),
but some partial convexity properties can still be salvaged. First,
convexity in the second argument does hold for all $\alpha$
\cite{Csiszar1995}:
\begin{theorem}\label{thm:Qconvex}
For any order $\alpha \in [0,\infty]$ R\'{e}nyi divergence is convex in
its second argument. That is, for any probability distributions $P$,
$Q_0$ and $Q_1$
\begin{equation}
\label{eqn:convexitylargealpha}D_{\alpha}(P\Vert(1-\lambda)Q_{0}+\lambda
Q_{1}) \leq(1-\lambda)D_{\alpha}(P\Vert Q_{0})+\lambda D_{\alpha}(P\Vert
Q_{1})
\end{equation}
for any $0 < \lambda < 1$. For finite $\alpha$, equality holds if and
only if
\begin{flalign*}
\text{$\alpha = 0$: } & \text{$D_0(P\Vert Q_{0}) = D_0(P\Vert Q_{1})$;} &\\
\text{$0 < \alpha < \infty$: } & \text{$q_0 = q_1$ ($P$-a.s.)}
\end{flalign*}
\end{theorem}

\begin{IEEEproof}
For $\alpha \in [0,1]$ this follows from the previous theorem. (For
$P_{0}=P_{1}$ the equality conditions reduce to the ones given here.)
For $\alpha \in (1,\infty)$, let $Q_{\lambda}=(1-\lambda)Q_{0}+\lambda
Q_{1}$ and define $f(x,Q_{\lambda})=(p(x)/q_{\lambda}(x))^{\alpha-1}$.
It is sufficient to show that
\begin{multline*}
\ln\E_{X\sim P}[f(X,Q_{\lambda})]\\
  \leq (1-\lambda)\ln\E_{X\sim P}[f(X,Q_{0})]
      +\lambda\ln\E_{X\sim P}[f(X,Q_{1})].
\end{multline*}
Noting that, for every $x\in\samplespace$, $f(x,Q)$ is log-convex in $Q$, this
is a consequence of the general fact that an expectation over log-convex
functions is itself log-convex, which can be shown using H\"{o}lder's
inequality:
\begin{align*}
\E_{P}[f(X,Q_{\lambda})]
  &\leq\E_{P}[f(X,Q_{0})^{1-\lambda
}f(X,Q_{1})^{\lambda}]\\
  &\leq\E_{P}[f(X,Q_{0})]^{1-\lambda}\E_{P}[f(X,Q_{1})]^{\lambda}.
\end{align*}
Taking logarithms completes the proof of \eqref{eqn:convexitylargealpha}.
Equality holds in the first inequality if and only if $q_{0}=q_{1}$
($P$-a.s.), which is also sufficient for equality in the second inequality.
Finally, \eqref{eqn:convexitylargealpha} extends to $\alpha=\infty$ by letting
$\alpha$ tend to $\infty$.
\end{IEEEproof}

And secondly, R\'enyi divergence is jointly \emph{quasi-convex} in both
arguments for all $\alpha$: 
\begin{theorem}\label{thm:PQquasiconvex}
For any order $\alpha \in [0,\infty]$ R\'{e}nyi divergence is jointly
quasi-convex in its arguments. That is, for any two pairs of probability
distributions $(P_{0},Q_{0})$ and $(P_{1},Q_{1})$, and any $\lambda \in
(0,1)$
\begin{equation}\label{eqn:PQquasiconvex}
  \begin{split}
D_{\alpha}\big(  (1-\lambda)&P_{0}+\lambda P_{1}\Vert(1-\lambda)Q_{0}+\lambda
Q_{1}\big)\\
&\leq \max\{D_{\alpha}(P_{0}\Vert Q_{0}),D_{\alpha}(P_{1}\Vert Q_{1})\}.
  \end{split}
\end{equation}
\end{theorem}

\begin{IEEEproof}
  For $\alpha \in [0,1]$, quasi-convexity is implied by convexity. For
  $\alpha \in (1,\infty)$, strict monotonicity of $x \mapsto
  \frac{1}{\alpha-1} \ln x$ implies that quasi-convexity is equivalent
  to quasi-convexity of the Hellinger integral $\int p^\alpha
  q^{1-\alpha} \intder \mu$. Since quasi-convexity is implied by
  ordinary convexity, it is sufficient to establish that the Hellinger
  integral is jointly convex in $P$ and $Q$. Let $p_\lambda =
  (1-\lambda)p_0 + \lambda p_1$ and $q_\lambda = (1-\lambda)q_0 +
  \lambda q_1$. Then joint convexity of the Hellinger integral is
  implied by the pointwise inequality
  \begin{equation*}
    (1-\lambda) p_0^\alpha q_0^{1-\alpha} + \lambda p_1^\alpha q_1^{1-\alpha}
    \geq p_\lambda^\alpha q_\lambda^{1-\alpha},
  \end{equation*}
  which holds by essentially the same argument as for
  \eqref{eqn:pointwise} in the proof of Theorem~\ref{thm:PQconvex}, with
  the convex function $f(x) = x^\alpha$.

  Finally, the case $\alpha=\infty$ follows by letting $\alpha$ tend to
  $\infty$:
  \begin{align*}
    D_{\infty}\big( &(1-\lambda)P_{0}+\lambda
      P_{1}\Vert(1-\lambda)Q_{0}+\lambda Q_{1}\big)\\
    &= \sup_{\alpha < \infty} D_\alpha\big(  (1-\lambda)P_{0}+\lambda
      P_{1}\Vert(1-\lambda)Q_{0}+\lambda Q_{1}\big)\\
    &\leq \sup_{\alpha < \infty} \max\{D_{\alpha}(P_{0}\Vert
    Q_{0}),D_{\alpha}(P_{1}\Vert Q_{1})\}\\
    &= \max\{\sup_{\alpha < \infty} D_{\alpha}(P_{0}\Vert
      Q_{0}),\sup_{\alpha < \infty} D_{\alpha}(P_{1}\Vert Q_{1})\}\\
    &= \max\{D_{\infty}(P_{0}\Vert Q_{0}),D_{\infty}(P_{1}\Vert
      Q_{1})\}.\qedhere
  \end{align*}
\end{IEEEproof}

\subsection{A Generalized Pythagorean Inequality}

An important result in statistical applications of information theory is
the Pythagorean inequality for Kullback-Leibler divergence
\cite{CoverThomas1991,Csiszar1975,Topsoe2007}. It states that, if
$\distset$ is a convex set of distributions, $Q$ is any distribution not
in $\distset$, and $D_{\text{min}}=\inf_{P\in\distset}D(P\Vert Q)$, then
there exists a distribution $P^{\ast}$ such that
\begin{equation}
D(P\Vert Q)\geq D(P\Vert P^{\ast})+D_{\text{min}}\qquad\text{for all
$P\in\distset$.}%
\end{equation}
The main use of the Pythagorean inequality lies in its implication that if
$P_{1},P_{2},\ldots$ is a sequence of distributions in $\distset$ such that
$D(P_{n}\Vert Q)\rightarrow D_{\text{min}}$, then $P_{n}$ converges to
$P^{\ast}$ in the strong sense that $D(P_{n}\Vert P^{\ast})\rightarrow0$.

For $\alpha\neq1$ R\'{e}nyi divergence does not satisfy the ordinary
Pythagorean inequality, but there does exist a generalization if we
replace convexity of $\distset$ by the following alternative notion of
convexity:
\begin{definition}
For $\alpha \in (0,\infty)$, we will call a set of distributions $\distset$
\emph{$\alpha$-convex} if, for any probability distribution $\lambda =
(\lambda_1,\lambda_2)$ and any two
distributions $P_1,P_2 \in \distset$, we also have $P_\lambda \in
\distset$, where $P_\lambda$ is the
$(\alpha,\lambda)$-mixture of $P_1$ and $P_2$, which will be defined
below.
\end{definition}
For $\alpha = 1$, the $(\alpha,\lambda)$-mixture is simply the ordinary
mixture $\lambda_1 P_1 + \lambda_2 P_2$, so that $1$-convexity is
equivalent to ordinary convexity. We generalize this to other $\alpha$
as follows:
\begin{definition}\label{def:alphamixture}
  Let $\alpha \in (0,\infty)$ and let $P_1,\ldots,P_m$ be any
  probability distributions. Then for any probability distribution
  $\lambda = (\lambda_1,\ldots,\lambda_m)$ we define the
  \emph{$(\alpha,\lambda)$-mixture} $P_\lambda$ of
  $P_1,\ldots,P_m$ as the distribution with density
  \begin{equation}\label{eqn:alphaconvex}
    p_\lambda =
      \frac{\big(\sum_{\theta=1}^m \lambda_\theta p_\theta^\alpha\big)^{1/\alpha}}
           {Z},
  \text{ where }
    Z = \int \Big(\sum_{\theta=1}^m \lambda_\theta p_\theta^\alpha\Big)^{1/\alpha} \der \mu
  \end{equation}
  is a normalizing constant.
\end{definition}
The normalizing constant $Z$ is always well defined:
\begin{lemma}\label{lem:normrange}
  The normalizing constant $Z$ in \eqref{eqn:alphaconvex} is bounded by
  \begin{equation}
    Z \in 
    \begin{cases}
    [m^{-(1-\alpha)/\alpha},1]
      &\text{for $\alpha \in (0,1]$,}\\
    [1,m^{(\alpha-1)/\alpha}]
      &\text{for $\alpha \in [1,\infty)$.}
    \end{cases}
  \end{equation}
\end{lemma}
\begin{IEEEproof}
  For $\alpha = 1$, we have $Z=1$, as required. So it remains to
  consider the simple orders. Let $f(y) = y^{1/\alpha}$ for $y \geq 0$,
  so that $Z = \int f\big(\sum_\theta \lambda_\theta
  p_\theta^\alpha\big) \der \mu$. Suppose first that $\alpha \in (0,1)$.
  Then $f$ is convex, which implies that $f(a+b)-f(a) \geq f(b) - f(0) =
  f(b)$ for any $a,b$, so that, by induction, $f(\sum_\theta a_\theta)
  \geq \sum_\theta f(a_\theta)$ for any $a_\theta$. Taking $a_\theta =
  \lambda_\theta p_\theta^\alpha$ and using Jensen's inequality, we
  find:
  \begin{equation*}
    \begin{split}
    \sum_\theta f\Big(\lambda_\theta p_\theta^\alpha\Big)
    &\leq f\Big(\sum_\theta \lambda_\theta p_\theta^\alpha\Big)
    \leq \sum_\theta \lambda_\theta f(p_\theta^\alpha)\\
    \sum_\theta \lambda_\theta^{1/\alpha} p_\theta
    &\leq \big(\sum_\theta \lambda_\theta p_\theta^\alpha\big)^{1/\alpha}
    \leq \sum_\theta \lambda_\theta p_\theta.
    \end{split}
  \end{equation*}
  Since every $p_\theta$ integrates to $1$, it follows that
  \begin{equation*}
    \sum_\theta \lambda_\theta^{1/\alpha}
      \leq Z_\lambda \leq 1.
  \end{equation*}
  The left-hand side is minimized at $\lambda = \nicefrac{1}{m}$, where it
  equals $m^{-(1-\alpha)/\alpha}$, which completes the proof for $\alpha
  \in (0,1)$. The proof for $\alpha \in (1,\infty)$ goes the same way,
  except that all inequalities are reversed because $f$ is concave.
\end{IEEEproof}

And, like for $\alpha = 1$, the set of $(\alpha,\lambda)$-mixtures is
closed under taking further mixtures of its elements:
\begin{lemma}\label{lem:closedundermixtures}
  Let $\alpha \in (0,\infty)$, let $P_1,\ldots,P_m$ be arbitrary
  probability distributions and let $P_{\lambda_1}$ and $P_{\lambda_2}$
  be their $(\alpha,\lambda_1)$- and $(\alpha,\lambda_2)$-mixtures for
  some distributions $\lambda_1,\lambda_2$. Then, for any distribution
  $\gamma = (\gamma_1,\gamma_2)$, the $(\alpha,\gamma)$-mixture of
  $P_{\lambda_1}$ and $P_{\lambda_2}$ is an $(\alpha,\nu)$-mixture of
  $P_1,\ldots,P_m$ for the distribution $\nu$ such that
  \begin{equation}
    \nu = \frac{\gamma_1
  }{Z_1^\alpha C} \lambda_{1} +\frac{\gamma_2}{Z_2^\alpha C}\lambda_{2},
  \end{equation}
  where $C = \frac{\gamma_1}{Z_1^\alpha}+\frac{\gamma_2}{Z_2^\alpha}$,
  and $Z_1$ and $Z_2$ are the normalizing constants of $P_{\lambda_1}$
  and $P_{\lambda_2}$ as defined in \eqref{eqn:alphaconvex}.
\end{lemma}

\begin{IEEEproof}
  Let $M_\gamma$ be the $(\alpha,\gamma)$-mixture of $P_{\lambda_1}$ and
  $P_{\lambda_2}$, and take $\lambda_i =
  (\lambda_{i,1,},\ldots,\lambda_{i,m})$. Then
  \begin{align*}
    m_\gamma &\propto \big(\gamma_1 p_{\lambda_1}^\alpha
                  + \gamma_2 p_{\lambda_2}^\alpha)^{1/\alpha}\\
      &= \Big(\frac{\gamma_1}{Z_1^\alpha} \sum_\theta \lambda_{1,\theta} p_\theta^\alpha
            + \frac{\gamma_2}{Z_2^\alpha} \sum_\theta \lambda_{2,\theta}
            p_\theta^\alpha\Big)^{1/\alpha}\\
      &\propto \Big(\sum_\theta \frac{\frac{\gamma_1
      \lambda_{1,\theta}}{Z_1^\alpha}+\frac{\gamma_2\lambda_{2,\theta}}{Z_2^\alpha}}{C}
      p_\theta^\alpha\Big)^{1/\alpha},
  \end{align*}
  from which the result follows.
\end{IEEEproof}
We are now ready to generalize the Pythagorean inequality to any $\alpha
\in (0,\infty)$:
\begin{theorem}[Pythagorean Inequality]\label{thm:pythagoras}
  Let $\alpha \in (0,\infty)$. Suppose that $\distset$ is an
  $\alpha$-convex set of distributions. Let $Q$ be an arbitrary
  distribution and suppose that the \emph{$\alpha$-information
  projection}
  \begin{equation}
    P^\ast = \argmin_{P \in \distset} D_\alpha(P\|Q)
  \end{equation}
  exists. Then we have the Pythagorean inequality
  \begin{equation}\label{eqn:pythagoras}
    D_\alpha(P\|Q) \geq D_\alpha(P\|P^\ast) + D_\alpha(P^\ast\|Q)
    \qquad \text{for all $P \in \distset$.}
  \end{equation}
\end{theorem}

This result is new, although the work of Sundaresan on a generalization
of R\'{e}nyi divergence might be related
\cite{Sundaresan2002,Sundaresan2006}. Our proof follows the same
approach as the proof for $\alpha = 1$ by Cover and Thomas
\cite{CoverThomas1991}.

\begin{IEEEproof}
  For $\alpha = 1$, this is just the standard Pythagorean inequality for
  Kullback-Leibler divergence. See, for example, the proof by Tops{\o}e
  \cite{Topsoe2007}. It remains to prove the theorem when $\alpha$ is a
  simple order.

  Let $P \in \distset$ be arbitrary, and let $P_\lambda$ be the
  $\big(\alpha,(1-\lambda,\lambda)\big)$-mixture of $P^\ast$ and $P$.
  Since $\distset$ is $\alpha$-convex and $P^\ast$ is the minimizer over
  $\distset$, we have $\frac{\der}{\der \lambda}
  D_\alpha(P_\lambda\|Q)\big|_{\lambda = 0} \geq 0$.

  This derivative evaluates to:
  \begin{multline*}
    \frac{\der}{\der \lambda} D_\alpha(P_\lambda\|Q)
      = \frac{1}{\alpha-1}
        \frac{\int p^\alpha q^{1-\alpha}\der\mu - \int (p^\ast)^\alpha q^{1-\alpha}\der\mu }
       {Z_\lambda^\alpha \int p_\lambda^\alpha q^{1-\alpha}
       \der \mu}\\
      - \frac{\alpha}{\alpha-1}
        \frac{(1-\lambda) \int (p^\ast)^\alpha q^{1-\alpha}\der\mu + \lambda
        \int p^\alpha q^{1-\alpha}\der\mu}
       {Z_\lambda^{\alpha+1} \int p_\lambda^\alpha q^{1-\alpha}
       \der \mu}
       \frac{\der}{\der \lambda} Z_\lambda.
  \end{multline*}
  Let $X_\lambda = \big((1-\lambda)(p^\ast)^\alpha + \lambda
  p^\alpha\big)^{1/\alpha}$, so that $Z_\lambda = \int X_\lambda \der
  \mu$. If $\alpha \in (0,1)$, then $X_\lambda$ is convex in $\lambda$
  so that $ \frac{X_\lambda - X_0}{\lambda}$ is nondecreasing in
  $\lambda$, and if $\alpha \in (0,\infty)$, then $X_\lambda$ is concave
  in $\lambda$ so that $\frac{X_\lambda - X_0}{\lambda}$ is
  nonincreasing. By Lemma~\ref{lem:normrange}, we also see
  that $\int \frac{X_\lambda - X_0}{\lambda} \der \mu = \frac{Z_\lambda
  - 1}{\lambda}$ is bounded by $0$ for $\lambda > 0$, from above if
  $\alpha \in (0,1)$ and from below if $\alpha \in (1,\infty)$. It
  therefore follows from the monotone convergence theorem that
  \begin{align*}
    \frac{\der}{\der \lambda} Z_\lambda\big|_{\lambda = 0}
      &= \lim_{\lambda \downarrow 0} \frac{Z_\lambda - Z_0}{\lambda}
      = \lim_{\lambda \downarrow 0} \int \frac{X_\lambda - X_0}{\lambda}
      \der \mu\\
      &= \int \lim_{\lambda \downarrow 0} \frac{X_\lambda - X_0}{\lambda}
      \der \mu
      = \int \frac{\der}{\der \lambda} X_\lambda\big|_{\lambda = 0} \der \mu,
  \end{align*}
  where
  \begin{equation*}
    \frac{\der}{\der \lambda} X_\lambda
      = \frac{1}{\alpha} \Big((1-\lambda)(p^\ast)^\alpha + \lambda
  p^\alpha\Big)^{1/\alpha-1}(p^\alpha - (p^\ast)^\alpha).
  \end{equation*}
  If $D_\alpha(P^\ast\|Q) = \infty$, then the theorem is trivially true, so
  we may assume without loss of generality that $D_\alpha(P^\ast\|Q) <
  \infty$, which implies that $0 < \int (p^\ast)^\alpha q^{1-\alpha} \der \mu
  < \infty$.

  Putting everything together, we therefore find
  \begin{align*}
    0 &\leq \frac{\der}{\der \lambda}
            D_\alpha(P_\lambda\|Q)\big|_{\lambda = 0}\\
      &= \frac{1}{\alpha-1}
        \frac{\int p^\alpha q^{1-\alpha}\der\mu - \int (p^\ast)^\alpha q^{1-\alpha}\der\mu }
       {\int (p^\ast)^\alpha q^{1-\alpha}
       \der \mu}\\
      &\quad- \frac{1}{\alpha-1}
        \int (p^\ast)^{1-\alpha}(p^\alpha - (p^\ast)^\alpha) \der \mu\\
      &= \frac{1}{\alpha-1}\Big(
        \frac{\int p^\alpha q^{1-\alpha}\der\mu}
       {\int (p^\ast)^\alpha q^{1-\alpha}
       \der \mu}
      - \int (p^\ast)^{1-\alpha}p^\alpha \der \mu\Big).
  \end{align*}
  Hence, if $\alpha > 1$ we have
  \begin{equation*}
    \int p^\alpha q^{1-\alpha}\der\mu
    \geq \int (p^\ast)^\alpha
      q^{1-\alpha} \der \mu \int (p^\ast)^{1-\alpha}p^\alpha \der \mu,
  \end{equation*}
  and if $\alpha < 1$ we have the converse of this inequality. In both
  cases, the Pythagorean inequality \eqref{eqn:pythagoras} follows upon
  taking logarithms and dividing by $\alpha -1$ (which flips the
  inequality sign for $\alpha < 1$).
\end{IEEEproof}

\subsection{Continuity}

In this section we study continuity properties of the R\'enyi divergence
$D_{\alpha}(P\Vert Q)$ of different orders in the pair of probability
distributions $(P,Q)$. It turns out that continuity depends on the order
$\alpha$ and the topology on the set of all probability distributions.

The set of probability distributions on $(\samplespace,\salgebra)$ may
be equipped with the topology of \emph{setwise convergence}, which is
the coarsest topology such that, for any event $A \in \salgebra$, the
function $P \mapsto P(A)$ that maps a distribution to its probability on
$A$, is continuous. In this topology, convergence of a sequence of
probability distributions $P_{1},P_{2},\ldots$ to a probability
distribution $P$ means that $P_{n}(A)\rightarrow P(A)$ for any
$A\in\salgebra$.

Alternatively, one might consider the topology defined by the
\emph{total variation distance}
\begin{equation}\label{eqn:totalvariation}
V(P,Q)=\int\lvert p-q\rvert \intder \mu=2\sup_{A\in\salgebra}\lvert
P(A)-Q(A)\rvert,
\end{equation}
in which $P_{n}\rightarrow P$ means that $V(P_{n},P)\rightarrow0$. The total
variation topology is stronger than the topology of setwise convergence in the
sense that convergence in total variation distance implies convergence on any
$A\in\salgebra$. The two topologies coincide if the sample space
$\samplespace$ is countable.

In general, R\'enyi divergence is lower semi-continuous for positive orders:

\begin{theorem}
\label{thm:lowersemicontinuity} For any order $\alpha \in (0,\infty]$,
$D_{\alpha}(P\Vert Q)$ is a lower semi-continuous function of the pair
$(P,Q)$ in the topology of setwise convergence.
\end{theorem}

\begin{IEEEproof}
Suppose $\samplespace=\{x_{1},\ldots,x_{k}\}$ is finite. Then for any simple
order $\alpha$
\[
D_{\alpha}(P\Vert Q)=\frac{1}{\alpha-1}\ln\sum_{i=1}^{k}p_{i}^{\alpha}%
q_{i}^{1-\alpha},
\]
where $p_{i}=P(x_{i})$ and $q_{i}=Q(x_{i})$. If $0<\alpha<1$, then
$p_{i}^{\alpha}q_{i}^{1-\alpha}$ is continuous in $(P,Q)$. For $1<\alpha
<\infty$, it is only discontinuous at $p_{i}=q_{i}=0$, but there
$p_{i}^{\alpha}q_{i}^{1-\alpha}=0=\min_{(P,Q)}p_{i}^{\alpha}q_{i}^{1-\alpha}$,
so then $p_{i}^{\alpha}q_{i}^{1-\alpha}$ is still lower semi-continuous.
These properties carry over to
$\sum_{i=1}^{k}p_{i}^{\alpha}q_{i}^{1-\alpha}$ and thus
$D_{\alpha}(P\Vert Q)$ is continuous for $0<\alpha<1$ and lower
semi-continuous for $\alpha>1$. A supremum over (lower semi-)continuous
functions is itself lower semi-continuous. Therefore, for simple orders
$\alpha$, Theorem~\ref{thm:simplesupfinite} implies that $D_{\alpha }(P\Vert
Q)$ is lower semi-continuous for arbitrary $\samplespace$. This property
extends to the extended orders $1$ and $\infty$ by $D_{\beta}(P\Vert
Q)=\sup_{\alpha<\beta}D_{\alpha}(P\Vert Q)$ for $\beta\in\{1,\infty\}$.
\end{IEEEproof}

Moreover, if $\alpha \in (0,1)$ and the total variation topology
is assumed, then Theorem~\ref{thm:continuity} below shows that
R\'{e}nyi divergence is uniformly continuous.

First we prove that the topologies induced by R\'{e}nyi divergences of orders
$\alpha \in (0,1)$ are all equivalent:
\begin{theorem}
\label{thm:equivalentopologies} For any $0<\alpha \leq \beta <1$
\begin{equation*}
  \frac{\alpha}{\beta} \frac{1-\beta}{1-\alpha} D_\beta(P \Vert Q)
    \leq D_\alpha(P \Vert Q) \leq D_\beta(P \Vert Q).
\end{equation*}
\end{theorem}

This follows from the following symmetry-like property, which may be
verified directly.

\begin{proposition}
[Skew Symmetry]\label{prop:simplesymmetry} For any $0 < \alpha< 1$
\[
D_{\alpha}(P\Vert Q)=\frac{\alpha}{1-\alpha} D_{1-\alpha}(Q\Vert P).
\]
\end{proposition}

Note that, in particular, R\'{e}nyi divergence is symmetric for $\alpha
=\nicefrac{1}{2}$, but that skew symmetry does not hold for $\alpha=0$ and
$\alpha=1$.

\begin{IEEEproof}
[Proof of Theorem~\ref{thm:equivalentopologies}] We have already
established the second inequality in
Theorem~\ref{thm:increasinginorder}, so it remains to prove the first
one. Skew symmetry implies that
\begin{align*}
  \frac{1-\alpha}{\alpha} D_{\alpha}(P\Vert Q)
    &= D_{1-\alpha}(Q\Vert P)\\
    &\geq D_{1-\beta}(Q\Vert P)
    = \frac{1-\beta}{\beta} D_{\beta}(P\Vert Q),
\end{align*}
from which the result follows.
\end{IEEEproof}

\begin{remark}
By \eqref{eqn:relationHellinger}, these results show that, for $\alpha
\in (0,1)$, $D_{\alpha}(P_{n}\Vert Q)\rightarrow0$ is equivalent to
convergence of $P_{n}$ to $Q$ in Hellinger distance, which is equivalent
to convergence of $P_n$ to $Q$ in total variation
\cite[p.\,364]{Shiryaev1996}.
\end{remark}
Next we shall prove a stronger result on the relation between R\'{e}nyi
divergence and total variation.
\begin{theorem}
\label{thm:continuity} For $\alpha \in (0,1)$, the R\'{e}nyi
divergence $D_{\alpha}(P\Vert Q)$ is a uniformly continuous function of $(P,Q)$ in
the total variation topology.
\end{theorem}

\begin{lemma}
\label{lem:continuitybound} Let $0<\alpha<1$. Then for all $x,y\geq0$
and $\varepsilon>0$
\[
\lvert x^{\alpha}-y^{\alpha}\rvert\leq\varepsilon^{\alpha}+\varepsilon
^{\alpha-1}\lvert x-y\rvert.
\]
\end{lemma}

\begin{IEEEproof}
If $x,y\leq\varepsilon$ or $x = y$ the inequality $\lvert x^{\alpha}%
-y^{\alpha}\rvert\leq\varepsilon^{\alpha}$ is obvious. So assume that $x > y$
and $x\geq\varepsilon$. Then
\[
\frac{\lvert x^{\alpha}-y^{\alpha}\rvert} {\lvert x-y\rvert} \leq\frac{\lvert
x^{\alpha}-0^{\alpha}\rvert} {\lvert x-0\rvert} = x^{\alpha-1} \leq
\varepsilon^{\alpha-1}.\qedhere
\]
\end{IEEEproof}

\begin{IEEEproof}
[Proof of Theorem~\ref{thm:continuity}]First note that R\'{e}nyi
divergence is a function of the \emph{power divergence} $d_{\alpha}%
(P,Q)=\int\left(  1-\left(  \frac{\der P}{\der Q}\right)  ^{\alpha
}\right)  \der Q:$
\[
D_{\alpha}(P\Vert Q)=\frac{1}{\alpha-1}\ln\left(  1-d_{\alpha}(P,Q)\right).
\]
Since $x\mapsto\frac{1}{\alpha-1}\ln(1-x)$ is continuous, it is
sufficient to prove that $d_{\alpha}(P,Q)$ is a uniformly continuous function of
$(P,Q)$. For any $\varepsilon>0$ and distributions $P_{1},P_{2}$ and
$Q$, Lemma~\ref{lem:continuitybound} implies that
\begin{align*}
\vert d_{\alpha}(P_{1},Q)-d_{\alpha}&(P_{2},Q)\vert \leq
\int\left\vert \left(  \frac{\der P_{1}}{\der Q}\right)  ^{\alpha
}-\left(  \frac{\der P_{2}}{\der Q}\right)  ^{\alpha}\right\vert
\der Q\\
&  \leq\int\left(  \varepsilon^{\alpha}+\varepsilon^{\alpha-1}\left\vert
\frac{\der P_{1}}{\der Q}-\frac{\der P_{2}}{\der Q}\right\vert \right)  \der Q\\
&  =\varepsilon^{\alpha}+\varepsilon^{\alpha-1}\int\left\vert \frac
{\der P_{1}}{\der Q}-\frac{\der P_{2}}{\der Q}\right\vert
\der Q\\
&  =\varepsilon^{\alpha}+\varepsilon^{\alpha-1}V(P_{1},P_{2}).
\end{align*}
As $d_{\alpha}(P,Q)=d_{1-\alpha}(Q,P)$, it also follows that $\left\vert
d_{\alpha}(P,Q_{1})-d_{\alpha}(P,Q_{2})\right\vert \leq\varepsilon^{1-\alpha
}+\varepsilon^{-\alpha}V(Q_{1},Q_{2})$ for any $Q_{1},Q_{2}$ and $P$.
Therefore
\begin{align*}
\vert d_{\alpha}(&P_{1},Q_{1})-d_{\alpha}(P_{2},Q_{2})\vert\\
  &\leq\left\vert
  d_{\alpha}(P_{1},Q_{1})-d_{\alpha}(P_{2},Q_{1})\right\vert\\
  &\quad+\left\vert d_{\alpha}(P_{2},Q_{1})-d_{\alpha}(P_{2},Q_{2})\right\vert \\
&  \leq\varepsilon^{\alpha}+\varepsilon^{\alpha-1}V(P_{1},P_{2})+\varepsilon
^{1-\alpha}+\varepsilon^{-\alpha}V(Q_{1},Q_{2}),
\end{align*}
from which the theorem follows.
\end{IEEEproof}

A partial extension to $\alpha=0$ follows:
\begin{corollary}\label{cor:uppersemicontinuity0}
The R\'enyi divergence $D_{0}(P\Vert Q)$ is an upper semi-continuous function
of $(P,Q)$ in the total variation topology.
\end{corollary}

\begin{IEEEproof}
This follows from Theorem~\ref{thm:continuity} because $D_{0}(P\Vert Q)$
is the infimum of the continuous functions $(P,Q) \mapsto
D_\alpha(P\Vert Q)$ for $\alpha \in (0,1)$.
\end{IEEEproof}

If we consider continuity in $Q$ only, then for any finite sample space
we obtain:
\begin{theorem}\label{thm:Qcontinuity}
  Suppose $\samplespace$ is finite, and let $\alpha \in [0,\infty]$.
  Then for any $P$ the R\'enyi divergence $D_\alpha(P\|Q)$ is continuous
  in $Q$ in the topology of setwise convergence.
\end{theorem}

\begin{IEEEproof}
  Directly from the closed-form expressions for R\'enyi divergence.
\end{IEEEproof}

Finally, we will also consider the \emph{weak topology}, which is weaker
than the two topologies discussed above. In the weak topology,
convergence of $P_1,P_2,\ldots$ to $P$ means that
\begin{equation}\label{eqn:weakconvergence}
  \int f(x) \intder P_n(x) \to \int f(x) \intder P(x)
\end{equation}
for any bounded, continuous function $f \colon \samplespace \to \reals$.
Unlike for the previous two topologies, the reference to continuity of
$f$ means that the weak topology depends on the topology of the sample
space $\samplespace$. We will therefore assume that $\samplespace$ is a
\emph{Polish space} (that is, it should be a complete separable metric
space), and we let $\salgebra$ be the Borel $\sigma$-algebra. Then
Prokhorov \cite{Prokhorov1956} shows that there exists a metric that
makes the set of finite measures on $\samplespace$ a Polish space as
well, and which is such that convergence in the metric is equivalent to
\eqref{eqn:weakconvergence}. The weak topology then, is the topology
induced by this metric.

\begin{theorem}\label{thm:weaklowersemicontinuity}
  Suppose that $\samplespace$ is a Polish space. Then for any order
  $\alpha \in (0,\infty]$, $D_\alpha(P\|Q)$ is a lower semi-continuous
  function of the pair $(P,Q)$ in the weak topology.
\end{theorem}

The proof is essentially the same as the proof for $\alpha = 1$ by
Posner \cite{Posner1975}.

\begin{IEEEproof}
  Let $P_1,P_2,\ldots$ and $Q_1,Q_2,\ldots$ be sequences of
  distributions that weakly converge to $P$ and $Q$, respectively. We
  need to show that
  \begin{equation}\label{eqn:weaklowersemi}
    \liminf_{n \to \infty} D_\alpha(P_n\|Q_n) \geq D_\alpha(P\|Q).
  \end{equation}
  For any set $A \in \salgebra$, let $\partial A$ denote its boundary,
  which is its closure minus its interior, and let $\salgebra_0
  \subseteq \salgebra$ consist of the sets $A \in \salgebra$ such
  that $P(\partial A) = Q(\partial A) = 0$. Then $\salgebra_0$ is an algebra
  by Lemma~1.1 of Prokhorov \cite{Prokhorov1956}, applied to the measure
  $P + Q$, and the Portmanteau theorem implies that $P_n(A) \to P(A)$
  and $Q_n(A) \to Q(A)$ for any $A \in \salgebra_0$
  \cite{VanDerVaartWellner1996}.

  Posner \cite[proof of Theorem~1]{Posner1975} shows that $\salgebra_0$ generates
  $\salgebra$ (that is, $\sigma(\salgebra_0) = \salgebra$). By the
  translator's proof of Theorem~2.4.1 in Pinsker's book
  \cite{Pinsker1964}, this implies that, for any finite partition
  $\{A_1,\ldots,A_k\} \subseteq \salgebra$ and any $\gamma > 0$, there
  exists a finite partition $\{A'_1,\ldots,A'_k\} \subseteq \salgebra_0$
  such that $P(A_i \symdiff A'_i) \leq \gamma$ and $Q(A_i \symdiff A'_i)
  \leq \gamma$ for all $i$, where $A_i \symdiff A'_i = (A_i \setminus
  A'_i) \union (A'_i \setminus A_i)$ denotes the symmetric set
  difference. By the data processing inequality and lower
  semi-continuity in the topology of setwise convergence, this implies
  that \eqref{eqn:supfinite} still holds when the supremum is restricted
  to finite partitions $\partition$ in $\salgebra_0$ instead of
  $\salgebra$.
  
  Thus, for any $\varepsilon > 0$, we can find a finite partition
  $\partition \subseteq \salgebra_0$ such that
  \begin{equation*}
    D_\alpha(P_{\lvert\partition}\Vert Q_{\lvert\partition})
      \geq D_{\alpha}(P\Vert Q) - \varepsilon.
  \end{equation*}
  The data processing inequality and the fact that $P_n(A) \to P(A)$ and
  $Q_n(A) \to Q(A)$ for all $A \in \partition$, together with lower
  semi-continuity in the topology of setwise convergence, then imply
  that
  \begin{multline*}
    D_\alpha(P_n\Vert Q_n)
      \geq D_\alpha\big((P_n)_{\lvert\partition}\Vert
      (Q_n)_{\lvert\partition}\big)\\
      \geq D_{\alpha}(P_{\lvert\partition}\Vert Q_{\lvert\partition}) - \varepsilon
      \geq D_{\alpha}(P\Vert Q) - 2\varepsilon
  \end{multline*}
  for all sufficiently large $n$. Consequently,
  \begin{equation*}
    \liminf_{n \to \infty} D_\alpha(P_n\Vert Q_n)
      \geq D_{\alpha}(P\Vert Q) - 2\varepsilon
  \end{equation*}
  for any $\varepsilon > 0$, and \eqref{eqn:weaklowersemi} follows by
  letting $\varepsilon$ tend to $0$.
\end{IEEEproof}

\begin{theorem}[Compact Sublevel Sets]\label{thm:compactsublevelsets}
Suppose $\samplespace$ is a Polish space, let $Q$ be arbitrary, and let
$c \in [0,\infty)$ be a constant. Then the sublevel set
\begin{equation}
  \sublevelset = \{P \mid D_\alpha(P\|Q) \leq c\}
\end{equation}
is convex and compact in the topology of weak convergence for any order
$\alpha \in [1,\infty]$.
\end{theorem}

\begin{IEEEproof}
  Convexity follows from quasi-convexity of R\'enyi divergence in its
  first argument.

  Suppose that $P_1,P_2,\ldots \in \sublevelset$ converges to a finite
  measure $P$. Then \eqref{eqn:weakconvergence}, applied to the constant
  function $f(x) = 1$, implies that $P(\samplespace) = 1$, so that $P$
  is also a probability distribution. Hence by lower semi-continuity
  (Theorem~\ref{thm:weaklowersemicontinuity}) $\sublevelset$ is closed.
  It is therefore sufficient to show that $\sublevelset$ is relatively
  compact.

  For any event $A \in \salgebra$, let $\comp{A} = \samplespace
  \setminus A$ denote its complement. Prokhorov \cite[Theorem
  1.12]{Prokhorov1956} shows that $\sublevelset$ is relatively compact
  if, for any $\varepsilon > 0$, there exists a compact set $A \subseteq
  \samplespace$ such that $P(\comp{A}) < \varepsilon$ for all $P \in
  \sublevelset$.
  
  Since $\samplespace$ is a Polish space, for any $\delta > 0$ there
  exists a compact set $B_\delta \subseteq \samplespace$ such that
  $Q(B_\delta) \geq 1-\delta$
  \cite[Lemma~1.3.2]{VanDerVaartWellner1996}. For any distribution $P$,
  let $P_{\lvert B_\delta}$ denote the restriction of $P$ to the binary
  partition $\{B_\delta,\comp{B}_\delta\}$. Then, by monotonicity in
  $\alpha$ and the data processing inequality, we have, for any $P \in
  \sublevelset$,
  \begin{align*}
    c &\geq D_\alpha(P\|Q)
      \geq D_1(P\|Q)
      \geq D_1(P_{\lvert B_\delta}\|Q_{\lvert B_\delta})\\
      &= P(B_\delta) \ln \frac{P(B_\delta)}{Q(B_\delta)}
        + P(\comp{B}_\delta) \ln \frac{P(\comp{B}_\delta)}{Q(\comp{B}_\delta)}\\
      &\geq P(B_\delta) \ln P(B_\delta)
        + P(\comp{B}_\delta) \ln P(\comp{B}_\delta)
        + P(\comp{B}_\delta) \ln \frac{1}{Q(\comp{B}_\delta)}\\
      &\geq \frac{-2}{\e} 
        + P(\comp{B}_\delta) \ln \frac{1}{Q(\comp{B}_\delta)},
  \end{align*}
  where the last inequality follows from $x \ln x \geq -1/\e$.
  Consequently,
  \begin{equation*}
    P(\comp{B}_\delta) \leq \frac{c + 2/\e}{\ln\big(1/Q(\comp{B}_\delta)\big)},
  \end{equation*}
  and since $Q(\comp{B}_\delta) \to 0$ as $\delta$ tends to $0$ we can
  satisfy the condition of Prokhorov's theorem by taking $A$ equal to
  $B_\delta$ for any sufficiently small $\delta$ depending on
  $\varepsilon$.
\end{IEEEproof}

\subsection{Limits of $\sigma$-Algebras}

As shown by Theorem~\ref{thm:simplesupfinite}, there exists a sequence
of finite partitions $\partition_{1},\partition_{2},\ldots$ such that
\begin{equation}
\label{eqn:limitfinitepartitions}D_{\alpha}(P_{\lvert\partition_{n}}\Vert
Q_{\lvert\partition_{n}}) \uparrow D_{\alpha}(P\Vert Q).
\end{equation}
Theorem~\ref{thm:LimitAlgebraIncreasing} below elaborates on this result. It
implies that \eqref{eqn:limitfinitepartitions} holds for any increasing
sequence of partitions $\partition_{1} \subseteq\partition_{2}
\subseteq\cdots$ that generate $\sigma$-algebras converging to
$\salgebra$, in the sense that $\salgebra = \sigma\left(
\bigcup_{n=1}^{\infty}\partition_{n}\right)$. An analogous result
holds for infinite sequences of increasingly coarse partitions, which is
shown by Theorem~\ref{thm:LimitAlgebraDecreasing}. For the special case
$\alpha = 1$, information-theoretic proofs of
Theorems~\ref{thm:LimitAlgebraIncreasing} and
\ref{thm:LimitAlgebraDecreasing} are given by Barron \cite{Barron2000}
and Harremo\"es and Holst \cite{HarremoesHolst2009}.
Theorem~\ref{thm:LimitAlgebraIncreasing} may also be derived from
general properties of $f$-divergences \cite{LieseVajda2006}.

\begin{theorem}
[Increasing]\label{thm:LimitAlgebraIncreasing} Let $\salgebra_{1}
\subseteq\salgebra_{2}\subseteq\cdots\subseteq\salgebra$ be an
increasing family of $\sigma$-algebras, and let
$\salgebra_\infty=\sigma\left(\bigcup_{n=1}^{\infty}\salgebra_{n}\right)$
be the smallest $\sigma$-algebra containing them. Then for any order
$\alpha \in (0,\infty]$
\begin{equation}\label{eqn:LimitAlgebraIncreasing}
  \lim_{n\rightarrow\infty} D_{\alpha}(P_{\lvert\salgebra_{n}}\Vert Q_{\lvert\salgebra_{n}})
    =D_{\alpha}(P_{\lvert\salgebra_{\infty}}\Vert Q_{\lvert\salgebra_{\infty}}).
\end{equation}
\end{theorem}

For $\alpha= 0$, \eqref{eqn:LimitAlgebraIncreasing} does not hold. A
counterexample is given after Example~\ref{ex:gaussiandichotomy} below.

\begin{lemma}
\label{lem:UniformIntegrabilityIncreasing} Let $\salgebra_{1}\subseteq
\salgebra_{2}\subseteq\cdots\subseteq\salgebra$ be an increasing
family of $\sigma$-algebras, and suppose that $\mu$ is a probability
distribution. Then the family of random variables $\{p_{n}\}_{n\geq1}$
with members $p_{n} = \E\left[ \left.  p \right\vert \salgebra_{n}
\right]$ is uniformly integrable (with respect to $\mu$).
\end{lemma}

The proof of this lemma is a special case of part of the proof of
L\'{e}vy's upward convergence theorem in Shiryaev's textbook
\cite[p.\,510]{Shiryaev1996}. We repeat it here for completeness.

\begin{IEEEproof}
For any constants $b,c > 0$
\begin{align*}
\int_{p_{n} > b} p_{n}\intder\mu &  = \int_{p_{n} > b} p\intder\mu\\
&  \leq\int_{p_{n} > b, p \leq c} p\intder\mu+ \int_{p > c}
p\intder\mu\\
&  \leq c \cdot\mu\left(  p_{n} > b\right)  + \int_{p > c} p\intder\mu\\
&  \overset{(*)}{\leq} \frac{c}{b} \E[p_{n}] + \int_{p > c}
p\intder\mu= \frac{c}{b} + \int_{p > c} p\intder\mu,
\end{align*}
in which the inequality marked by $(*)$ is Markov's. Consequently
\begin{multline*}
\lim_{b \to\infty} \sup_{n} \int_{p_{n} > b} \lvert p_{n} \rvert
\intder\mu  = \lim_{c \to\infty} \lim_{b \to\infty} \sup_{n}
\int_{p_{n} > b} \lvert p_{n} \rvert\intder\mu\\
  \leq\lim_{c \to\infty} \lim_{b \to\infty} \frac{c}{b} + \lim_{c \to\infty}
\int_{p > c} p\intder\mu= 0,
\end{multline*}
which proves the lemma.
\end{IEEEproof}

\begin{IEEEproof}
[Proof of Theorem~\ref{thm:LimitAlgebraIncreasing}]As by the data processing
inequality $D_{\alpha}(P_{\lvert\salgebra_{n}}\Vert Q_{\lvert\salgebra%
_{n}})\leq D_{\alpha}(P\Vert Q)$ for all $n$, we only need to show that
$\lim_{n\rightarrow\infty}D_{\alpha}(P_{\lvert\salgebra_{n}}\Vert
Q_{\lvert\salgebra_{n}})\geq D_{\alpha}(P_{\lvert\salgebra_{\infty}}\Vert
Q_{\lvert\salgebra_{\infty}})$. To this end, assume without loss of
generality that $\salgebra=\salgebra_{\infty}$ and that $\mu$ is a
probability distribution (i.e.\ $\mu=(P+Q)/2$). Let $p_{n}=\E\left[
\left.  p\right\vert \salgebra_{n}\right]  $ and $q_{n}=\E\left[
\left.  q\right\vert \salgebra_{n}\right]  $, and define the distributions
$\tilde{P}_{n}$ and $\tilde{Q}_{n}$ on $(\samplespace,\salgebra)$ by
\[
\tilde{P}_{n}(A)=\int_{A}p_{n}\intder\mu,\quad\tilde{Q}_{n}(A)=\int
_{A}q_{n}\intder\mu\qquad(A\in\salgebra),
\]
such that, by the Radon-Nikod\'{y}m theorem and
Proposition~\ref{prop:SubAlgebraExpectation}, $\frac{\der\tilde{P}_{n}%
}{\der\mu}=p_{n}=\frac{\der P_{\lvert\salgebra_{n}}}%
{\der\mu_{\lvert\salgebra_{n}}}$ and $\frac{\der\tilde{Q}_{n}%
}{\der\mu}=q_{n}=\frac{\der Q_{\lvert\salgebra_{n}}}%
{\der\mu_{\lvert\salgebra_{n}}}$ ($\mu$-a.s.) It follows that
\[
D_{\alpha}(\tilde{P}_{n}\Vert\tilde{Q}_{n})=D_{\alpha}(P_{\lvert
\salgebra_{n}}\Vert Q_{\lvert\salgebra_{n}})
\]
for $0<\alpha<\infty$ and therefore by continuity also for $\alpha=\infty$. We
will proceed to show that $(\tilde{P}_{n},\tilde{Q}_{n})\rightarrow(P,Q)$ in
the topology of setwise convergence. By lower semi-continuity of R\'{e}nyi divergence this
implies that
$\lim_{n\rightarrow\infty}D_{\alpha}(\tilde{P}_{n}\Vert\tilde
{Q}_{n})\geq D_{\alpha}(P\Vert Q)$, from which the theorem follows. By
L\'{e}vy's upward convergence theorem \cite[p.\,510]{Shiryaev1996},
$\lim_{n\rightarrow\infty}p_{n}=p$ ($\mu$-a.s.) Hence uniform
integrability of the family $\{p_{n}\}$ (by
Lemma~\ref{lem:UniformIntegrabilityIncreasing}) implies that for any
$A\in\salgebra$
\[
\lim_{n\rightarrow\infty}\tilde{P}_{n}(A)=\lim_{n\rightarrow\infty}\int
_{A}p_{n}\intder\mu=\int_{A}p\intder\mu=P(A)
\]
\cite[Thm.~5, p.\,189]{Shiryaev1996}. Similarly $\lim_{n\rightarrow\infty
}\tilde{Q}_{n}(A)=Q(A)$, so we find that $(\tilde{P}_{n},\tilde{Q}%
_{n})\rightarrow(P,Q)$, which completes the proof.
\end{IEEEproof}

\begin{theorem}
[Decreasing]\label{thm:LimitAlgebraDecreasing} Let $\salgebra \supseteq
\salgebra_{1}\supseteq\salgebra_{2}\supseteq\cdots$ be a decreasing
family of $\sigma$-algebras, and let $\salgebra_{\infty}=\bigcap
_{n=1}^{\infty}\salgebra_{n}$ be the largest $\sigma$-algebra contained
in all of them. Let $\alpha \in [0,\infty)$. If $\alpha \in [0,1)$ or
there exists an $m$ such that $D_{\alpha}(P_{\lvert\salgebra_{m}}\Vert
Q_{\lvert \salgebra_{m}}) < \infty$, then
\[
\lim_{n\rightarrow\infty} D_{\alpha}(P_{\lvert\salgebra_{n}}\Vert
Q_{\lvert\salgebra_{n}}) = D_{\alpha}(P_{\lvert\salgebra_{\infty}} \Vert
Q_{\lvert\salgebra_{\infty}}).
\]

\end{theorem}

The theorem cannot be extended to the case $\alpha=\infty$.

\begin{lemma}
\label{lem:UniformIntegrabilityDecreasing} Let $\salgebra\supseteq
\salgebra_{1}\supseteq\salgebra_{2}\supseteq\cdots$ be a decreasing
family of $\sigma$-algebras. Let $\alpha \in (0,\infty)$, $p_{n}= 
\frac{\der P_{\lvert\salgebra_{n}}}{\der\mu_{\lvert\salgebra_{n}}},
q_{n} = \frac{\der
Q_{\lvert\salgebra_{n}}}{\der\mu_{\lvert\salgebra_{n}}}$ and
$X_{n}=f(\frac{p_{n}}{q_{n}})$,
where $f(x)=x^{\alpha}$ if $\alpha\neq1$ and $f(x)=x\ln x+\e^{-1}$ if
$\alpha=1$. If $\alpha \in (0,1)$, or $\E_{Q}[X_{1}]<\infty$ and $P\ll Q$,
then the family $\{X_{n}\}_{n\geq1}$ is uniformly integrable (with respect to
$Q$).
\end{lemma}

\begin{IEEEproof}
Suppose first that $\alpha \in (0,1)$. Then for any $b>0$
\begin{align*}
\int_{X_{n}>b}X_{n}\intder Q  &  \leq\int_{X_{n}>b}X_{n}\left(
\frac{X_{n}}{b}\right)  ^{(1-\alpha)/\alpha}\intder Q\\
&  \leq b^{-(1-\alpha)/\alpha}\int X_{n}^{1/\alpha}\intder Q\leq
b^{-(1-\alpha)/\alpha},
\end{align*}
and, as $X_{n}\geq0$, $\lim_{b\rightarrow\infty}\sup_{n}\int_{\lvert
X_{n}\rvert>b}\lvert X_{n}\rvert\intder Q=0$, which was to be shown.

Alternatively, suppose that $\alpha \in [1,\infty)$. Then $\frac{p_{n}}{q_{n}}%
=\frac{\der P_{\lvert\salgebra_{n}}}{\der Q_{\lvert\salgebra%
_{n}}}$ ($Q$-a.s.) and hence by Proposition~\ref{prop:SubAlgebraExpectation}
and Jensen's inequality for conditional expectations
\begin{equation*}
  X_n=f\left(  \E\left[  \left.  \frac{\der P}{\der Q}\right\vert \salgebra_n\right] \right)
     \leq\E\left[  \left.f\left(\frac{\der P}{\der Q}\right)  \right\vert \salgebra_n\right]
     =\E\left[\left.X_1\right\vert \salgebra_n\right]
\end{equation*}
($Q$-a.s.) As $\min_{x}\,x\ln x=-\e^{-1}$, it follows that $X_{n}\geq0$
and for any $b,c>0$
\begin{align*}
\int_{\lvert X_{n}\rvert>b}\lvert X_{n}\rvert&\intder Q
    =\int_{X_{n} >b}X_{n}\intder Q\\
  &\leq\int_{X_{n}>b}\E\left[  \left.
X_{1}\right\vert \salgebra_{n}\right]  \intder Q
  =\int_{X_{n}>b} X_{1}\intder Q\\
&  =\int_{X_{n}>b,X_{1}\leq c}X_{1}\intder Q+\int_{X_{n}>b,X_{1}>c}%
X_{1}\intder Q\\
&  \leq c\cdot Q(X_{n}>b)+\int_{X_{1}>c}X_{1}\intder Q\\
&  \leq\frac{c}{b}\E_{Q}[X_{n}]+\int_{X_{1}>c}X_{1}\intder
Q\\
  &\leq\frac{c}{b}\E_{Q}[X_{1}]+\int_{X_{1}>c}X_{1}\intder Q,
\end{align*}
where $\E_{Q}[X_{n}]\leq\E_{Q}[X_{1}]$ in the last inequality
follows from the data processing inequality. Consequently,
\begin{multline*}
\lim_{b\rightarrow\infty}\sup_{n}\int_{\lvert X_{n}\rvert>b}\lvert X_{n}%
\rvert\intder Q=\lim_{c\rightarrow\infty}\lim_{b\rightarrow\infty}\sup
_{n}\int_{\lvert X_{n}\rvert>b}\lvert X_{n}\rvert\intder Q\\
\leq\lim_{c\rightarrow\infty}\lim_{b\rightarrow\infty}\frac{c}{b}%
\E_{Q}[X_{1}]+\lim_{c\rightarrow\infty}\int_{X_{1}>c}X_{1}%
\intder Q=0,
\end{multline*}
and the lemma follows.
\end{IEEEproof}

\begin{IEEEproof}
[Proof of Theorem~\ref{thm:LimitAlgebraDecreasing}]First suppose that
$\alpha>0$ and, for $n=1,2,\ldots,\infty$, let $p_{n}=\frac{\der%
P_{\lvert\salgebra_{n}}}{\der\mu_{\lvert\salgebra_{n}}},q_{n}%
=\frac{\der Q_{\lvert\salgebra_{n}}}{\der\mu_{\lvert
\salgebra_{n}}}$ and $X_{n}=f\left(  \frac{p_{n}}{q_{n}}\right)  $ with
$f(x)=x^{\alpha}$ if $\alpha\neq1$ and $f(x)=x\ln x+\e^{-1}$ if $\alpha=1$, as
in Lemma~\ref{lem:UniformIntegrabilityDecreasing}. If $\alpha\geq1$, then
assume without loss of generality that $\salgebra=\salgebra_{1}$ and
$m=1$, such that $D_{\alpha}(P_{\lvert\salgebra_{m}}\Vert Q_{\lvert
\salgebra_{m}})<\infty$ implies $P\ll Q$. Now, for any $\alpha>0$, it is
sufficient to show that
\begin{equation}
\E_{Q}[X_{n}]\rightarrow\E_{Q}[X_{\infty}].
\label{eqn:DecreasingToProveQ}%
\end{equation}
By Proposition~\ref{prop:SubAlgebraExpectation}, $p_{n}=\E_{\mu
}\left[  \left.  p\right\vert \salgebra_{n}\right]  $ and $q_{n}%
=\E_{\mu}\left[  \left.  q\right\vert \salgebra_{n}\right]  $.
Therefore by a version of L\'{e}vy's theorem for decreasing sequences of
$\sigma$-algebras \cite[Theorem~6.23]{Kallenberg1997},
\[%
\begin{split}
p_{n}  &  =\E_{\mu}\left[  \left.  p\right\vert \salgebra%
_{n}\right]  \rightarrow\E_{\mu}\left[  \left.  p\right\vert
\salgebra_{\infty}\right]  =p_{\infty},\\
q_{n}  &  =\E_{\mu}\left[  \left.  q\right\vert \salgebra%
_{n}\right]  \rightarrow\E_{\mu}\left[  \left.  q\right\vert
\salgebra_{\infty}\right]  =q_{\infty},
\end{split}
\quad(\text{$\mu$-a.s.})
\]
and hence $X_{n}\rightarrow X_{\infty}$ ($\mu$-a.s.\ and therefore $Q$-a.s.)
If $0<\alpha<1$, then
\[
\E_{Q}[X_{n}]=E_{\mu}\left[  p_{n}^{\alpha}q_{n}^{1-\alpha}\right]
\leq\E_{\mu}\left[  \alpha p_{n}+(1-\alpha)q_{n}\right]  =1<\infty.
\]
And if $\alpha\geq1$, then by the data processing inequality $D_{\alpha
}(P_{\lvert\salgebra_{n}}\Vert Q_{\lvert\salgebra_{n}})<\infty$ for all
$n$, which implies that also in this case $\E_{Q}[X_{n}]<\infty$.
Hence uniform integrability (by Lemma~\ref{lem:UniformIntegrabilityDecreasing}%
) of the family of nonnegative random variables $\{X_{n}\}$ implies
\eqref{eqn:DecreasingToProveQ} \cite[Thm.~5, p.\,189]{Shiryaev1996}, and the
theorem follows for $\alpha>0$. The remaining case, $\alpha=0$, is proved by
\begin{align*}
\lim_{n\rightarrow\infty}&D_{0}(P_{\lvert\salgebra_{n}}\Vert Q_{\lvert
\salgebra_{n}})\\
  &=\inf_{n}\inf_{\alpha>0}D_{\alpha}(P_{\lvert
\salgebra_{n}}\Vert Q_{\lvert\salgebra_{n}})
  =\inf_{\alpha>0}\inf_{n}D_{\alpha}(P_{\lvert\salgebra_{n}}\Vert Q_{\lvert\salgebra_{n}})\\
  &=\inf_{\alpha>0}D_{\alpha}(P_{\lvert\salgebra_{\infty}}\Vert Q_{\lvert\salgebra_{\infty}})
  =D_{0}(P_{\lvert\salgebra_{\infty}}\Vert Q_{\lvert\salgebra_{\infty}}).\qedhere
\end{align*}
\end{IEEEproof}

\subsection{Absolute Continuity and Mutual Singularity}

\label{sec:absolutecontinuity}

Shiryaev \cite[pp.\ 366, 370]{Shiryaev1996} relates Hellinger integrals
to absolute continuity and mutual singularity of probability
distributions. His results may more elegantly be expressed in terms of
R\'{e}nyi divergence. They then follow from the observations that
$D_{0}(P\Vert Q)=0$ if and only if $Q$ is absolutely continuous with
respect to $P$ and that $D_{0}(P\Vert Q)=\infty$ if and only if $P$ and
$Q$ are mutually singular, together with right-continuity of
$D_{\alpha}(P\Vert Q)$ in $\alpha$ at $\alpha=0$. As illustrated in the
next section, these properties give a convenient mathematical tool to
establish absolute continuity or mutual singularity of infinite product
distributions.

\begin{theorem}
[{\cite[Theorem 2, p.\,366]{Shiryaev1996}}]\label{thm:abscont} The following
conditions are equivalent:
\begin{enumerate}[(i)]
\item $Q\ll P$,\label{it:abscont1}
\item $Q(p > 0) = 1$,\label{it:abscont2}
\item $D_{0}(P\|Q) = 0$,
\item $\lim_{\alpha\downarrow0} D_{\alpha}(P\Vert Q)=0$.
\end{enumerate}
\end{theorem}

\begin{IEEEproof}
Clearly \eqref{it:abscont2} is equivalent to $Q(p=0)=0$, which is equivalent
to \eqref{it:abscont1}. The other cases follow by $\lim_{\alpha\downarrow
0}D_{\alpha}(P\Vert Q)=D_{0}(P\Vert Q)=-\ln Q(p>0)$.
\end{IEEEproof}

\begin{theorem}
[{\cite[Theorem 3, p.\,366]{Shiryaev1996}}]\label{thm:mutsing} The following
conditions are equivalent:
\begin{enumerate}[(i)]
\item $P \perp Q$,\label{it:mutsing1}
\item $Q(p > 0) = 0$,\label{it:mutsing2}
\item $D_{\alpha}(P\|Q) = \infty$ for some $\alpha \in [0,1)$,\label{it:mutsing3}
\item $D_{\alpha}(P\|Q) = \infty$ for all $\alpha \in [0,\infty]$.\label{it:mutsing4}
\end{enumerate}
\end{theorem}

\begin{IEEEproof}
Equivalence of \eqref{it:mutsing1}, \eqref{it:mutsing2} and $D_{0}(P\Vert Q) =
\infty$ follows from definitions. Equivalence of $D_{0}(P\Vert Q) = \infty$
and \eqref{it:mutsing4} follows from the fact that R\'enyi divergence is
continuous on $[0,1]$ and nondecreasing in $\alpha$. Finally,
\eqref{it:mutsing3} for some $\alpha \in (0,1)$ is equivalent to
\[
\int p^{\alpha}q^{1-\alpha} \intder\mu= 0,
\]
which holds if and only if $pq = 0$ ($\mu$-a.s.). It follows that in this case
\eqref{it:mutsing3} is equivalent to \eqref{it:mutsing1}.
\end{IEEEproof}

\emph{Contiguity} and \emph{entire separation} are asymptotic versions
of absolute continuity and mutual singularity \cite{VanDerVaart1998}. As
might be expected, analogues of Theorems~\ref{thm:abscont} and
\ref{thm:mutsing} also hold for these asymptotic concepts.

Let $(\samplespace_n,\salgebra_n)_{n=1,2,\ldots}$ be a sequence of
measurable spaces, and let $(P_n)_{n=1,2,\ldots}$ and
$(Q_n)_{n=1,2,\ldots}$ be sequences of distributions on these spaces.
Then the sequence $(P_n)$ is \emph{contiguous} with respect to the
sequence $(Q_n)$, denoted $(P_n) \contiguous (Q_n)$, if for all
sequences of events $(A_n \in \salgebra_n)_{n=1,2,\ldots}$ such that
$Q_n(A_n) \to 0$ as $n \to \infty$, we also have $P_n(A_n) \to 0$. If
both $(P_n) \contiguous (Q_n)$ and $(Q_n) \contiguous (P_n)$, then the
sequences are called \emph{mutually contiguous} and we write $(P_n)
\mutcontiguous (Q_n)$. The sequences $(P_n)$ and $(Q_n)$ are
\emph{entirely separated}, denoted $(P_n) \entiresep (Q_n)$, if there
exist a sequence of events $(A_n \in \salgebra_n)_{n=1,2,\ldots}$ and a
subsequence $(n_k)_{k=1,2,\ldots}$ such that $P_{n_k}(A_{n_k}) \to 0$
and $Q_{n_k}(\samplespace_{n_k} \setminus A_{n_k}) \to 0$ as $k \to
\infty$.

Contiguity and entire separation are related to absolute continuity and
mutual singularity in the following way \cite[p.\,369]{Shiryaev1996}: if
$\samplespace_n=\samplespace$, $P_n=P$ and $Q_n=Q$ for all $n$, then
\begin{equation}\label{eqn:contiguitycontinuity}
  \begin{split}
  (P_n) \contiguous (Q_n) \qquad
    & \Leftrightarrow \qquad P\ll Q,\\
  (P_n) \mutcontiguous (Q_n) \qquad
    & \Leftrightarrow\qquad P\sim Q,\\
  (P_n) \entiresep (Q_n) \qquad
    & \Leftrightarrow\qquad P\perp Q.
  \end{split}
\end{equation}
Theorems~1 and 2 by Shiryaev \cite[p.\,370]{Shiryaev1996} imply the
following two asymptotic analogues of Theorems~\ref{thm:abscont} and
\ref{thm:mutsing}:
\begin{theorem} \label{thm:contiguity} The following conditions are
equivalent:
\begin{enumerate}[(i)]
\item $(Q_n) \contiguous (P_n)$,\label{it:contig1}
\item $\displaystyle \lim_{\alpha \downarrow 0}
  \limsup_{n\to\infty} D_\alpha(P_n\Vert Q_n) = 0$.\label{it:contig2}
\end{enumerate}
\end{theorem}

\begin{theorem}\label{thm:entireseparation}
The following conditions are equivalent:
\begin{enumerate}[(i)]
\item $(P_n) \entiresep (Q_n)$,
\item $\displaystyle \lim_{\alpha \downarrow 0}
  \limsup_{n\to\infty} D_\alpha(P_n\Vert Q_n) = \infty$,
\item $\displaystyle\limsup_{n\to\infty}
  D_\alpha(P_n\Vert Q_n) = \infty$ for some $\alpha\in(0,1)$.
\item $\displaystyle\limsup_{n\to \infty}
  D_\alpha(P_n\Vert Q_n) = \infty$ for all $\alpha\in(0,\infty]$.
\end{enumerate}
\end{theorem}
If $P_n$ and $Q_n$ are the restrictions of $P$ and $Q$ to an increasing
sequence of sub-$\sigma$-algebras that generates $\salgebra$, then the
equivalences in \eqref{eqn:contiguitycontinuity} continue to hold,
because we can relate Theorems~\ref{thm:abscont} and
\ref{thm:contiguity} and Theorems~\ref{thm:mutsing} and
\ref{thm:entireseparation} via Theorem~\ref{thm:LimitAlgebraIncreasing}.

\subsection{Distributions on Sequences}

Suppose $(\samplespace^{\infty},\salgebra^{\infty})$ is the \emph{direct
product} of an infinite sequence of measurable spaces $(\samplespace%
_{1},\salgebra_{1}),(\samplespace_{2},\salgebra_{2}),\ldots$ That is,
$\samplespace^{\infty}= \samplespace_{1} \times\samplespace_{2} \times\cdots$ and
$\salgebra^{\infty}$ is the smallest $\sigma$-algebra containing all the
\emph{cylinder sets}
\[
S_{n}(A) = \{x^{\infty}\in\samplespace^{\infty}\mid x_{1},\ldots,x_{n} \in A\},
\qquad A \in\salgebra^{n},
\]
for $n = 1,2,\ldots$, where $\salgebra^{n} = \salgebra_{1} \otimes
\cdots\otimes\salgebra_{n}$. Then a sequence of probability distributions
$P^{1},P^{2},\ldots$, where $P^{n}$ is a distribution on $\samplespace^{n} =
\samplespace_{1} \times\cdots \times \samplespace_{n}$, is called \emph{consistent} if
\[
P^{n+1}(A \times\samplespace_{n+1}) = P^{n}(A), \qquad A \in\salgebra^{n}.
\]
For any such consistent sequence there exists a distribution $P^{\infty}$ on
$(\samplespace^{\infty},\salgebra^{\infty})$ such that its marginal
distribution on $\samplespace^{n}$ is $P^{n}$, in the sense that
\[
P^{\infty}(S_{n}(A)) = P^{n}(A), \qquad A \in\salgebra^{n}.
\]
If $P^{1},P^{2},\ldots$ and $Q^{1},Q^{2},\ldots$ are two consistent sequences
of probability distributions, then it is natural to ask whether the R\'enyi
divergence $D_{\alpha}(P^{n}\Vert Q^{n})$ converges to $D_{\alpha}(P^{\infty
}\Vert Q^{\infty})$. The following theorem shows that it does for $\alpha> 0$.

\begin{theorem}
[Consistent Distributions]\label{thm:consistentdistributions} Let $P^{1}%
,P^{2},\ldots$ and $Q^{1},Q^{2},\ldots$ be consistent sequences of probability
distributions on $(\samplespace^{1},\salgebra^{1}),(\samplespace%
^{2},\salgebra^{2}),\ldots$, where, for $n=1,\ldots,\infty$, $(\samplespace%
^{n},\salgebra^{n})$ is the direct product of the first $n$ measurable
spaces in the infinite sequence $(\samplespace_{1},\salgebra_{1}%
),(\samplespace_{2},\salgebra_{2}),\ldots$ Then for any $\alpha \in
(0,\infty]$
\[
D_{\alpha}(P^{n}\Vert Q^{n})\rightarrow D_{\alpha}(P^{\infty}\Vert Q^{\infty
})
\]
as $n\rightarrow\infty$.
\end{theorem}

\begin{IEEEproof}
Let $\salgebraAlt^{n}=\left\{  S_{n}(A)\mid A\in\salgebra^{n}\right\}  $.
Then
\[
D_{\alpha}(P^{n}|Q^{n})=D_{\alpha}(P_{\lvert\salgebraAlt^{n}}^{\infty}\Vert
Q_{\lvert\salgebraAlt^{n}}^{\infty})\rightarrow D_{\alpha}(P^{\infty}\Vert
Q^{\infty})
\]
by Theorem~\ref{thm:LimitAlgebraIncreasing}.
\end{IEEEproof}

As a special case, we find that finite additivity of R\'enyi divergence, which
is easy to verify, extends to countable additivity:

\begin{theorem}
[Additivity]\label{thm:additivity} For $n=1,2,\ldots$, let $(P_{n},Q_{n})$ be
pairs of probability distributions on measurable spaces $(\samplespace%
_{n},\salgebra_{n})$. Then for any $\alpha \in [0,\infty]$ and any
$N\in\{1,2,\ldots\}$
\begin{equation}
\sum_{n=1}^{N}D_{\alpha}(P_{n}\Vert Q_{n})=D_{\alpha}(P_{1}\times\cdots\times
P_{N}\Vert Q_{1}\times\cdots\times Q_{N}), \label{eqn:finiteadditivity}%
\end{equation}
and, except for $\alpha=0$, also
\begin{equation}
\sum_{n=1}^{\infty}D_{\alpha}(P_{n}\Vert Q_{n})=D_{\alpha}(P_{1}\times
P_{2}\times\cdots\Vert Q_{1}\times Q_{2}\times\cdots).
\label{eqn:countableadditivity}%
\end{equation}

\end{theorem}

Countable additivity as in \eqref{eqn:countableadditivity} does not hold for
$\alpha=0$. A counterexample is given following
Example~\ref{ex:gaussiandichotomy} below.

\begin{IEEEproof}
For simple orders $\alpha$, \eqref{eqn:finiteadditivity} follows from
independence of $P_{n}$ and $Q_{n}$ between different $n$, which implies that
\[
\prod_{n=1}^{N} \int\left(  \frac{\der Q_{n}}{\der P_{n}}\right)
^{1-\alpha} \der P_{n} = \int\left(  \frac{\der \prod_{n=1}^{N}
Q_{n}} {\der \prod_{n=1}^{N} P_{n}}\right)  ^{1-\alpha} \der%
\prod_{n=1}^{N} P_{n}.
\]
As $N$ is finite, this extends to the extended orders by continuity in
$\alpha$. Finally, \eqref{eqn:countableadditivity} follows from
Theorem~\ref{thm:consistentdistributions} by observing that the sequences
$P^{N} = P_{1} \times\cdots\times P_{N}$ and $Q^{N} = Q_{1} \times\cdots\times
Q_{N}$, for $N=1,2,\ldots$, are consistent.
\end{IEEEproof}

Theorems~\ref{thm:abscont} and \ref{thm:mutsing} can be used to
establish absolute continuity or mutual singularity of infinite product
distributions, as illustrated by the following proof by Shiryaev
\cite{Shiryaev1996} of the \emph{Gaussian dichotomy}
\cite{Feldman1958,Hajek1958,Thelen1989}.

\begin{example}
[Gaussian Dichotomy]\label{ex:gaussiandichotomy} Let $P=P_{1}\times
P_{2}\times\cdots$ and $Q=Q_{1}\times Q_{2}\times\cdots$, where $P_{n}$ and
$Q_{n}$ are Gaussian distributions with densities
\[
p_{n}(x)=\tfrac{1}{\sqrt{2\pi}}\e^{-\frac{1}{2}(x-\mu_{n})^{2}},\quad
q_{n}(x)=\tfrac{1}{\sqrt{2\pi}}\e^{-\frac{1}{2}(x-\nu_{n})^{2}}.
\]
Then
\[
D_{\alpha}(P_{n}\Vert Q_{n})=\frac{\alpha}{2}(\mu_{n}-\nu_{n})^{2},
\]
and by additivity for $\alpha > 0$
\begin{equation}\label{eqn:InfiniteGaussians}
D_{\alpha}(P\Vert Q)=\frac{\alpha}{2}\sum_{n=1}^{\infty}(\mu_{n}-\nu_{n})^{2}.
\end{equation}
Consequently, by Theorems~\ref{thm:abscont} and \ref{thm:mutsing} and
symmetry in $P$ and $Q$:
\begin{align}
Q\ll P\quad\Leftrightarrow\quad P\ll Q\quad &  \Leftrightarrow\quad\sum
_{n=1}^{\infty}(\mu_{n}-\nu_{n})^{2}<\infty,\\
Q\perp P\quad &  \Leftrightarrow\quad\sum_{n=1}^{\infty}(\mu_{n}-\nu_{n}%
)^{2}=\infty.
\end{align}
The observation that $P$ and $Q$ are either equivalent (both $P\ll Q$ and
$Q\ll P$) or mutually singular is called the \emph{Gaussian dichotomy}.
\end{example}

By letting $\alpha$ tend to $0$, Example~\ref{ex:gaussiandichotomy}
shows that countable additivity does not hold for $\alpha=0$: if
$\sum_{n=1}^{\infty}(\mu_{n}-\nu_{n})^{2}=\infty$, then
\eqref{eqn:InfiniteGaussians}
implies that $D_{0}(P\Vert Q)=\infty$, while
$\sum_{n=1}^{N}D_{0}(P_{n}\Vert Q_{n})=0$ for all $N$. In light of the
proof of Theorem~\ref{thm:additivity} this also provides a
counterexample to \eqref{eqn:LimitAlgebraIncreasing} for $\alpha=0$.

The Gaussian dichotomy raises the question of whether the same dichotomy holds
for other product distributions. Let $P \sim Q$ denote that $P$ and $Q$ are
\emph{equivalent} (both $P \ll Q$ and $Q \ll P$). Suppose that $P = P_{1}
\times P_{2} \times\cdots$ and $Q = Q_{1} \times Q_{2} \times\cdots$, where
$P_{n}$ and $Q_{n}$ are arbitrary distributions on arbitrary measurable
spaces. Then if $P_{n} \not \sim Q_{n}$ for some $n$, $P$ and $Q$ are not
equivalent either. The question is therefore answered by the following theorem:

\begin{theorem}
[Kakutani's Dichotomy]\label{thm:kakutani} Let $\alpha \in (0,1)$ and let $P =
P_1 \times P_2 \times\cdots$ and $Q = Q_1 \times Q_2 \times\cdots$,
where $P_n$ and $Q_n$ are distributions on arbitrary measurable spaces
such that $P_n \sim Q_n$. Then
\begin{align}
Q \sim P \quad &  \Leftrightarrow\quad\sum_{n=1}^{\infty}D_{\alpha}(P_{n}\Vert
Q_{n}) < \infty,\\
Q \perp P \quad &  \Leftrightarrow\quad\sum_{n=1}^{\infty}D_{\alpha}%
(P_{n}\Vert Q_{n}) = \infty.
\end{align}
\end{theorem}

\begin{IEEEproof}
If $\sum_{n=1}^{\infty}D_{\alpha}(P_{n}\Vert Q_{n}) = \infty$, then
$D_{\alpha}(P\Vert Q) = \infty$ and $Q \perp P$ follows by
Theorem~\ref{thm:mutsing}.

On the other hand, if $\sum_{n=1}^{\infty} D_{\alpha}(P_{n}\Vert Q_{n})
< \infty$, then for every $\varepsilon> 0$ there exists an $N$ such that
\[
\sum_{n=N+1}^{\infty}D_{\alpha}(P_{n}\Vert Q_{n}) \leq\varepsilon,
\]
and consequently by additivity and monotonicity in $\alpha$:
\begin{multline*}
D_{0}(P \Vert Q) = \lim_{\alpha\downarrow0} D_{\alpha}(P\Vert Q)\\
  \leq
\lim_{\alpha\downarrow0} D_{\alpha}(P_{1} \times\cdots\times P_{N} \Vert Q_{1}
\times\cdots\times Q_{N}) + \varepsilon= \varepsilon.
\end{multline*}
As this holds for any $\varepsilon> 0$, $D_{0}(P \Vert Q)$ must equal $0$,
and, by Theorem~\ref{thm:abscont}, $Q \ll P$. As $Q \ll P$ implies $Q
\not \perp P$, Theorem~\ref{thm:mutsing} implies that $D_{\alpha}(Q \Vert P) <
\infty$, and by repeating the argument with the roles of $P$ and $Q$ reversed
we find that also $P \ll Q$, which completes the proof.
\end{IEEEproof}

Theorem~\ref{thm:kakutani} (with $\alpha=\nicefrac{1}{2}$) is equivalent to
a classical result by Kakutani \cite{Kakutani1948}, which was stated in
terms of Hellinger integrals rather than R\'{e}nyi divergence, and
according to Gibbs and Su \cite{GibbsSu2002} might be responsible for
popularising Hellinger integrals. As shown by R\'enyi \cite{Renyi1967},
Kakutani's result is related to the amount of information that a
sequence of observations contains about the parameter of a statistical
model.

\subsection{Taylor Approximation for Parametric
Models}\label{sec:taylorparametric}

Suppose $\{P_{\theta}\mid\theta\in\Theta\subseteq\reals\}$ is
a parametric statistical model. Then it is well known that, for
sufficiently regular parametrisations, a second order Taylor
approximation of $D(P_{\theta}\Vert P_{\theta^{\prime}})$ in
$\theta^{\prime}$ at $\theta$ in the interior of $\Theta$ yields
\begin{equation}
  \lim_{\theta^{\prime}\rightarrow\theta}
    \frac{1}{(\theta-\theta^{\prime})^{2}}
    D(P_{\theta}\Vert P_{\theta^{\prime}})
  =\frac{1}{2}J(\theta),
\end{equation}
where $J(\theta)=\E\left[  (\frac{\der}{\der\theta}\ln
p_{\theta})^{2}\right]$ denotes the \emph{Fisher information} at
$\theta$ (see e.g.\ \cite[Problem~12.7]{CoverThomas1991} or
\cite{Kullback1959}). Haussler and Opper \cite{HausslerOpper1997} argue
that this property generalizes to
\begin{equation}
  \lim_{\theta^{\prime}\rightarrow\theta}
    \frac{1}{(\theta-\theta^{\prime})^{2}}
    D_{\alpha}(P_{\theta}\Vert P_{\theta^{\prime}})
  =\frac{\alpha}{2}J(\theta)
\end{equation}
for any $\alpha \in (0,\infty)$, but we are not aware of a reference
that spells out the exact technical conditions on the parametrisation
that are needed.

\section{Minimax results}\label{sec:minimax}

\subsection{Hypothesis Testing and Chernoff Information}

R\'{e}nyi divergence appears in bounds on the error probabilities when
testing a probabilistic hypothesis $Q$ against an alternative $P$
\cite{Nemetz1974,Csiszar1995,RachedAlajajiCampbell2001}. This can be
explained by the fact that $(1-\alpha)D_{\alpha}(P\Vert Q)$ equals the
\emph{cumulant generating function} for the random variable $\ln(p/q)$
under the distribution $Q$ (provided $\alpha \in (0,1)$ or $P \ll Q$)
\cite{Csiszar1995}. The following theorem relates this cumulant
generating function to two Kullback-Leibler divergences that involve the
distribution $P_{\alpha}$ with density
\begin{equation}\label{eqn:exponentialfamily}
  p_{\alpha}=\frac{q^{1-\alpha}p^{\alpha}}
                  {\int q^{1-\alpha}p^\alpha \intder\mu},
\end{equation}
which is well defined if and only if $0 < \int p^\alpha
q^{1-\alpha}\intder\mu < \infty$.

\begin{theorem}\label{thm:MomentGeneratingFunction}
For any simple order $\alpha$
\begin{equation}\label{eqn:MomentGeneratingFunction}
  (1-\alpha)D_{\alpha}(P\Vert Q)
    =\inf_{R}\left\{  \alpha D(R\Vert P)+(1-\alpha)D(R\Vert Q)\right\},
\end{equation}
with the convention that $\alpha D(R\Vert P)+(1-\alpha)D(R\Vert Q) =
\infty$ if it would otherwise be undefined. Moreover, if the
distribution $P_\alpha$ with density \eqref{eqn:exponentialfamily} is
well defined and $\alpha \in (0,1)$ or $D(P_\alpha \Vert P) < \infty$,
then the infimum is uniquely achieved by $R = P_\alpha$.
\end{theorem}

This result gives an interpretation of R\'enyi divergence as a trade-off
between two Kullback-Leibler divergences.

\begin{remark}
Theorem~\ref{thm:MomentGeneratingFunction} was formulated and proved for
distributions on finite sets by Shayevitz \cite{Shayevitz2010}, but
appeared in the above formulation already in \cite{VanErven2010}. Prior
to either of these, the identity \eqref{eqn:csiszarsidentity} below,
which forms the heart of the proof, has been used by Csisz{\'a}r
\cite{Csiszar2003}.
\end{remark}

\begin{IEEEproof}[Proof of Theorem~\ref{thm:MomentGeneratingFunction}]
  First suppose that $P_\alpha$ is well defined or, equivalently, that
  $D_\alpha(P \Vert Q) < \infty$. Then for $\alpha \in (0,1)$ or
  $D(R\Vert P) < \infty$, we have
  \begin{equation}\label{eqn:csiszarsidentity}
    \alpha D(R \Vert P) + (1-\alpha) D(R \Vert Q)
      = D(R \Vert P_\alpha) -\ln \int p^\alpha q^{1-\alpha} \
      \der\mu.
  \end{equation}
  Hence, if $0 < \alpha < 1$ or $D(P_\alpha \Vert P) < \infty$, the
  infimum over $R$ is uniquely achieved by $R = P_\alpha$, for which it
  equals $(1-\alpha)D_\alpha(P \Vert Q)$ as required. If, on the other
  hand, $\alpha > 1$ and $D(P_\alpha \Vert P) = \infty$, then we still
  have
  \begin{equation}\label{eqn:MomentGeneratingFunctionUpper}
    \inf_R \big\{ \alpha D(R \Vert P) + (1-\alpha) D(R \Vert Q)\big\}
      \geq (1-\alpha)D_\alpha(P \Vert Q).
  \end{equation}

  Secondly, suppose $\alpha \in (0,1)$ and $D_\alpha(P \Vert Q) =
  \infty$. Then $P \perp Q$, and consequently either $D(R \Vert P) =
  \infty$ or $D(R \Vert Q) = \infty$ for all $R$, which means that
  \eqref{eqn:MomentGeneratingFunction} holds.

  Next, consider the case that $\alpha > 1$ and $P \not \ll Q$. Then
  $D_\alpha(P\Vert Q) = \infty$ and the infimum over $R$ is achieved by
  $R = P$, for which it equals $-\infty$, and again
  \eqref{eqn:MomentGeneratingFunction} holds.

  Finally, we prove \eqref{eqn:MomentGeneratingFunction} for the
  remaining cases: $\alpha > 1,P \ll Q$ and either: (1)
  $D_\alpha(P \Vert Q) < \infty$, but $D(P_\alpha \Vert P) = \infty$; or
  (2) $D_\alpha(P \Vert Q) = \infty$. To this end, let $P_c = P(\cdot
  \mid p \leq cq)$ for all $c$ that are sufficiently large that $P(p
  \leq cq) > 0$. The reader may verify that $D_\alpha(P_c \Vert Q) <
  \infty$ and $D(S \Vert P_c) < \infty$ for $s = p_c^\alpha
  q^{1-\alpha}/\int p_c^\alpha q^{1-\alpha}\ \der\mu$, so that we
  have already proved that \eqref{eqn:MomentGeneratingFunction} holds if
  $P$ is replaced by $P_c$. Hence, observing that for all $R$
  \begin{equation*}
    D(R \Vert P_c) =
      \begin{cases}
        \infty & \text{if $R \not \ll P_c$,}\\
        D(R \Vert P) + \ln P(p \leq pc) & \text{otherwise,} 
      \end{cases}
  \end{equation*}
  we find that
  \begin{align*}
    \inf_R &\big\{ \alpha D(R \Vert P)
                   + (1-\alpha) D(R \Vert Q)\big\}\\
    &\leq \limsup_{c \to \infty} \Big(-\alpha \ln P(p \leq cq)\\
    &\qquad\qquad +\inf_R \big\{ \alpha D(R \Vert P_c)
                   + (1-\alpha) D(R \Vert Q)\big\}\Big)\\
    &\leq \limsup_{c \to \infty}\ (1-\alpha)D_\alpha(P_c \Vert Q)
    \leq (1-\alpha)D_\alpha(P \Vert Q),
  \end{align*}
  where the last inequality follows by lower semi-continuity of
  $D_\alpha$ (Theorem~\ref{thm:lowersemicontinuity}). In case~2,
  \eqref{eqn:MomentGeneratingFunction} follows immediately. In case~1,
  \eqref{eqn:MomentGeneratingFunction} follows by combining this
  inequality with its converse
  \eqref{eqn:MomentGeneratingFunctionUpper}.
\end{IEEEproof}

Theorem~\ref{thm:MomentGeneratingFunction} shows that
$(1-\alpha)D_{\alpha}(P\Vert Q)$ is the infimum over a set of functions
that are linear in $\alpha$, which implies the following corollary:
\begin{corollary}\label{cor:concavealpha}
  The function $(1-\alpha)D_{\alpha}(P\Vert Q)$ is concave in $\alpha$
  on $[0,\infty]$, with the conventions that it is $0$ at $\alpha = 1$
  even if $D(P\Vert Q) = \infty$ and that it is $0$ at $\alpha=\infty$
  if $P=Q$.
\end{corollary}

\begin{IEEEproof}
  Suppose first that $D(P \Vert Q) < \infty$. Then
  \eqref{eqn:MomentGeneratingFunction} also holds at $\alpha=1$. Hence
  $(1-\alpha)D_{\alpha}(P\Vert Q)$ is a point-wise infimum over linear
  functions on $(0,\infty)$, and thus concave. This extends to $\alpha
  \in \{0,\infty\}$ by continuity.

  Alternatively, suppose that $D(P \Vert Q) = \infty$. Then
  $(1-\alpha)D_{\alpha}(P\Vert Q)$ is still concave on $[0,1)$, where it
  is also nonnegative. And by monotonicity of R\'enyi divergence, we
  have that $D_\alpha(P \Vert Q) = \infty$ for all $\alpha \geq 1$.
  Consequently, $(1-\alpha)D_{\alpha}(P\Vert Q)$ is nonnegative and
  concave for $\alpha \in [0,1)$, at $\alpha = 1$ it is $0$ (by
  convention) and for $\alpha \in (1,\infty]$ it is $-\infty$. It then
  follows that $(1-\alpha)D_{\alpha}(P\Vert Q)$ is concave on all of
  $[0,\infty]$, as required.
\end{IEEEproof}

In addition, Theorem~\ref{thm:MomentGeneratingFunction} can be used to
prove Gilardoni's extension of Pinsker's inequality from the case
$\alpha = 1$ to any $\alpha \in (0,1]$ \cite{Gilardoni2010}, which was
mentioned in the introduction.

\begin{theorem}[Pinsker's Inequality]\label{thm:pinsker}
  Let $V(P,Q)$ be the total variation distance, as defined in
  \eqref{eqn:totalvariation}. Then, for any $\alpha \in (0,1]$,
  \begin{equation*}
    \frac{\alpha}{2} V^2(P,Q) \leq D_\alpha(P\|Q).
  \end{equation*}
\end{theorem}

\begin{IEEEproof}
  We omit the proof for $\alpha = 1$, which is the standard version of
  Pinsker's inequality (see \cite{FedotovHarremoesTopsoe2003} for a
  survey of its history). For $\alpha \in (0,1)$, consider first the
  case of two distributions $P = (p,1-p)$ and $Q=(q,1-q)$ on a binary
  alphabet. Then $V^2(P,Q) = 4(p-q)^2$ and by
  Theorem~\ref{thm:MomentGeneratingFunction} and the result for $\alpha
  = 1$, we find
  \begin{align*}
  (1-\alpha) D_{\alpha}(P\Vert Q)
    &= \inf_{R}\left\{\alpha D(R\Vert P)+(1-\alpha) D(R\Vert
    Q)\right\}\\
    &\geq \inf_r \left\{2 \alpha (r-p)^2 + 2(1-\alpha) (r-q)^2\right\}.
  \end{align*}
  The minimum is achieved by $r = \alpha p + (1-\alpha) q$, from which
  \begin{equation*}
    D_{\alpha}(P\Vert Q) \geq 2 \alpha (p-q)^2 = \frac{\alpha}{2}V^2(P,Q).
  \end{equation*}
  The general case of distributions $P$ and $Q$ on any sample space
  $\samplespace$ reduces to the binary case by the data processing
  inequality: for any event $A$, let $P_{\lvert A}$ and $Q_{\lvert A}$
  denote the restrictions of $P$ and $Q$ to the binary partition
  $\partition = \{A, \samplespace \setminus A\}$. Then
  \begin{align*}
    \tfrac{2}{\alpha} D_\alpha(P \Vert Q)
      &\geq \sup_A \tfrac{2}{\alpha} D_\alpha(P_{\lvert A} \Vert Q_{\lvert A})
      \geq \sup_A V^2(P_{\lvert A}, Q_{\lvert A})\\
      &= \sup_A 4 \big(P(A)-Q(A)\big)^2
      = V^2(P,Q),
  \end{align*}
  as required.
\end{IEEEproof}

As one might expect from continuity of $D_\alpha(P\|Q)$, the terms on
the right-hand side of \eqref{eqn:MomentGeneratingFunction} are
continuous in $\alpha$, at least on $(0,1)$:
\begin{lemma}\label{lem:continuousPalpha}
  If $D(P \Vert Q) < \infty$ or $D(Q \Vert P) < \infty$, then both
  $D(P_\alpha \Vert Q)$ and $D(P_\alpha \Vert P)$ are finite and
  continuous in $\alpha$ on $(0,1)$.
\end{lemma}

\begin{IEEEproof}
  The lemma is symmetric in $P$ and $Q$, so suppose without loss of
  generality that $D(P \Vert Q) < \infty$. Then $D_\alpha(P \Vert Q)
  \leq D(P \Vert Q) < \infty$ implies that $P_\alpha$ is well defined
  and finiteness of both $D(P_\alpha \Vert Q)$ and $D(P_\alpha \Vert P)$
  follows from Theorem~\ref{thm:MomentGeneratingFunction}. Now observe
  that
  \begin{multline*}
    D(P_\alpha \Vert Q)
      = \frac{1}{\int p^\alpha q^{1-\alpha}\intder \mu}
        \E_Q \left[
          \left(\frac{p}{q}\right)^\alpha \ln \left(\frac{p}{q}\right)^\alpha 
        \right]\\
      + (1-\alpha)D_\alpha(P \Vert Q).
  \end{multline*}
  Then by continuity of $D_\alpha(P \Vert Q)$ and hence of $\int
  p^\alpha q^{1-\alpha}\intder \mu$ in $\alpha$, it is sufficient to
  verify continuity of $\E_Q [(p/q)^\alpha \ln (p/q)^\alpha]$. To this
  end, observe that
  \begin{equation*}
    \lvert (p/q)^\alpha \ln (p/q)^\alpha \rvert \leq 
      \begin{cases}
        1/\e    & \text{if $p < q$,}\\
        (p/q)\ln (p/q) & \text{if $p \geq q$.}
      \end{cases}
  \end{equation*}
  As $D(P \Vert Q) < \infty$ implies $\E_Q[ \ind_{\{p \geq q\}} (p/q)\ln
  (p/q) ] < \infty$, we may apply the dominated convergence theorem to
  obtain
  \begin{equation*}
    \lim_{\alpha \to \alpha^*}
        \E_Q \left[
            \left(\frac{p}{q}\right)^\alpha \ln \left(\frac{p}{q}\right)^\alpha 
          \right]
    = \E_Q \left[
            \left(\frac{p}{q}\right)^{\alpha^*} \ln
            \left(\frac{p}{q}\right)^{\alpha^*}
          \right]
  \end{equation*}
  for any $\alpha^* \in (0,1)$, which proves continuity of $D(P_\alpha
  \Vert Q)$. Continuity of $D(P_\alpha \Vert P)$ now follows from
  Theorem~\ref{thm:MomentGeneratingFunction} and continuity of
  $(1-\alpha)D_\alpha(P \Vert Q)$.
\end{IEEEproof}

\begin{theorem}\label{thm:ChernoffInformationDivergence}
Suppose that $D(P \Vert Q) < \infty$. Then the following minimax
identity holds:
\begin{multline}\label{eqn:Chernoff}
\sup_{\alpha \in (0,\infty)}\inf_{R}\left\{  \alpha D(R\Vert P)+(1-\alpha)D(R\Vert
Q)\right\}\\  =\inf_{R}\sup_{\alpha \in (0,\infty)}\left\{  \alpha D(R\Vert P)+(1-\alpha
)D(R\Vert Q)\right\},
\end{multline}
with the convention that $\alpha D(R\Vert P)+(1-\alpha)D(R\Vert Q) =
\infty$ if it would otherwise be undefined. Moreover,
\eqref{eqn:Chernoff} still holds if $\alpha$ is restricted to $(0,1)$ on
its left-hand side; and if there exists an $\alpha^* \in (0,1)$ such
that $D(P_{\alpha^*}\Vert P)=D(P_{\alpha^*}\Vert Q)$, then
$(\alpha^*,P_{\alpha^*})$ is a saddle-point for \eqref{eqn:Chernoff} and
both sides of \eqref{eqn:Chernoff} are equal to
\begin{equation}\label{eqn:Chernoffvalue}
  \begin{split}
  (1-\alpha^*) D_{\alpha^*}(P \Vert Q)
    &= \sup_{\alpha \in (0,1)} (1-\alpha) D_\alpha(P \Vert Q)\\
    &= D(P_{\alpha^*} \Vert P) = D(P_{\alpha^*} \Vert Q).
  \end{split}
\end{equation}
\end{theorem}

The minimax value defined in \eqref{eqn:Chernoff} is the \emph{Chernoff
information}, which gives an asymptotically tight bound on both the type
1 and the type 2 errors in tests of $P$ vs.\ $Q$. The same connection
between Chernoff information and $D(P_{\alpha^{\ast}}\Vert P)$ is
discussed by Cover and Thomas \cite[Section~12.9]{CoverThomas1991}, with a
different proof.

\begin{IEEEproof}[Proof of
Theorem~\ref{thm:ChernoffInformationDivergence}]
Let $f(\alpha,R)=\alpha D(R\Vert P)+(1-\alpha)D(R\Vert Q)$. 
For $\alpha \in (0,1)$, $D_\alpha(P \Vert Q) \leq D(P\Vert Q) < \infty$
implies that $P_\alpha$ is well defined. Suppose there exists $\alpha^*
\in (0,1)$ such that $D(P_{\alpha^*} \Vert P)= D(P_{\alpha^*} \Vert Q)$.
Then Theorem~\ref{thm:MomentGeneratingFunction} implies that
$(\alpha^*,P_{\alpha^*})$ is a saddle-point for $f(\alpha,R)$, so that
\eqref{eqn:Chernoff} holds \cite[Lemma~36.2]{Rockafellar1970}, and
Theorem~\ref{thm:MomentGeneratingFunction} also implies that all
quantities in \eqref{eqn:Chernoffvalue} are equal to
$f(\alpha^*,P_{\alpha^*})$.

Let $\alphaset$ be either $(0,1)$ or $(0,\infty)$. As the $\sup \inf$ is
never bigger than the $\inf \sup$ \cite[Lemma~36.1]{Rockafellar1970}, we
have that
\begin{equation*}
  \sup_{\alpha \in \alphaset} \inf_R f(\alpha,R) \leq 
  \sup_{\alpha \in (0,\infty)} \inf_R f(\alpha,R) \leq \inf_R \sup_{\alpha
  \in (0,\infty)} f(\alpha,R),
\end{equation*}
so it remains to prove the converse inequality.

By Lemma~\ref{lem:continuousPalpha} we know that both $D(P_\alpha \Vert
P)$ and $D(P_\alpha \Vert Q)$ are finite and continuous in $\alpha$ on
$(0,1)$. By the intermediate value theorem, there are therefore three
possibilities: (1) there exists $\alpha^* \in (0,1)$ such that
$D(P_{\alpha^*} \Vert P) = D(P_{\alpha^*} \Vert Q)$, for which we have already
proved \eqref{eqn:Chernoff}; (2) $D(P_\alpha \Vert P) < D(P_\alpha \Vert
Q)$ for all $\alpha \in (0,1)$; and (3) $D(P_\alpha \Vert P) >
D(P_\alpha \Vert Q)$ for all $\alpha \in (0,1)$.

We proceed with case (2), observing that
\begin{align*}
  \inf_R \sup_{\alpha \in (0,\infty)} f(\alpha,R)
    &= \inf_{R \colon D(R \Vert Q) < \infty}
       \sup_{\alpha \in (0,\infty)} f(\alpha,R)\\
    &= \inf_{R \colon D(R \Vert Q) < \infty}
       \Big\{ D(R \Vert Q)\\
    &\quad+ \sup_{\alpha \in (0,\infty)}
        \alpha \big(D(R \Vert P) - D(R\Vert Q)\big)\Big\}\\
    &= \inf_{R \colon D(R \Vert P) \leq D(R \Vert Q) < \infty}
       D(R \Vert Q)\\
    &\leq \inf_{0 < \alpha < 1}
       D(P_\alpha \Vert Q).
\end{align*}
Now by Theorem~\ref{thm:MomentGeneratingFunction}
\begin{align*}
  \inf_{0 < \alpha < 1}
       &D(P_\alpha \Vert Q)
  \leq \liminf_{\alpha \downarrow 0} D(P_\alpha \Vert Q)\\
  &= \liminf_{\alpha \downarrow 0} \Big\{D_\alpha(P \Vert Q) -
  \frac{\alpha}{1-\alpha} D(P_\alpha \Vert P)\Big\}\\
  &\leq \lim_{\alpha \downarrow 0} D_\alpha(P \Vert Q)
  =\lim_{\alpha \downarrow 0} (1-\alpha) D_\alpha(P \Vert Q)\\
  &=\lim_{\alpha \downarrow 0} \inf_R f(\alpha,R)
  \leq \sup_{\alpha \in \alphaset} \inf_R f(\alpha,R),
\end{align*}
as required. It remains to consider case (3), which turns out to be
impossible by the following argument: two applications of
Theorem~\ref{thm:MomentGeneratingFunction} give
\begin{align*}
  D_{\nicefrac{1}{2}}&(P \Vert Q)
    = \inf_{0 < \alpha < 1} \Big\{D(P_\alpha \Vert P) + D(P_\alpha \Vert Q)\Big\}\\
    &\leq 2 \inf_{0 < \alpha < 1} D(P_\alpha \Vert P)
    \leq 2 \limsup_{\alpha \uparrow 1} D(P_\alpha \Vert P)\\
    &= 2 \limsup_{\alpha \uparrow 1} \Big\{
      \frac{1-\alpha}{\alpha} D_\alpha(P \Vert Q)
      - \frac{1-\alpha}{\alpha} D(P_\alpha \Vert P)\Big\}\\
    &\leq 2 \limsup_{\alpha \uparrow 1}
      \frac{1-\alpha}{\alpha} D_\alpha(P \Vert Q)
    = 0.
\end{align*}
It follows that $P = Q$, which contradicts the assumption that
$D(P_\alpha \Vert P) > D(P_\alpha \Vert Q)$ for any $\alpha \in
(0,1)$.
\end{IEEEproof}

\subsection{Channel Capacity and Minimax Redundancy}\label{sec:channelcapacity}

Consider a non-empty family $\{P_{\theta} \mid \theta\in\Theta\}$ of
probability distributions on a sample space $\samplespace$. We may think
of $\theta$ as a parameter in a statistical model or as an input letter
of an information channel. In the main results of this section we will
only consider discrete sample spaces $\samplespace$, which are either
finite with $n$ elements or countably infinite. Whenever distributions
on $\Theta$ are involved, we also implicitly assume that $\Theta$ is a
topological space that is equipped with the Borel $\sigma$-algebra, that
$\{\theta\}$ is a closed set for every $\theta$, and that the map
$\theta \mapsto P_\theta$ is measurable.

We will study
\begin{equation}
  C_\alpha = \sup_{\pi}\inf_{Q} \int D_{\alpha}\left(  P_{\theta}\Vert
  Q\right)\intder \pi(\theta),
\end{equation}
which has been proposed as the appropriate generalization of the
\emph{channel capacity} from $\alpha=1$ to general $\alpha$
\cite{Csiszar1995,Shayevitz2011}.

If $\samplespace$ is finite, then the channel capacity is also finite:
\begin{theorem}
  If $\samplespace$ has $n$ elements, then $C_\alpha \leq \ln n$ for any
  $\alpha \in [0,\infty]$.
\end{theorem}

\begin{IEEEproof}
  Let $U$ denote the uniform distribution on $\samplespace$. Then
  \begin{align*}
    \sup_{\pi}\inf_{Q} &\int D_{\alpha}\left(  P_{\theta}\Vert
        Q\right)\intder \pi(\theta)
      \leq \sup_{\pi} \int D_{\alpha}\left(  P_{\theta}\Vert
        U\right)\intder \pi(\theta)\\
      &= \sup_\theta D_{\alpha}\left(  P_{\theta}\Vert
        U\right)
      \leq \sup_\theta D_{\infty}\left(  P_{\theta}\Vert
        U\right)\\
      &=\sup_\theta \ln \max_x \frac{P_\theta(x)}{1/n}
      \leq \ln n.\qedhere
  \end{align*}
\end{IEEEproof}

For $\alpha=1$, it is a classical result by Gallager and Ryabko
\cite{Ryabko1981} that the channel capacity equals the \emph{minimax
redundancy}:
\begin{equation}
  R_\alpha = \inf_Q \sup_{\theta \in \Theta} D_\alpha(P_\theta \| Q).
\end{equation}
For finite $\Theta$, Csisz\'ar \cite{Csiszar1995} has shown that this
result in fact extends to any $\alpha \in (0,\infty)$, noting that
the minimax redundancy $R_\alpha$ (and therefore the channel capacity
$C_\alpha$) may be geometrically interpreted as the ``radius'' of the
family of distributions $\{P_\theta \mid \theta \in \Theta\}$ with
respect to the R\'enyi divergence of order $\alpha$. It turns out that
Csisz\'ar's result extends to general $\Theta$ and all orders $\alpha$:
\begin{theorem}\label{thm:radius}
  Suppose $\samplespace$ is finite. Then for any $\alpha \in [0,\infty]$
  the channel capacity equals the minimax redundancy:
  \begin{equation}\label{eqn:radius}
    C_\alpha = R_\alpha.
  \end{equation}
\end{theorem}
For $\alpha=1$, Haussler \cite{Haussler1997} has extended this result to
infinite sample spaces $\samplespace$. It seems plausible that his
approach might extend to other orders $\alpha$ as well.

Equation~\ref{eqn:radius} is equivalent to the minimax identity
\begin{equation}\label{eqn:minmaxradius}
  \sup_{\pi}\inf_{Q} \psi_\alpha(\pi,Q)
    = \inf_{Q}\sup_{\pi} \psi_\alpha(\pi,Q),
\end{equation}
where
\begin{equation}
  \psi_\alpha(\pi,Q)
    = \int D_{\alpha}\left(  P_{\theta}\Vert Q\right)\intder
      \pi(\theta).
\end{equation}
We will prove this identity using Sion's minimax theorem
\cite{Sion1958,Komiya1988}, which we state with its arguments exchanged
to make them line up with the arguments of $\psi_\alpha$:
\begin{theorem}[Sion's Minimax Theorem]\label{thm:Sion}
  Let $A$ be a convex subset of a linear topological space and $B$ a
  compact convex subset of a linear topological space. Let $f \colon A
  \times B \to \reals$ be such that
  \begin{enumerate}[(i)]
  \item $f(\cdot,b)$ is upper semi-continuous and quasi-concave on $A$
  for each $b \in B$;
  \item $f(a,\cdot)$ is lower semi-continuous and quasi-convex on $B$
  for each $a \in A$.
  \end{enumerate}
  Then
  \begin{equation*}
    \sup_{a \in A}\min_{b \in B} f(a,b)
      = \min_{b \in B} \sup_{a \in A} f(a,b).
  \end{equation*}
\end{theorem}
\begin{IEEEproof}[Proof of Theorem~\ref{thm:radius}]
  Sion's minimax theorem cannot be applied directly, because
  $\psi_\alpha$ may be infinite. For $\lambda \in (0,1)$, we therefore
  introduce the auxiliary function
  \begin{equation*}
    \psi_\alpha^\lambda(\pi,Q)
      = \psi_\alpha\big(\pi, (1-\lambda)U + \lambda Q\big),
  \end{equation*}
  where $U$ is the uniform distribution on $\samplespace$. Finiteness of
  $\psi_\alpha^\lambda$ follows from
  \begin{equation}\label{eqn:boundedness}
    \begin{split}
    D_\alpha\big(P_\theta\| (1-\lambda)U + \lambda Q\big)
      \leq D_\alpha\big(P_\theta\| U\big) -\ln (1-\lambda)\\
      \leq D_\infty\big(P_\theta\| U\big) -\ln (1-\lambda)
      \leq \ln n -\ln (1-\lambda),
    \end{split}
  \end{equation}
  where $n$ denotes the number of elements in $\samplespace$.

  To verify the other conditions of Theorem~\ref{thm:Sion}, we observe
  that $\psi_\alpha^\lambda(\cdot, Q)$ is linear, and hence continuous
  and concave. Convexity of $\psi_\alpha^\lambda(\pi,\cdot)$ follows
  from convexity of $\psi_\alpha(\pi,\cdot)$, which holds because
  $\psi_\alpha(\pi,\cdot)$ is a linear combination of convex functions.
  Continuity of $\psi_\alpha^\lambda(\pi,\cdot)$ follows by the
  dominated convergence theorem (which applies by
  \eqref{eqn:boundedness}) and continuity of
  $D_\alpha(P_\theta\|\cdot)$. Thus we may apply Sion's minimax theorem.

  By
  \begin{equation*}
    D_\alpha\big(P_\theta\| (1-\lambda)U + \lambda Q\big)
      \leq D_\alpha\big(P_\theta\| Q\big) -\ln \lambda,
  \end{equation*}
  we also have $\psi_\alpha^\lambda(\pi,Q) \leq
  \psi_\alpha(\pi,Q) -\ln \lambda$, and hence we may reason as follows:
  \begin{multline*}
    \sup_\pi \inf_Q \psi_\alpha(\pi,Q) - \ln \lambda
      \geq \sup_\pi \inf_Q \psi_\alpha^\lambda(\pi,Q)\\
      = \inf_Q \sup_\pi \psi_\alpha^\lambda(\pi,Q)
      \geq \inf_Q \sup_\pi \psi_\alpha(\pi,Q).
  \end{multline*}
  By letting $\lambda$ tend to $1$ we find
  \begin{equation*}
    \sup_\pi \inf_Q \psi_\alpha(\pi,Q)
      \geq \inf_Q \sup_\pi \psi_\alpha(\pi,Q).
  \end{equation*}
  As the $\sup \inf$ never exceeds the $\inf \sup$
  \cite[Lemma~36.1]{Rockafellar1970}, the converse inequality also
  holds, and the proof is complete.
\end{IEEEproof}

A distribution $\piopt$ on the parameter space $\Theta$ is a \emph{capacity
achieving} input distribution if
\begin{equation}
  \inf_{Q}\int D_{\alpha}\left(  P_{\theta}\Vert Q\right)  \intder
  \piopt(\theta)
    =C_{\alpha}.
\end{equation}
A distribution $\Qopt$ on $\samplespace$ may be called a
\emph{redundancy achieving} distribution if
\begin{equation}
  \sup_\theta D_{\alpha}\left(  P_{\theta}\Vert \Qopt\right)
    =R_{\alpha}.
\end{equation}

If the sample space is finite, then a redundancy achieving distribution
always exists:
\begin{lemma}\label{lem:supredundancyfunction}
  Suppose $\samplespace$ is finite and let $\alpha \in [0,\infty]$. Then
  the function $Q \mapsto \sup_\theta D_\alpha(P_\theta\|Q)$ is
  continuous and convex, and has at least one minimum. Consequently, a
  redundancy achieving distribution $\Qopt$ exists.
\end{lemma}

\begin{IEEEproof}
  Denote the number of elements in $\samplespace$ by $n$, let
  $\simplex_n = \{(p_1,\ldots,p_n)\mid \sum_{i=1}^n p_i = 1, p_i \geq
  0\}$ denote the probability \emph{simplex} on $n$ outcomes, and let
  $f(Q) = \sup_\theta D_\alpha(P_\theta\|Q)$. Since $f$ is the supremum
  over continuous, convex functions, it is lower semi-continuous and
  convex itself. As the domain of $f$ is $\simplex_n$, which is compact,
  this implies that it attains its minimum. Moreover, convexity on a
  simplex implies upper semi-continuity
  \cite[Theorem~10.2]{Rockafellar1970}, so that $f$ is both lower and
  upper semi-continuous, which means that it is continuous.
\end{IEEEproof}

\begin{theorem}\label{thm:center}
  Suppose $\samplespace$ is finite and let $\alpha\in [0,\infty]$. If
  there exists a (possibly non-unique) capacity achieving input
  distribution $\piopt$, then $\int D_\alpha(P_\theta\| Q) \intder
  \piopt(\theta)$ is minimized by $Q=\Qopt$ and $D_\alpha(P_\theta\|
  \Qopt) = R_\alpha$ almost surely under $\piopt$.
\end{theorem}

If $R_\alpha$ is regarded as the radius of $\{P_\theta \mid \theta \in
\Theta\}$, then this theorem shows how $\Qopt$ may be interpreted as its
center.

\begin{IEEEproof}
Since $\piopt$ is capacity achieving,
\begin{align*}
C_{\alpha}  &  =\inf_{Q}\int D_{\alpha}\left(  P_{\theta}\Vert Q\right)
\intder \piopt(\theta)\\
&  \leq\int D_{\alpha}\left(  P_{\theta}\Vert \Qopt\right)  \intder%
\piopt(\theta)\\
&  \leq\int R_{\alpha}\intder \piopt(\theta)
  = R_\alpha = C_{\alpha}.
\end{align*}
The result follows because both inequalities must be equalities.
\end{IEEEproof}

Three orders $\alpha$ for the channel capacity $C_\alpha$ and minimax
redundancy $R_\alpha$ are of particular interest. The classical ones are
$\alpha=1$, because it corresponds to the original definition of channel
capacity by Shannon, and $\alpha = 0$ because $C_0$ gives an upper bound
on the zero error capacity, which also dates back to Shannon.

Now let us look at the case $\alpha=\infty$, assuming for simplicity
that $\samplespace$ is countable. We find that
\begin{align}
  \sup_\theta D_\infty(P_\theta\|Q)
    &= \sup_\theta \ln \sup_x \frac{P_\theta(x)}{Q(x)}\notag\\
    &= \sup_x\ln\frac{\sup_{\theta}P_{\theta}(x)}{Q(x)}
\end{align}
is the \emph{worst-case regret} of $Q$ relative to $\{P_\theta \mid
\theta \in \Theta\}$ \cite{Grunwald2007}. As is well known
\cite{Shtarkov1987,Grunwald2007}, the distribution that minimizes
the worst-case regret is uniquely given by the \emph{normalized maximum
likelihood} or \emph{Shtarkov} distribution
\begin{equation}\label{eqn:shtarkov}
  S(x) = \frac{\sup_\theta P_\theta(x)}{\sum_x \sup_\theta P_\theta(x)},
\end{equation}
provided that the normalizing sum is finite, so that $S$ is
well defined.
\begin{theorem}\label{thm:capacityequalityInfty}
  Suppose that $\samplespace$ is countable and that the minimax
  redundancy $R_\infty$ is finite. Then $S$ is well defined and the
  worst-case regret of any distribution $Q$ satisfies
  \begin{equation}\label{eqn:regret}
    \sup_\theta D_\infty(P_\theta\| Q)
      = R_\infty
       +D_\infty(S\| Q).
  \end{equation}
  In particular, $\Qopt = S$ is unique and
  \begin{equation}\label{eqn:regretvalue}
    R_\infty = \ln \sum_x \sup_\theta P_\theta(x) < \infty.
  \end{equation}
\end{theorem}

\begin{IEEEproof}
  Since $R_\infty < \infty$, for any finite $C > R_\infty$ there must
  exist a distribution $Q_C$ such that
  $\sup_x\ln\frac{\sup_{\theta}P_{\theta}(x)}{Q_C(x)} \leq C$. Hence
  \begin{equation*}
    \sum_x \sup_{\theta}P_{\theta}(x)
      \leq \sum_x Q_C(x) \e^C
      = \e^C < \infty,
  \end{equation*}
  so that $S$ is well defined.
  
  Now for any arbitrary distribution $Q$, we have
  \begin{align*}
    \sup_x \ln\frac{\sup_\theta P_\theta(x)}{Q(x)} 
      &=\sup_x \Big(\ln\frac{\sup_\theta P_\theta(x)}{S(x)}
        +\ln\frac{S(x)}{Q(x)}\Big)\\
      &=\ln \sum_x \sup_\theta P_\theta(x)
        +\sup_x \ln\frac{S(x)}{Q(x)}\\
      &=\sup_x \ln\frac{\sup_\theta P_\theta(x)}{S(x)}
        +\sup_x \ln\frac{S(x)}{Q(x)}.
  \end{align*}
  Since $\sup_x \ln\frac{S(x)}{Q(x)} = D_\infty(S\|Q) \geq 0$, with
  strict inequality unless $Q = S$, this establishes \eqref{eqn:regret}
  and $\Qopt = S$. Finally, \eqref{eqn:regretvalue} follows by
  evaluating $\sup_x \ln\frac{\sup_\theta P_\theta(x)}{S(x)}$.
\end{IEEEproof}

We conjecture that the previous result generalizes to any positive order
$\alpha$ as a one-sided inequality:
\begin{conjecture}\label{conj:capacityinequality}
  Let $\alpha \in (0,\infty]$ and suppose that $R_\alpha < \infty$. Then
  we conjecture that there exists a unique redundancy achieving
  distribution
  \begin{equation}
    \Qopt = \argmin_Q \sup_\theta D_\alpha(P_\theta\|Q),
  \end{equation}
  and that for all $Q$
  \begin{equation}\label{eqn:capacityinequality}
    \sup_\theta D_\alpha(P_\theta \|Q)
      \geq R_\alpha + D_\alpha(\Qopt\|Q).
  \end{equation}
\end{conjecture}
This conjecture is reminiscent of Sibson's identity
\cite{Csiszar1995,Sibson1969}. It would imply that any distribution $Q$ that
is close to achieving the minimax redundancy in the sense that
\begin{equation}
  \sup_\theta D_\alpha(P_\theta \|Q)
    \leq R_\alpha + \delta,
\end{equation}
must be close to $\Qopt$ in the sense that
\begin{equation}
  D_\alpha(\Qopt\|Q) \leq \delta.
\end{equation}
As shown in Example~\ref{ex:Qopt} below,
Conjecture~\ref{conj:capacityinequality} does not hold for $\alpha = 0$.
For $\alpha > 0$, it can be expressed as a minimax identity for the
function
\begin{equation}
  \phi_\alpha(R,Q) =
    \sup_{\theta \in \Theta} D_\alpha(P_\theta\|Q) - D_\alpha(R\|Q),
\end{equation}
where we adopt the convention that $\phi_\alpha(R,Q) = \infty$ if both
$\sup_{\theta \in \Theta} D_\alpha(P_\theta\|Q)$ and $D_\alpha(R\|Q)$
are infinite. However, we cannot use Sion's minimax theorem
(Theorem~\ref{thm:Sion}) to prove the conjecture, because in
general $\phi_\alpha$ is not quasi-convex in its second
argument\footnote{We mistakenly claimed this in an earlier draft of this
paper.}.

A distribution $\pi$ on the parameter space $\Theta$ is called a
\emph{barycentric input distribution} if
\begin{equation}
  \Qopt = \int P_\theta \intder \pi(\theta).
\end{equation}

\begin{example}\label{ex:Qopt}
Take $\alpha \in (0,\infty]$ and consider the distributions
\begin{equation}
  P_1 = \big(\nicefrac{1}{2}, 0, \nicefrac{1}{2}\big),
  \qquad P_2 = \big(0, \nicefrac{1}{2}, \nicefrac{1}{2}\big)
\end{equation}
on a three-element set. Then by symmetry and convexity of R\'enyi
divergence in its second argument, there must exist a redundancy
achieving distribution of the form
\begin{equation}
  \Qoptalpha = (q, q, 1-2q).
\end{equation}
If $\alpha$ is a simple order, then for $\theta \in \{1,2\}$ the
divergence is
\begin{align}
  D_\alpha(P_\theta \| &\Qoptalpha)\notag\\
    &= \frac{1}{\alpha-1}
      \ln\Big(
        \big(\nicefrac{1}{2}\big)^\alpha q^{1-\alpha}
        +\big(\nicefrac{1}{2}\big)^\alpha
       (1-2q)^{1-\alpha}\Big)\notag\\
    &= \frac{\alpha\ln2}{1-\alpha}
       +\frac{1}{\alpha-1}\ln
        \big(q^{1-\alpha}+(1-2q)^{1-\alpha}\big).
\end{align}
To find $q$, we therefore we have to extremize
\begin{equation}
  f(q) = q^{1-\alpha} + (1-2q)^{1-\alpha},
\end{equation}
which leads to
\begin{equation}\label{eqn:exqopt}
  q = \frac{1}{2+2^{\nicefrac{1}{\alpha}}}.
\end{equation}
The reader may verify that \eqref{eqn:exqopt} also holds for $\alpha =
1$, giving $\Qoptone=(\frac{1}{4},\frac{1}{4},\frac{1}{2})$,
and for $\alpha = \infty$, leading to
$\Qoptinfty=(\frac{1}{3},\frac{1}{3},\frac{1}{3})$. Note
that only for $\alpha=1$ is $\Qoptalpha$ a convex combination of $P_1$
and $P_2$, with unique barycentric input distribution $\pi =
(\nicehalf,\nicehalf)$.

Finally, consider $\alpha = 0$. In this case
\eqref{eqn:exqopt} still holds, giving $\Qoptzero = (0,0,1)$. Now let $Q =
(\nicehalf,\nicehalf,0)$. Then, for $\theta \in \{1,2\}$, we see that the first
two terms in \eqref{eqn:capacityinequality} are well behaved:
\begin{align*}
  \lim_{\alpha \downarrow 0} \sup_\theta D_\alpha(P_\theta \| Q)
    = \sup_\theta D_0(P_\theta \|Q) = \ln 2,\\
  \lim_{\alpha \downarrow 0} \sup_\theta D_\alpha(P_\theta \| \Qoptalpha)
    = 0
    = \sup_\theta D_0(P_\theta \| \Qoptzero).
\end{align*}
The last term, however, evaluates to $D_0(\Qoptzero\|Q) = \infty$, so we
obtain a counterexample to \eqref{eqn:capacityinequality}. The
difference in behaviour between $\alpha = 0$ and $\alpha > 0$ may be
understood by observing that $\lim_{\alpha \downarrow 0}
D_\alpha(\Qoptalpha\|Q) = \ln 2 \neq D_0(\Qoptzero\|Q)$.
\end{example}

\begin{theorem}\label{thm:piopt}
  Suppose that $\samplespace$ is finite and that there exists a maximum
  likelihood function $\mltheta \colon \samplespace \to \Theta$ (that
  is, $P_\theta(x) \leq P_{\mltheta(x)}(x)$ for all $x \in
  \samplespace$). Then, for $\alpha = \infty$, the distribution
  \begin{equation}
    \piopt(\theta) = S(\{x \mid \mltheta(x) = \theta\})
  \end{equation}
  is a capacity achieving input distribution, where $S$ is as defined in
  \eqref{eqn:shtarkov}.
\end{theorem}

\begin{IEEEproof}
  As $\samplespace$ is finite, there can be at most a finite set
  $\Theta_\samplespace \subset \Theta$ of $\theta$ on which $\piopt(\theta) >
  0$. Hence, for any $Q$,
  \begin{align*}
    \int D_\infty(P_\theta\|Q) \der \piopt(\theta)
      &= \sum_{\theta \in \Theta_\samplespace}
        D_\infty(P_\theta\|Q) \piopt(\theta)\\
      &= \sum_{\theta \in \Theta_\samplespace}
        D_\infty(P_\theta\|Q) \sum_x S(x) \ind_{\{\mltheta(x) = \theta\}}\\
      &= \sum_x S(x) D_\infty(P_{\mltheta(x)}\|Q)\\
      &= \sum_x S(x) \max_y \ln \frac{P_{\mltheta(x)}(y)}{Q(y)}\\
      &\geq \sum_x S(x) \ln \frac{P_{\mltheta(x)}(x)}{Q(x)}\\
      &= D(S\|Q) + \sum_x S(x) \ln \frac{P_{\mltheta(x)}(x)}{S(x)}\\
      &= D(S\|Q) + \sum_x S(x) R_\infty\\
      &= D(S\|Q) + R_\infty.
  \end{align*}
  By taking the infimum over $Q$ on both sides we get
  \begin{equation*}
    \inf_Q \int D_\infty(P_\theta\|Q) \der \piopt(\theta)
      \geq R_\infty.
  \end{equation*}
  Since the reverse inequality is trivial and $R_\infty = C_\infty$, we
  find that $\piopt$ is a capacity achieving input distribution, as
  required.
\end{IEEEproof}

\begin{example}
Let $\theta \in [0,1]$ denote the success probability of a binomial
distribution $P_\theta = \bin(2,\theta)$ on $\samplespace = \{0,1,2\}$.
Then for $\alpha=\infty$ the redundancy achieving distribution is
$S=(\frac{2}{5},\frac{1}{5},\frac{2}{5})$ and the minimax redundancy is
$R_\infty = \ln \frac{5}{2}$.

In this case there are many barycentric input distributions. For
example, the distribution $\pi = \frac{1}{5} M_0 + \frac{3}{5} U +
\frac{1}{5} M_1$ is a barycentric input distribution, where $M_\theta$
is a point-mass on $\theta$ and $U$ is the uniform distribution on
$[0,1]$. Another example is the distribution $\pi =
(\frac{3}{10},\frac{2}{5},\frac{3}{10})$ on the maximum likelihood
parameters $\Psi = \{0,\half,1\}$ for the elements of $\samplespace$.
By Theorem~\ref{thm:piopt}, there also exists a capacity achieving input
distribution $\piopt$, and it is supported on $\Psi$, with probabilities
\begin{equation*}
  \big(\piopt(0),\piopt(\thalf),\piopt(1)\big)
    = \big(S(0),S(1),S(2)\big) = (\tfrac{2}{5},\tfrac{1}{5},\tfrac{2}{5}).
\end{equation*}
\end{example}

\section{Negative Orders}
\label{sec:negativeorders}

Until now we have only discussed R\'{e}nyi divergence of nonnegative
orders. However, using formula \eqref{eqn:commondefinition} for $\alpha
\in (-\infty,0)$ (reading $\frac{q^{1-\alpha}}{p^{-\alpha}}$ for
$p^\alpha q^{1-\alpha}$), it may also be defined for these negative
orders. This definition extends to $\alpha = -\infty$ by
\begin{equation}
D_{-\infty}(P\Vert Q)=\lim_{\alpha\downarrow-\infty}D_{\alpha}(P\Vert
Q).
\end{equation}

According to R\'enyi \cite{Renyi1961}, only positive orders can be
regarded as measures of information, and negative orders indeed seem to
be hardly used in applications. Nevertheless, for completeness we will
also study R\'{e}nyi divergence of negative orders. As will be seen
below, our results for positive orders carry over to the negative
orders, but most properties are reversed. People may have avoided
negative orders because of these reversed properties. Avoiding negative
orders is always possible, because they are related to orders $\alpha>1$
by an extension of skew symmetry:
\begin{lemma}[Skew Symmetry]\label{lem:symmetry} For any
$\alpha \in (-\infty,\infty)$, $\alpha\not \in \{0,1\}$
\begin{equation}\label{eqn:symmetry}
  D_{\alpha}(P\Vert Q)=\frac{\alpha}{1-\alpha}D_{1-\alpha}(Q\Vert P).
\end{equation}
Furthermore
\begin{align}
D_{-\infty}(P\Vert Q)
  &=-D_{\infty}(Q\Vert P)\notag\\
  &=\ln\inf_{A \in \salgebra}\frac{P(A)}{Q(A)}
  =\ln\left(\essentialinf_Q\frac{p}{q}\right),
\end{align}
with the conventions that $\nicefrac{0}{0} = 0$ and $\nicefrac{x}{0} = \infty$ for $x > 0$.
\end{lemma}

\begin{IEEEproof}
  The identity \eqref{eqn:symmetry} follows directly from definitions.
  It implies $D_{-\infty}(P\Vert Q) =-D_{\infty}(Q\Vert P)$, because
  $\frac{\alpha}{1-\alpha}$ tends to $-1$ as $\alpha \to -\infty$. The
  remaining identities follow from the closed-form expressions for
  $D_{\infty}(Q\Vert P)$ in Theorem~\ref{thm:AlphaInfty}.
\end{IEEEproof}

Skew symmetry gives a kind of symmetry between the orders
$\nicefrac{1}{2}+\alpha$ and $\nicefrac{1}{2}-\alpha$. In applications in
physics this symmetry is related to the use of so-called \emph{escort
probabilities} \cite{Naudts2004}.

Whereas the nonnegative orders generally satisfy the same or similar
properties for different values of $\alpha$, the fact that
$\tfrac{\alpha}{1-\alpha}<0$ for $\alpha<0$, implies that properties for
negative orders are often inverted. For example, R\'{e}nyi divergence
for negative orders is nonpositive, concave in its first argument and
upper semi-continuous in the topology of setwise convergence. In
addition, the data processing inequality holds with its inequality
reversed and for $\alpha \in (-\infty,0)$
Theorem~\ref{thm:simplesupfinite} applies with an infimum instead of a
supremum.

Not all properties are inverted, however. Most notably, it does remain true
that R\'enyi divergence is nondecreasing and continuous in $\alpha$ (see
also Figure~\ref{fig:VaryAlpha}):
\begin{theorem}
\label{thm:increasinginordernegative} For $\alpha \in [-\infty,\infty]$,
the R\'enyi divergence $D_{\alpha}(P\Vert Q)$ is nondecreasing in
$\alpha$.
\end{theorem}

\begin{IEEEproof}
For $\alpha< 0$, $D_{\alpha}(P\Vert Q) \leq0$ and for $\alpha\geq0$,
$D_{\alpha}(P\Vert Q) \geq0$, so the divergence for negative orders never
exceeds the divergence for nonnegative orders. The remainder of the proof
follows from Theorem~\ref{thm:increasinginorder} and skew symmetry.
\end{IEEEproof}

\begin{theorem}
\label{thm:negativecontinuity} The R\'enyi divergence $D_{\alpha}(P\|Q)$ is
continuous in $\alpha$ on $\alphaset = \{\alpha \in [-\infty,\infty]\mid0
\leq\alpha\leq1\text{ or } \lvert D_{\alpha}(P\Vert Q)\rvert< \infty\}$.
\end{theorem}

\begin{IEEEproof}
R\'enyi divergence is nondecreasing in $\alpha$, nonnegative for $\alpha\geq0$
and nonpositive for $\alpha< 0$. Therefore the required continuity follows
directly from Theorem~\ref{thm:extendedcontinuity} and skew symmetry, except
for the case
\[
\lim_{\alpha\uparrow0} D_{\alpha}(P\|Q) = D_{0}(P\|Q),
\]
which is required to hold if there exists a value $\beta< 0$ such that
$D_{\beta }(P\|Q) > -\infty$. In this case $D_{1-\beta}(Q\|P) =
\frac{1-\beta}{\beta} D_\beta(P\|Q) < \infty$, which implies: (a) that $Q
\ll P$, so $D_0(P\|Q) = 0$; and (b) that $D(Q\|P) < \infty$ and by
Theorem~\ref{thm:AlphaOne}
\begin{equation*}
  \lim_{\alpha \uparrow 0} D_\alpha(P\|Q)
    = \lim_{\alpha \uparrow 0} \frac{\alpha}{1-\alpha} D_{1-\alpha}(Q\|P)
    = 0 \cdot D(Q\|P) = 0.\qedhere
\end{equation*}
\end{IEEEproof}

\section{Counterexamples}\label{sec:counterexamples}

Some useful properties that are satisfied by other divergences, are not
satisfied by R\'enyi divergence. Here we give counterexamples for a few
important ones.

\subsection{Convexity in $P$ does not hold for $\alpha>1$}
\label{sec:notconvex}

R\'{e}nyi divergence for $\alpha \in (1,\infty)$ is not convex in its first argument.
Consider the following counterexample: let $0<p_{0}<p_{1}<1$ be any two
numbers, and let $p_{\nicefrac{1}{2}} =\frac{p_{0}+p_{1}}{2}$. Let $\varepsilon>0$
be arbitrary, and let $0<q<1$ be small enough that
\[
\max_{i\in\{0,1\}}\frac{(1-p_{i})^{\alpha}(1-q)^{1-\alpha}}{p_{i}^{\alpha
}q^{1-\alpha}}\leq\varepsilon.
\]
Then convexity of $D_{\alpha}$ in its first argument would imply that
\begin{multline*}
\frac{1}{2}\ln\left(  p_{0}^{\alpha}q^{1-\alpha}+(1-p_{0})^{\alpha
}(1-q)^{1-\alpha}\right)\\
  +\frac{1}{2}\ln\left(  p_{1}^{\alpha}q^{1-\alpha
}+(1-p_{1})^{\alpha}(1-q)^{1-\alpha}\right) \\
\geq\ln\left(  p_{\nicefrac{1}{2}}^{\alpha}q^{1-\alpha}+(1-p_{\nicefrac{1}{2}}%
)^{\alpha}(1-q)^{1-\alpha}\right),
\end{multline*}
which implies
\begin{align*}
\frac{1}{2}\ln\big(p_{0}^{\alpha}q^{1-\alpha}(1+\varepsilon)\big)
  +\frac{1}{2}\ln\big(p_{1}^{\alpha}q^{1-\alpha}&(1+\varepsilon)\big)\\
  &\geq\ln\big(  p_{\nicefrac{1}{2}}^{\alpha}q^{1-\alpha}\big)\\
\frac{1}{2}\ln\big(p_{0}^{\alpha}(1+\varepsilon)\big)
  +\frac{1}{2}\ln\big(  p_{1}^{\alpha}(1+\varepsilon)\big)
  &\geq\ln\big(p_{\nicefrac{1}{2}}^{\alpha}\big).
\end{align*}
As this expression holds for all $\varepsilon>0$, we get
\begin{align*}
\frac{1}{2}\ln p_{0}^{\alpha}+\frac{1}{2}\ln p_{1}^{\alpha}  &  \geq\ln
p_{\nicefrac{1}{2}}^{\alpha}\\
\frac{1}{2}\ln p_{0}+\frac{1}{2}\ln p_{1}  &  \geq\ln
\frac{p_0 + p_1}{2},
\end{align*}
which is a contradiction, because the natural logarithm is strictly
concave.

\subsection{R\'{e}nyi divergence is not continuous}

In general the R\'{e}nyi divergence of order $\alpha \in (0,1)$ is not
continuous in the topology of setwise convergence. To construct a
counterexample, let $P_{n}$ denote the
probability distribution on $[0,2\pi]$ with density
$\frac{1+\sin(nx)}{2\pi}$ and let $Q_{n}$ denote the probability
distribution on $[0,2\pi]$ with density $\frac{1-\sin\left(  nx\right)
}{2\pi}$ for $n=1,2,\ldots$ Then $D_{\alpha }(P_{n}\Vert Q_{n}) > 0$
does not depend on $n$, and both $P_{n}$ and $Q_{n}$ converge to the
uniform distribution $U$ on $\left[ 0,2\pi\right]  $ in the topology of
setwise convergence. Consequently,
$\lim_{n\rightarrow\infty}D_{\alpha}\left(  P_{n}\Vert Q_{n}\right)
\neq0=D_{\alpha}\left(  U\Vert U\right)  $, so in general $D_{\alpha}$
is not continuous in the topology of setwise convergence.

\subsection{Not a metric}

Except for the order $\alpha=\nicefrac{1}{2}$, R\'{e}nyi divergence is not
symmetric and cannot be a metric. For $\alpha=\nicefrac{1}{2}$, R\'{e}nyi
divergence is symmetric and by \eqref{eqn:relationHellinger} it locally
behaves like the square of a metric. Therefore one may wonder whether it
actually is the square of a metric itself. Consider the following three
distributions on two points:
\begin{align*}
P  &  =\left(  0,1\right),&
Q  &  =\left(  \nicefrac{1}{2},\nicefrac{1}{2}\right),&
R  &  =\left(  1,0\right).
\end{align*}
Then
\begin{align*}
D_{\nicefrac{1}{2}}\left(P\| Q\right)   &  =\ln2,&
D_{\nicefrac{1}{2}}\left(Q\| R\right)   &  =\ln2,&
D_{\nicefrac{1}{2}}\left(P\| R\right)   &  =\infty.
\end{align*}
As the square roots of these divergences violate the triangle
inequality, $D_{\nicefrac{1}{2}}$ cannot be the square of a metric.

\section{Summary}

We have reviewed and derived the most important properties of R\'{e}nyi
divergence and Kullback-Leibler divergence. These include convexity and
continuity properties, a generalization of the Pythagorean inequality to
general orders, limits of $\sigma$-algebras, additivity for product
distributions on infinite sequences, and the relation of the special
order $0$ to absolute continuity and mutual singularity of such
distributions.

We have also derived several key minimax identities. In particular,
Theorems~\ref{thm:MomentGeneratingFunction} and
\ref{thm:ChernoffInformationDivergence} illuminate the relation between
R\'enyi divergence, Kullback-Leibler divergence and Chernoff information
in hypothesis testing. And Theorem~\ref{thm:radius} extends the known
equivalence of channel capacity and minimax redundancy to continuous
channel inputs (for all orders).

\section*{Acknowledgments}

The authors would like to thank Peter Gr\"unwald, Wouter Koolen and two
anonymous referees for useful comments. Part of the research was done
while both authors were with the Centrum Wiskunde \& Informatica in
Amsterdam, the Netherlands, and while Tim van Erven was with the VU
University, also in Amsterdam. This work was supported in part by NWO
Rubicon grant 680-50-1112.

\bibliographystyle{ieeetr}
\bibliography{renyi}

\begin{thebibliography}{10}

\bibitem{Renyi1961}
A.~R{\'e}nyi, ``On measures of entropy and information,'' in {\em Proceedings
  of the Fourth Berkeley Symposium on Mathematical Statistics and Probability},
  vol.~1, pp.~547--561, 1961.

\bibitem{Harremoes2006}
P.~Harremo{\"e}s, ``Interpretations of {R\'e}nyi entropies and divergences,''
  {\em Physica A: Statistical Mechanics and its Applications}, vol.~365, no.~1,
  pp.~57--62, 2006.

\bibitem{Grunwald2007}
P.~D. Gr{\"u}nwald, {\em The Minimum Description Length Principle}.
\newblock The MIT Press, 2007.

\bibitem{Csiszar1995}
I.~Csisz{\'a}r, ``Generalized cutoff rates and {R\'e}nyi's information
  measures,'' {\em IEEE Transactions on Information Theory}, vol.~41, no.~1,
  pp.~26--34, 1995.

\bibitem{Zhang2006}
T.~Zhang, ``From $\epsilon$-entropy to {KL}-entropy: Analysis of minimum
  information complexity density estimation,'' {\em The Annals of Statistics},
  vol.~34, no.~5, pp.~2180--2210, 2006.

\bibitem{HausslerOpper1997}
D.~Haussler and M.~Opper, ``Mutual information, metric entropy and cumulative
  relative entropy risk,'' {\em The Annals of Statistics}, vol.~25, no.~6,
  pp.~2451--2492, 1997.

\bibitem{VanErven2010}
T.~van Erven, {\em When Data Compression and Statistics Disagree: Two
  Frequentist Challenges for the Minimum Description Length Principle}.
\newblock PhD thesis, Leiden University, 2010.

\bibitem{LeCam1973}
L.~Le{~}Cam, ``Convergence of estimates under dimensionality restrictions,''
  {\em The Annals of Statistics}, vol.~1, no.~1, pp.~38--53, 1973.

\bibitem{Birge1986}
L.~Birg{\'e}, ``On estimating a density using {H}ellinger distance and some
  other strange facts,'' {\em Probability Theory and Related Fields}, vol.~71,
  pp.~271--291, 1986.

\bibitem{VanDeGeer1993}
S.~van~de Geer, ``{H}ellinger-consistency of certain nonparametric maximum
  likelihood estimators,'' {\em The Annals of Statistics}, vol.~21, no.~1,
  pp.~14--44, 1993.

\bibitem{MoralesPardoVajda2000}
D.~Morales, L.~Pardo, and I.~Vajda, ``{R\'e}nyi statistics in directed families
  of exponential experiments,'' {\em Statistics}, vol.~34, pp.~151--174, 2000.

\bibitem{MansourMohriRostamizadeh2009}
Y.~Mansour, M.~Mohri, and A.~Rostamizadeh, ``Multiple source adaptation and the
  {R\'e}nyi divergence,'' in {\em Proceedings of the Twenty-Fifth Conference on
  Uncertainty in Artificial Intelligence (UAI)}, pp.~367--374, 2009.

\bibitem{HeroMaMichelGorman2003}
A.~O. Hero, B.~Ma, O.~Michel, and J.~D. Gorman, ``Alpha-divergence for
  classification, indexing and retrieval (revised),'' Tech. Rep. CSPL-334,
  Communications and Signal Processing Laboratory, The University of Michigan,
  2003.

\bibitem{AczelDaroczy1975}
J.~Acz{\'e}l and Z.~Dar{\'o}czy, {\em On Measures of Information and Their
  Characterizations}.
\newblock Academic Press, 1975.

\bibitem{Ben-BassatRaviv1978}
M.~Ben-Bassat and J.~Raviv, ``{R}enyi's entropy and the probability of error,''
  {\em IEEE Transactions on Information Theory}, vol.~24, no.~3, pp.~324--330,
  1978.

\bibitem{VanErvenHarremoes2010:isit}
T.~van Erven and P.~Harremo{\"e}s, ``R{\'e}nyi divergence and majorization,''
  in {\em Proceedings of the IEEE International Symposium on Information Theory
  (ISIT)}, 2010.

\bibitem{Shayevitz2010}
O.~Shayevitz, ``A note on a characterization of {R}{\'e}nyi measures and its
  relation to composite hypothesis testing.'' arXiv:1012.4401v1, Dec. 2010.

\bibitem{Shayevitz2011}
O.~Shayevitz, ``On {R\'e}nyi measures and hypothesis testing,'' in {\em IEEE
  International Symposium on Information Theory Proceedings}, pp.~800--804,
  2011.

\bibitem{Huzurbazar1955}
V.~S. Huzurbazar, ``Exact forms of some invariants for distributions admitting
  sufficient statistics,'' {\em Biometrika}, vol.~42, no.~3/4, pp.~pp.
  533--537, 1955.

\bibitem{LieseVajda1987}
F.~Liese and I.~Vajda, {\em Convex Statistical Distances}.
\newblock Leipzig: Teubner, 1987.

\bibitem{Gil2011}
M.~Gil, ``On {R\'e}nyi divergence measures for continuous alphabet sources,''
  Master's thesis, Queen's University, 2011.

\bibitem{GilAlajajiLinder2013}
M.~Gil, F.~Alajaji, and T.~Linder, ``{R\'e}nyi divergence measures for commonly
  used univariate continuous distributions,'' {\em Information Sciences},
  vol.~249, pp.~124--131, 2013.

\bibitem{AldousDiaconis1987}
D.~Aldous and P.~Diaconis, ``Strong uniform times and finite random walks,''
  {\em Advances in Applied Mathematics}, vol.~8, pp.~69--97, 1987.

\bibitem{GibbsSu2002}
A.~L. Gibbs and F.~E. Su, ``On choosing and bounding probability metrics,''
  {\em International Statistical Review}, vol.~70, pp.~419--435, 2002.

\bibitem{Gilardoni2010}
G.~L. Gilardoni, ``On {P}insker's and {V}ajda's type inequalities for
  {C}sisz{\'a}r's {$f$}-divergences,'' {\em IEEE Transactions on Information
  Theory}, vol.~56, no.~11, pp.~5377--5386, 2010.

\bibitem{Pollard2002}
D.~Pollard, {\em A User's Guide to Measure Theoretic Probability}.
\newblock Cambridge University Press, 2002.

\bibitem{LieseVajda2006}
F.~Liese and I.~Vajda, ``On divergences and informations in statistics and
  information theory,'' {\em IEEE Transactions on Information Theory}, vol.~52,
  no.~10, pp.~4394--4412, 2006.

\bibitem{Shiryaev1996}
A.~N. Shiryaev, {\em Probability}.
\newblock Springer-Verlag, 1996.

\bibitem{AliSilvey1966}
S.~M. Ali and S.~D. Silvey, ``A general class of coefficients of divergence of
  one distribution from another,'' {\em Journal of the Royal Statistical
  Society, series B}, vol.~28, no.~1, pp.~131--142, 1966.

\bibitem{CoverThomas1991}
T.~M. Cover and J.~A. Thomas, {\em Elements of Information Theory}.
\newblock Wiley, 1991.

\bibitem{Csiszar1975}
I.~Csisz{\'a}r, ``I-divergence geometry of probability distributions and
  minimization problems,'' {\em The Annals of Probability}, vol.~3, no.~1,
  pp.~146--158, 1975.

\bibitem{Topsoe2007}
F.~Tops{\o}e, {\em Entropy, Search, Complexity}, vol.~16 of {\em Bolyai Society
  Mathematical Studies}, ch.~8, Information Theory at the Service of Science,
  pp.~179--207.
\newblock Springer, 2007.

\bibitem{Sundaresan2002}
R.~Sundaresan, ``A measure of discrimination and its geometric properties,'' in
  {\em Proceedings of the IEEE International Symposium on Information Theory
  (ISIT)}, 2002.

\bibitem{Sundaresan2006}
R.~Sundaresan, ``Guessing under source uncertainty with side information,'' in
  {\em Proceedings of the IEEE International Symposium on Information Theory
  (ISIT)}, 2006.

\bibitem{Prokhorov1956}
Y.~V. Prokhorov, ``Convergence of random processes and limit theorems in
  probability theory,'' {\em Theory of Probability and Its Applications},
  vol.~I, no.~2, pp.~157--214, 1956.

\bibitem{Posner1975}
E.~C. Posner, ``Random coding strategies for minimum entropy,'' {\em IEEE
  Transactions on Information Theory}, vol.~21, no.~4, pp.~388--391, 1975.

\bibitem{VanDerVaartWellner1996}
A.~W. van~der Vaart and J.~A. Wellner, {\em Weak Convergence and Empirical
  Processes: With Applications to Statistics}.
\newblock Springer, 1996.
\newblock (Corrected second printing, 2000).

\bibitem{Pinsker1964}
M.~S. Pinsker, {\em Information and Information Stability of Random Variables
  and Processes}.
\newblock Holden-Day, 1964.
\newblock Translated by A. Feinstein.

\bibitem{Barron2000}
A.~R. Barron, ``Limits of information, {M}arkov chains and projections,'' in
  {\em Proceedings of the IEEE International Symposium on Information Theory
  (ISIT)}, p.~25, 2000.

\bibitem{HarremoesHolst2009}
P.~Harremo{\"e}s and K.~K. Holst, ``Convergence of {M}arkov chains in
  information divergence,'' {\em Journal of Theoretical Probability}, vol.~22,
  pp.~186--202, 2009.

\bibitem{Kallenberg1997}
O.~Kallenberg, {\em Foundations of Modern Probability}.
\newblock Springer, 1997.

\bibitem{VanDerVaart1998}
A.~W. van~der Vaart, {\em Asymptotic Statistics}.
\newblock Cambridge University Press, 1998.

\bibitem{Feldman1958}
J.~Feldman, ``Equivalence and perpendicularity of {G}aussian processes,'' {\em
  Pacific Journal of Mathematics}, vol.~8, no.~4, pp.~699--708, 1958.

\bibitem{Hajek1958}
J.~H{\'a}jek, ``On a property of normal distributions of any stochastic
  process,'' {\em Czechoslovak Mathematical Journal}, vol.~8, no.~4,
  pp.~610--618, 1958.
\newblock In Russian with English summary.

\bibitem{Thelen1989}
B.~J. Thelen, ``Fisher information and dichotomies in equivalence/contiguity,''
  {\em The Annals of Probability}, vol.~17, no.~4, pp.~1664--1690, 1989.

\bibitem{Kakutani1948}
S.~Kakutani, ``On equivalence of infinite product measures,'' {\em The Annals
  of Mathematics}, vol.~49, no.~1, pp.~214--224, 1948.

\bibitem{Renyi1967}
A.~R{\'e}nyi, ``On some basic problems of statistics from the point of view of
  information theory,'' in {\em Proceedings of the Fifth Berkeley Symposium on
  Mathematical Statistics and Probability}, vol.~1: Statistics, pp.~531--543,
  1967.

\bibitem{Kullback1959}
S.~Kullback, {\em Information theory and statistics}.
\newblock Wiley, 1959.

\bibitem{Nemetz1974}
T.~Nemetz, ``On the {$\alpha$}-divergence rate for {M}arkov-dependent
  hypotheses,'' {\em Problems of Control and Information Theory}, vol.~3,
  no.~2, pp.~147--155, 1974.

\bibitem{RachedAlajajiCampbell2001}
Z.~Rached, F.~Alajaji, and L.~L. Campbell, ``{R\'e}nyi's divergence and entropy
  rates for finite alphabet {M}arkov sources,'' {\em IEEE Transactions on
  Information Theory}, vol.~47, no.~4, pp.~1553--1561, 2001.

\bibitem{Csiszar2003}
I.~Csisz{\'a}r, ``Information projections revisited,'' {\em IEEE Transactions
  on Information Theory}, vol.~49, no.~6, pp.~1474--1490, 2003.

\bibitem{FedotovHarremoesTopsoe2003}
A.~A. Fedotov, P.~Harremo{\"e}s, and F.~Tops{\o}e, ``Refinements of {P}insker's
  inequality,'' {\em IEEE Transactions on Information Theory}, vol.~49, no.~6,
  pp.~1491--1498, 2003.

\bibitem{Rockafellar1970}
R.~T. Rockafellar, {\em Convex Analysis}.
\newblock Princeton University Press, 1970.

\bibitem{Ryabko1981}
B.~Ryabko, ``Comments on ``a source matching approach to finding minimax
  codes'' by {D}avisson, {L}. {D}. and {L}eon-{G}arcia, {A}.,'' {\em IEEE
  Transactions on Information Theory}, vol.~27, no.~6, pp.~780--781, 1981.
\newblock Including also the ensuing Editor's Note.

\bibitem{Haussler1997}
D.~Haussler, ``A general minimax result for relative entropy,'' {\em IEEE
  Transactions on Information Theory}, vol.~43, no.~4, pp.~1276--1280, 1997.

\bibitem{Sion1958}
M.~Sion, ``On general minimax theorems,'' {\em Pacific Journal of Mathematics},
  vol.~8, no.~1, pp.~171--176, 1958.

\bibitem{Komiya1988}
H.~Komiya, ``Elementary proof for {S}ion's minimax theorem,'' {\em Kodai
  Mathematical Journal}, vol.~11, no.~1, pp.~5--7, 1988.

\bibitem{Shtarkov1987}
Y.~M. Shtar'kov, ``Universal sequential coding of single messages,'' {\em
  Problems of Information Transmission}, vol.~23, no.~3, pp.~175--186, 1987.

\bibitem{Sibson1969}
R.~Sibson, ``Information radius,'' {\em Z. Warscheinlichkeitstheorie verw.
  Geb.}, vol.~14, pp.~149--160, 1969.

\bibitem{Naudts2004}
J.~Naudts, ``Estimators, escort probabilities, and {$\phi$}-exponential
  families in statistical physics,'' {\em Journal of Inequalities in Pure and
  Applied Mathematics}, vol.~5, no.~4, 102, 2004.

\end{thebibliography}

\begin{IEEEbiography}[%
{\includegraphics[width=1in,clip,keepaspectratio]{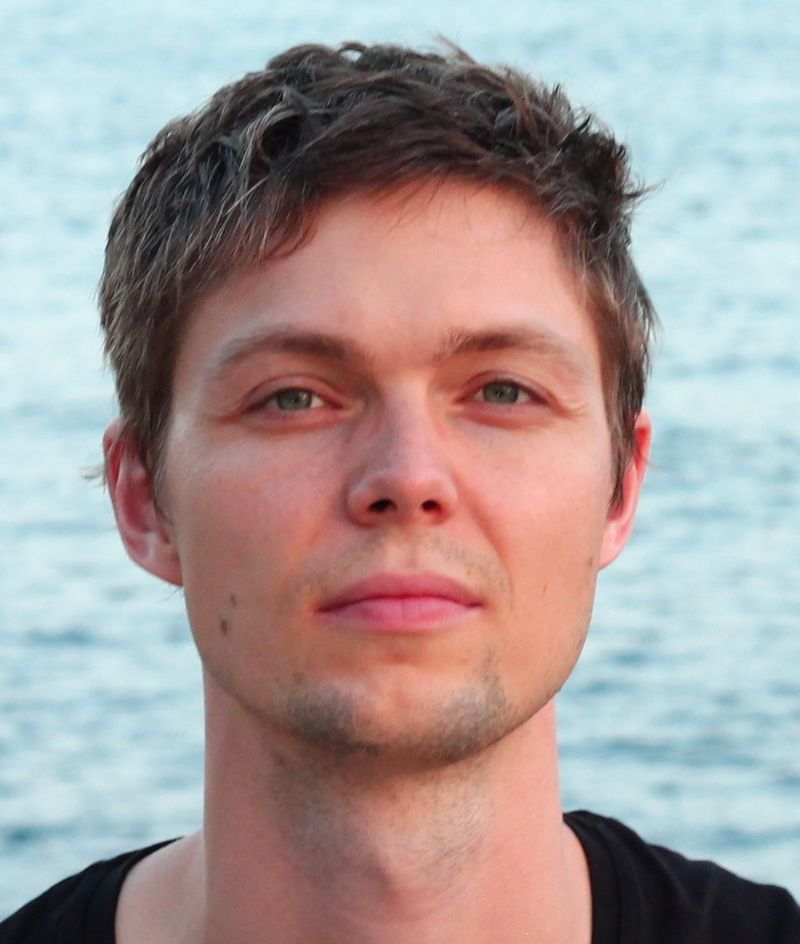}}]%
{Tim van Erven}
  is originally from the Netherlands. He performed his PhD
  research at the Centrum Wiskunde \& Informatica (CWI) in Amsterdam,
  and received his PhD degree from Leiden University in 2010. After
  postdoc positions at the CWI and the Vrije Universiteit in Amsterdam,
  he obtained a Rubicon grant from the Netherlands Organisation for
  Scientific Research (NWO) to do a two-year postdoc at the Universit\'e
  Paris-Sud in France.

  His interests are in topics related to information theory and
  statistics, including minimum description length (MDL) learning,
  sequential prediction with individual sequences (online learning), and
  statistical learning theory. The present paper was partially motivated
  by the importance of R\'enyi divergence in stating sufficient
  conditions for convergence of the MDL estimator, which he studied in
  his PhD thesis \cite[Chapter~5]{VanErven2010}.
\end{IEEEbiography}

\begin{IEEEbiography}[%
%{\includegraphics[width=1in,height=1.25in,clip,keepaspectratio]{photo-PeterHarremoes}}]%
{\includegraphics[width=1in,clip,keepaspectratio]{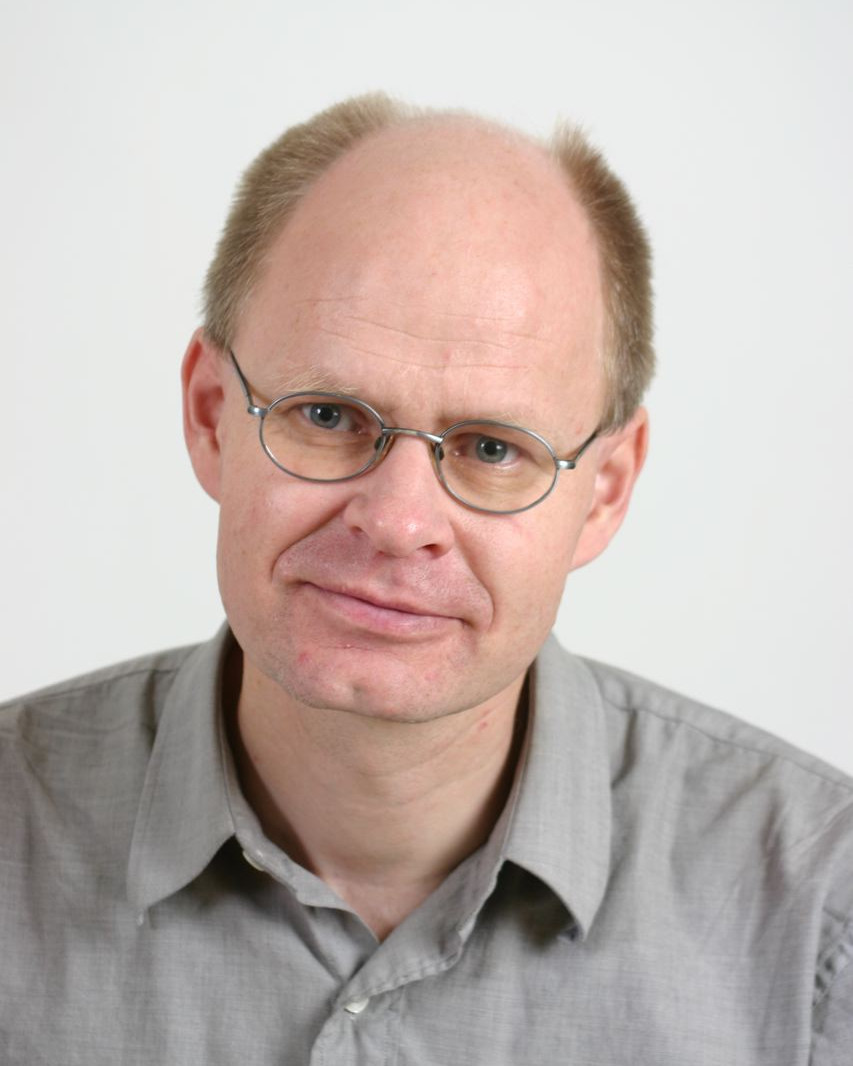}}]%
{Peter Harremo\"es}
Peter Harremo\"es (M'00) received the BSc degree in mathematics in 1984,
the Exam.\ Art.\ degree in archaeology in 1985, and the MSc degree in
mathematics in 1988, all from the University of Copenhagen, Denmark. In
1993 he received his PhD degree in the natural sciences from Roskilde
University, Denmark.

From 1993 to 1998, he worked as a mountaineer. From 1998 to 2000, he held
various teaching positions in mathematics. From 2001 to 2006, he was
Postdoctoral Fellow with the University of Copenhagen, with an extended
visit to the Zentrum f\"ur Interdisziplin\"are Forschung, Bielefeld,
Germany in 2003. From 2006 to 2009, he was affiliated with the Centrum
Wiskunde \& Informatica, Amsterdam, The Netherlands, under the European
Pascal Network of Excellence. Since then he has been affiliated with
Niels Brock, Copenhagen Business College, in Denmark.

From 2007 to 2011 Peter Harremo\"es has been Editor-in-Chief of the
journal \emph{Entropy}. He is currently an editor for that journal.
\end{IEEEbiography}

%\newpage

\vfill


\end{document}